\documentclass[a4paper,11pt]{article}
\pdfoutput=1 % if your are submitting a pdflatex (i.e. if you have
\usepackage{jcappub} % for details on the use of the package, please
\usepackage[T1]{fontenc} % if needed

\usepackage{multirow}

\title{\boldmath Field-level simulation-based inference with galaxy catalogs: the impact of systematic effects}

\author[1,2]{Natalí S. M. de Santi,}
\author[1,3]{Francisco Villaescusa-Navarro,}
\author[2]{L. Raul Abramo}
\author[3,1]{Helen Shao}
\author[1,3]{Lucia A. Perez}
\author[4,5,6,7]{Tiago Castro}
\author[8, 9]{Yueying Ni}
\author[10]{Christopher C. Lovell}
\author[11]{Elena Hernández-Martínez}
\author[12]{Federico Marinacci}
\author[1]{David N. Spergel}
\author[11,13]{Klaus Dolag}
\author[14]{Lars Hernquist}
\author[15, 16]{Mark Vogelsberger}

\affiliation[1]{Center for Computational Astrophysics, Flatiron Institute,\\ 162 5th Avenue, New York, NY, 10010, USA}
\affiliation[2]{Instituto de Física, Universidade de São Paulo,\\ R. do Matão 1371, 05508-900, São Paulo, Brasil}
\affiliation[3]{Department of Astrophysical Sciences, Princeton University,\\ 4 Ivy Lane, Princeton, NJ 08544 USA}
\affiliation[4]{INAF, Osservatorio Astronomico di Trieste,\\ Via G. B. Tiepolo 11, I-34143 Trieste, Italy}
\affiliation[5]{INFN, Sezione di Trieste,\\ I-34100 Trieste, Italy}
\affiliation[6]{IFPU, Institute for Fundamental Physics of the Universe,\\ via Beirut 2, 34151, Trieste, Italy}
\affiliation[7]{ICSC - Italian Research Center on High Performance Computing,\\ Big Data and Quantum Computing, Italy}
\affiliation[8]{Harvard-Smithsonian Center for Astrophysics,\\ 60 Garden Street, Cambridge, MA 02138, US}
\affiliation[9]{McWilliams Center for Cosmology, Department of Physics, Carnegie Mellon University,\\ Pittsburgh, PA 15213, US}
\affiliation[10]{Institute of Cosmology and Gravitation, University of Portsmouth,\\ Burnaby Road, Portsmouth, PO1 3FX, UK}
\affiliation[11]{Universit\"ats-Sternwarte, Fakult\"at f\"ur Physik, Ludwig-Maximilians-Universit\"at M\"unchen,\\ Scheinerstr. 1, 81679 M\"unchen, Germany}
\affiliation[12]{Department of Physics and Astronomy "Augusto Righi", University of Bologna,\\ via Gobetti 93/2, 40129 Bologna, Italy}
\affiliation[13]{Max-Planck-Institut f\"ur Astrophysik,\\ Karl-Schwarzschild-Stra{\ss}e 1, 85741 Garching, Germany}
\affiliation[14]{Harvard-Smithsonian Center for Astrophysics,\\ 60 Garden Street, Cambridge, MA}
\affiliation[15]{Kavli Institute for Astrophysics and Space Research, Department of Physics, MIT,\\ Cambridge, MA 02139, USA}
\affiliation[16]{The NSF AI Institute for Artificial Intelligence and Fundamental Interactions,\\ Massachusetts Institute of Technology, Cambridge MA 02139, USA}

\emailAdd{natalidesanti@gmail.com}

\abstract{It has been recently shown that a powerful way to constrain cosmological parameters from galaxy redshift surveys is to train graph neural networks to perform field-level likelihood-free inference without imposing cuts on scale. 
In particular, \citet{deSanti2023} developed models that could accurately infer the value of $\Omega_{\rm m}$ from catalogs that only contain the positions and radial velocities of galaxies that are robust to different astrophysics and subgrid models. 
However, observations are affected by many effects, including 1) masking, 2) uncertainties in peculiar velocities and radial distances, and 3) different galaxy population selections.
Moreover, observations only allow us to measure redshift, which entangles the galaxy radial positions and velocities. In this paper we train and test our models on galaxy catalogs, created from thousands of state-of-the-art hydrodynamic simulations run with different codes from the CAMELS project, that incorporate these observational effects. 
We find that while such effects degrade the precision and accuracy of the models, the fraction of galaxy catalogs for which the models retain high performance and robustness is over $90\%$, demonstrating the potential for applying them to real data.}


\begin{document}
\maketitle
\flushbottom

%%%%%%%%%%%%%%%%%%%%%%%%%%%%%%%%%%%%%%%%%%%%%%%
% Introduction
%%%%%%%%%%%%%%%%%%%%%%%%%%%%%%%%%%%%%%%%%%%%%%%
%%%
\section{Introduction}
\label{sec:intro}
%%%

Parameter inference is key to probing cosmology and testing our models.
It underlies our attempts to solve puzzles that ranging from the nature of dark energy and dark matter (DM) \citep{Feng2010} to the attempt to addressing some of the outstanding tensions such as the value of the Hubble parameter ($H_0$), or the relation between the matter content ($\Omega_{\rm m}$) and the power spectrum amplitude at $8 h^{- 1}$ Mpc ($\sigma_8$), given by $S_8 = \sigma_8 \left( \Omega_{\rm m}/0.3 \right)^{1/2}$
\citep{Riess2022, DIVALENTINO2021, Planck2014}.
Due to these existing puzzles and tensions, we seek to constrain these cosmological parameters with the highest possible accuracy.

Efforts are being made on the fronts of both observations and simulations to collect large amounts of high-quality data and all the mock data required to analyze them.
In the former front, surveys are observing deeper into the Universe, with better image quality and photometric accuracy--see DESI \citep{DESI},
J-PAS \citep{Benitez2014}, examples of wider surveys, such as Euclid \citep{Euclid2022-Tiago_Castro, Euclid2016} and Roman \citep{Roman2015}, or even going deeper, e.g. JWST \citep{JWST}.

On the front of galaxy simulation, many diverse hydrodynamical codes with unique subgrid physical models for galaxy formation and evolution have been used to generate large cosmological simulations \citep{Vogelsberger2020}. Notable in the context of this work are Astrid \citep{Astrid2022}, SIMBA \citep{SIMBA2019},
IllustrisTNG \citep{Pillepich2018, Vogelsberger2014Nature, Vogelsberger2014MNRAS}, Magneticum \citep{MAGNETICUM2014}, and SWIFT-EAGLE \citep{Schaye2015}.

Additionally, methods such as semi-analytical modeling \citep{Somerville1999} and subhalo abundance matching \citep{Contreras2021, Klypin2015} leverage information in halo catalogs and/or merger histories presenting as faster analogs to hydrodynamic subgrid models to populate DM halos with galaxies. Finally, much work has applied machine learning (ML) techniques to better understand or simulate the halo--galaxy connection \citep{Wu2023, Cristian2023, Rodrigues2023, Lovell2023, deSanti2022MNRAS, Christian2022, Lovell2022, Shao2021, vonMarttens2021, Wadekar2020, Jo2019, Yip2019, Zhang2019, Kamdar2016}.

The question of how to optimally constrain the value of the cosmological parameters remains in active study. Traditionally, observations are compressed into summary statistics for which a likelihood is considered, and parameters are inferred by sampling the likelihood with methods such as Markov-Chain Monte Carlo (MCMC) \cite{MCMC2010}, variational inference (VI) \cite{VI2022} and nested sampling (NS) \cite{NS2006}. 

There are three main problems in this traditional approach. First, it is not known which summary statistic best compresses the information contained within the data \cite{Gualdi2021, Banerjee2021, Hahn2020, Uhlemann2020}.
Second, these methods require numerical estimates of the covariance matrices that demand a large number of computationally expensive simulations 
\cite{CARPool2022, deSanti2022JCAP, Alan2018, Taylor2013}. 
Lastly, parameter estimations are often linked to some particular model, which may not be informative in terms of constraints on the underlying theory.

In light of this, there has been a shift toward applying ML methods to actually infer the full cosmological posteriors.
For instance, it is possible to train neural networks to perform likelihood-free inference on the value of the cosmological parameters from both standard and new summary statistics (such as power spectrum, count-in-cells, and others \cite{Wang2023, lucia2022, Nicola2022}). 

In particular, great progress has come with the development of field-level likelihood-free inference (also called simulation-based inference or implicit likelihood inference \cite{Lemos2023, Cranmer2020}), which do not require summary statistics and the resulting loss of information.
The use of convolutional neural networks with multifields has been achieving competitive results to constrain cosmology
\cite{Paco2022, Paco2021} and even to break the degeneracy of $\sigma_8$ and $\Omega_{\rm m}$, as shown by Reference \cite{Kacprzak2022}.

Even more impressive are the predictions coming directly from galaxy or halo catalogs, whose properties can be converted into graphs and used to feed graph neural networks (GNNs), as showed in References 
\cite{Massara2023, deSanti2023, helen2023, pablo-galaxies-2022, helen-halos-2022, lucas2022, Anagnostidis_2022}.
In the context of cosmology, this kind of analysis does not impose any cut on scale\footnote{Note that we still have the limitation given by the scale resolution of the simulations.} and can easily incorporate different
physical symmetries, as well as already being permutationally invariant.

More specifically, authors from Reference \cite{pablo-galaxies-2022} showed that only using galaxy positions, stellar mass, radius, and
metallicity the GNNs were able to infer $\Omega_{\rm m}$ within $\sim [4, 8] \%$ of accuracy. 
However, their model was not robust: models trained in galaxy catalogs from one simulation did not work when tested on galaxy catalogs created from simulations run with a different subgrid physical model. 
Subsequently, authors from Reference \cite{helen-halos-2022} showed instead that the positions, velocities, and masses of DM halos were enough to
infer $\Omega_{\rm m}$ and $\sigma_8$ with a mean relative error of $\sim 6 \%$. Moreover, their method was robust to
numerics in N-body codes and variations in astrophysical parameters. The only caveat of that method relies on the
fact that it still needs DM halos and some of their properties, which are not directly observable.
In an attempt to yield a physical explanation for the success of the GNNs predictions, authors from Reference \cite{helen2023} provided
analytical equations to predict $\Omega_{\rm m}$, showing the network employs a non-trivial combination of the main components
of all these analyses: the positions and velocity  moduli of DM halos. That model even proved to be robust across
different N-body simulations and, more surprisingly, also worked when using galaxy catalogs from different hydrodynamic simulations, including one additional parameter, different for each subgrid model.
Finally, we have shown in Reference \cite{deSanti2023} a ML suite that was able to infer the value of $\Omega_{\rm m}$ directly from galaxy catalogs with $\sim 12 \%$ of precision, using
only positions and the $z$ component velocity. That work showed that the model was robust (without any additional parameters) to cosmology, astrophysics, and subgrid models, as it was tested using thousands of galaxy catalogs produced from hydrodynamic simulations run with $5$ different codes

In particular, our previous paper \cite{deSanti2023} presented the first steps towards using GNNs as an alternative method to extract
cosmological information from real galaxy observations. 
First, we showed that these models can be trained to robustly infer cosmology independent of the subgrid physical model used to generate the simulated galaxies. 
Second, we provided plausible explanations for what makes our models robust, both in terms of the properties that were employed 
(for instance, positions and velocities -- something already found to be useful in the context of DM halos by Reference \cite{helen-halos-2022}), 
as well as the intrinsic diversity of the data sets used for training the models.
Third, we were able to represent the data as graphs, which are structures that arise naturally from the underlying data (the cosmic web). 

However, while GNNs can be powerful tools, in order to deploy {them on} real data we should take into account several observational and systematic effects. 
The question now becomes: are those methods still robust when applied to realistic data sets that include observational systematics? 

The first systematic we consider is masking: galaxy surveys are not homogeneous, and can be affected by proximity effects, such as those caused by bright stars.
This effect can be addressed by masking (i.e., excluding) the typical regions where those stars are located \citep{Coupon2018, Coupon2009, Heymans2012}. 
The second systematic we consider is the measurement uncertainty in observed galaxy peculiar velocities, with relative uncertainties that can be as high as $\sim 15\%$ 
\citep{SLOAN2022, SLOAN_catalog-2022, Kourkchi2020, Howlett2017}. 
The third effect we investigate is the impact of galaxy selection effects, by imposing criteria based on colors, stellar masses or star formation rate \citep{Madgwick2003, Zehavi2002}. 
One of our goals here is to check whether the method presented in \citet{deSanti2023} is able to handle these systematic effects while keeping its robustness, and how much of the original accuracy and precision are lost as a result.
This is a key milestone before GNNs can be used for inferring cosmology from real galaxy catalogs.

The manuscript is organized as follows: in Section \ref{sec:data}, we present the data set we use to train and test the models and how we implement the systematic effects. In Section \ref{sec:methodology}, we discuss the method of constructing graphs from the galaxy catalogs and how we process them through GNNs.
In Section \ref{sec:results}, we present the main results and, in Section \ref{sec:disc_and_conc}, we discuss and summarize the results of this work.

%%%%%%%%%%%%%%%%%%%%%%%%%%%%%%%%%%%%%%%%%%%%%%%%%%%
% Data
%%%%%%%%%%%%%%%%%%%%%%%%%%%%%%%%%%%%%%%%%%%%%%%%%%% 
%%%
\section{Data}
\label{sec:data}
%%%

In this section, we describe the data set used in this work and the manner employed to simulate the systematic effects. 

\begin{table*}
 \caption{\label{tab:summary} Characteristics of the hydrodynamical simulations used in this work.}
 \begin{center}
 \resizebox{\textwidth}{!}{
  \begin{tabular}{ccccc}
   \hline\hline
   \multirow{3}{*}{\textbf{Model}} & \multirow{3}{*}{\textbf{Usage}} & 
   {\bf Number of} & \textbf{Mean number} & \multirow{3}{*}{{\bf Reference}} \\
   & & {\bf simulations} & \textbf{of galaxies} & \\
   & & {\bf used} & \textbf{per catalog} & \\
   \hline\hline
   Astrid       & Train, validate \& test & 1000(LH) + 27(CV)            & 1114 & \cite{Astrid2022}    \\
   SIMBA        &       Test              & 1000(LH) + 27(CV)            & 1093 & \cite{SIMBA2019}     \\
   IllustrisTNG &       Test              & 1000(LH) + 27(CV) + 2048(SB) & 737  & \cite{Pillepich2018} \\
   Magneticum   &       Test              & 50(LH) + 27(CV)              & 3655 & \cite{MAGNETICUM2014}\\
   SWIFT-EAGLE  &       Test              & 64(LH)                       & 1266 & \cite{EAGLE2015}     \\
   \cline{1-5}
   \end{tabular}}
  \end{center}
\end{table*}

%%%%%%%%%%%%%%%%%%%%%%%%%
\subsection{Simulations}
\label{sec:raw_data}

The galaxy catalogs we are working with were constructed from the output of thousands of hydrodynamic simulations of the Cosmology and
Astrophysics with MachinE Learning Simulations -- CAMELS project\footnote{See the project website in \href{https://www.camel-simulations.org}{https://www.camel-simulations.org} and the most recent documentation in \href{https://camels.readthedocs.io/en/latest/}{https://camels.readthedocs.io/en/latest/}.}
\citep{Ni2023, Villaescusa-Navarro2022-relCAMELS, Paco2021-projCAMELS}. These simulations follow the evolution of $256^3$ DM particles and $256^3$ fluid
elements in periodic boxes of $25~h^{-1}{\rm Mpc}$ from $z = 127$ to $z = 0$. In this work, we focus our attention into the $z=0$ results

The CAMELS simulations can be classified into different sets and suites, depending on how the value of the parameters are organized and what code is used to run them.  In this work, we use $3$ different sets and $5$ suites. Table \ref{tab:summary} presents a summary of the simulations used.

The main characteristics of the different sets are:
\begin{itemize}
 \item {\bf Latin Hypercube (LH).} The boxes in this category have been run with different random seeds and have the value of their
 cosmological and astrophysical parameters arranged in a LH \citep{Paco2021-projCAMELS, LH2016} that spans:
 $\Omega_{\rm m} \in [0.1, 0.5]$ and $\sigma_8 \in [0.6, 1.0]$, $A_{\rm SN1} \in [0.25, 4.00]$,
 $A_{\rm SN2} \in [0.5, 2.0]$, $A_{\rm AGN1} \in [0.25, 4.00]$, and $A_{\rm AGN2} \in [0.5, 2.0]$. $A_{\rm SN}$ and
 $A_{\rm AGN}$ represent the astrophysical parameters that control the efficiency of supernova (SN) and active galactic
 nuclei (AGN) feedback, respectively (see References \cite{Yueying2023, Paco2021-projCAMELS} for a detailed discussion of them, in each different
 subgrid physical model). These simulations have been used for training, validating, and testing the GNNs.
 
 \item {\bf Cosmic Variance (CV).} These simulations are designed to quantify the effect of cosmic variance. They have been
 run with the same fiducial values for the cosmological and astrophysical parameters, varying only the initial conditions,
 i.e., having different initial random seeds. We have used these simulations for testing the models.

 \item {\bf Sobol Sequence (SB).} The catalogs in this set have their cosmological and astrophysical parameters being
 varied following a Sobol sequence. A total of $28$ parameters are varied: $5$ cosmological
 ($\Omega_{\rm m}$, $\Omega_{\rm b}$, $h$, $n_s$, $\sigma_8$) and $23$ astrophysical. The astrophysical parameters include
 the usual ones ($A_{\rm SN1}, A_{\rm SN2}, A_{\rm AGN1}, A_{\rm AGN2}$) and incorporate many others, responsible for
 controlling star formation effects, galactic winds, black hole (BH) growth, and quasar parameters. More details can be found
 in Reference \cite{Yueying2023}. We use these simulations for testing the models.
\end{itemize}

The CAMELS simulations can also be classified into suites, according to the code used to run them:

\begin{itemize}

 \item {\bf Astrid.} The code used to run these simulations was MP-Gadget \cite{MPGadget}, applying some modifications to
 the subgrid model employed in the Astrid simulation \citep{Yueying2023, Astrid-Y-2022, Astrid2022}. This set contains
 $1000$ LH and $27$ CV simulations.

 \item {\bf SIMBA.} The code used to run these simulations was \textsc{Gizmo} \cite{Hopkins2015}, employing the same subgrid
 physics as the SIMBA simulation \citep{SIMBA2019}. This subset contains $1000$ LH and $27$ CV simulations.

 \item {\bf IllustrisTNG.} The code used to run these simulations was \textsc{Arepo} \citep{Weinberger2020, springel2010},
 applying the same subgrid physics as the IllustrisTNG simulations \citep{Pillepich2018, Weinberger2016}. This suite
 contains $1000$ LH, $27$ CV, and $2048$ SB simulations. 
    
 \item {\bf Magneticum.} The code used to run these simulations was the parallel cosmological Tree-PM code P-Gadget3
 \citep{GADGET2}, following prescriptions of
 \cite{Tornatore2007, Springel2005MNRAS, DiMatteo2005, Springel2002, Springel2003MNRAS}, modified according to
 \cite{Steinborn2016MNRAS, Hirschmann2014MNRAS, Fabjan2011MNRAS, Dolag2006MNRAS, Dolag2005MNRAS, Dolag2004ApJ}.
 This set contains $50$ LH and $27$ CV simulations.
 
 \item {\bf SWIFT-EAGLE.} The code used to run these simulations was \textsc{Swift} \citep{SWIFT2023, Schaller2018, Schaller2016},
 using a new subgrid physics model based on the original Gadget-EAGLE simulations \citep{EAGLE2015, Crain2015}, with some
 parameter changes \citep{EAGLE-Borrow2022}. This subset contains $64$ LH simulations.
\end{itemize}
See Reference \cite{deSanti2023} for more details about Magneticum and SWIFT-EAGLE suites

We emphasize that the astrophysical parameters ($A_{\rm SN1}, A_{\rm SN2}, A_{\rm AGN1}, A_{\rm AGN2}$) taken into account
across the different subgrid physical models, despite their names, are different from simulation to simulation
\citep{Yueying2023, Paco2021-projCAMELS}. The main effect due to these differences is that the galaxy properties, as well
as their clustering properties, can be very distinct \citep{deSanti2023, lucia2022}.
However, by training on the LH subset, the models we build should be insensitive to the particular values of these parameters within each simulation.

%%%%%%%%%%%%%%%%%%%%%%
\subsection{Galaxy catalogs}
\label{sec:gal_cat}

Halos and subhalos were identified using $2$ different codes: \textsc{SubFind} \citep{Dolag2009, Springel2001}, for Astrid, SIMBA, IllustrisTNG, and Magneticum and \textsc{VELOCIraptor} \citep{velociraptop1, velociraptor2}, for SWIFT-EAGLE.
The most important difference in these finders is the use of the velocities of the DM particles, which are taken into account
by \textsc{VELOCIraptor}, together with their respective positions.
We aim to construct a model that remains robust across various methods for identifying halos and subhalos. 
Then, after training the ML algorithm in one suite which uses \textsc{SubFind}, we intend to test it on the SWIFT-EAGLE catalogs to meet this requirement.
The main impact of different choices for finders ends up in some changes in the number of galaxies, as shown in Reference \cite{fight-halo_finders-2022}, and which the first version of the model has already been proved to overcome
\citep{deSanti2023}. Galaxies are defined as subhalos that have
stellar masses above $1.30 \times 10^8 M_\odot/h$, following the same prescription as in Reference \cite{deSanti2023}.

A summary of the different data sets used in this work can be found in Table \ref{tab:summary}. We present characteristics
related to the usage of the subset, their number of catalogs, the mean number of galaxies per catalog, and the reference
for each of the original galaxy formation model. All of the presented numbers follows for a stellar mass cut of 
$M_{\star} = 1.95 \times 10^8 ~M_\odot/h$, and do not consider any of the selections which we will present in Section 
\ref{sec:observational_effects}.

%%%%%%%%%%%%%%%%%%%%%%
\subsection{Observational effects}
\label{sec:observational_effects}

We now describe the different observational/systematic effects that we consider, and how we simulate them. 

\begin{itemize}
 \item {\bf Masking.} In real surveys, some fraction of the galaxies are masked out due to a variety of reasons: bright stars, cosmic rays, bad pixels, etc.
 When we train a ML suite considering the entire sample of objects in the simulations, we are capturing all the information from the cosmic web. In contrast, if we deploy that machinery on masked catalogs, some of that information is erased, and we may end up with less trustworthy predictions.
 Here we simulate the effect of masks by randomly removing some percentage of the galaxies in the catalog.
 Since the fraction of survey areas which end up being masked out is typically below $10\%$ of the footprint
 \citep{Coupon2018, Heymans2012, Coupon2009}, we simulated masks that eliminate 5\% and 10\% of the galaxies in our samples.
 This mask is a very simple model for selection and incompleteness of a $z = 0$ galaxy catalog. 
 The impact of more realistic masks, which should depend on the details of the survey at hand and may include a dependence on angular positions, redshifts, colors and magnitudes, are a topic for future investigation.

 \item {\bf Peculiar velocity uncertainties.} 
 Peculiar velocity cannot be precisely measured, since observations are unable to distinguish between radial (line-of-sight) positions and radial velocities
 \citep{SLOAN_catalog-2022, SLOAN2022, Kourkchi2020, Howlett2017, Tonry1988, Tully1977, Hubble1929}.
 In our previous work \citep{deSanti2023} we built a robust model on the basis of exact values for the $3$D positions and velocities of all the galaxies. 
 However, this phase space information can become blurred or even biased if those positions and velocities are affected by measurement errors, which could then lead to inaccurate predictions.
 We simulate this effect by adding a random error to the line-of-sight peculiar velocity of each galaxy, $v_z$, in a catalog. This error is added in $2$ different ways:  
  \begin{itemize}
   \item {\bf Absolute error:}
     \begin{equation}
       v_z \rightarrow v_z + \mathcal{N} (\mu,\sigma) ,
     \end{equation}
   where $\mathcal{N}(\mu,\sigma)$ is a Gaussian distribution with mean $\mu$ and standard deviation $\sigma$. We use $\mu=0$, and
   both a low  ($\sigma=100$~km/s) as well as a large  ($\sigma=150$~km/s) uncertainty in the velocities.
   The magnitude of these velocity errors are commensurate with model observational uncertainties related to, e.g., the separation of the peculiar velocities from the Hubble flow, considering the typical velocity dispersion of galaxies inside groups and clusters, which are of order $\sim 300-500$km$/s$ \cite{Bahcall1996}.
   \item {\bf Relative error:}    
     \begin{equation}
       v_z \rightarrow v_z \left[ 1 + P  \mathcal{N} \left( 0, 1 \right) \right] ,
     \end{equation}
   where we consider $P = 0.15$ and $P = 0.25$, representing relative errors on the peculiar velocities of 15\% and 25\%,
   respectively.
   The idea is that these velocity errors come from uncertainties in the redshifts of the galaxies. 
   For a galaxy with peculiar velocity of $200$ km/s, this amounts to
     $30-50$ km$/s$, which is comparable to the intrinsic error of spectroscopic surveys such as DESI for galaxies and quasars \citep{Lan2023}.
  \end{itemize}

 \item {\bf Line-of-sight distance uncertainties.} 
 In galaxy redshift surveys the radial components of the peculiar velocities are degenerate with the radial positions of the galaxies.
 We account for this observational constraint by removing the line-of-sight component from the position vector -- i.e., we project the galaxies onto a $2$D plane.
 The main modification brought about by this particular test is that, compared with our previous work \citep{deSanti2023}, the graphs are now in $2$ dimensions for the edge indices. 
 More details are given in Appendix
 \ref{sec:GNN_and_graph_details}.

 \item {\bf Galaxy selection.} 
 In real surveys, galaxies are selected according to some criteria: e.g., objects brighter than some threshold, and/or those whose colors lie within certain ranges. 
 Moreover, some of these criteria are correlated with the  clustering and environmental properties (e.g. galaxy colors are related to their positions inside the halos).
 By folding these selection effects into the training and testing processes we are able to estimate their impact on our predictions for parameters such as $\Omega_{\rm m}$.
 We simulate galaxy selection in our catalogs by means of $2$ different criteria:
 \begin{itemize}
     \item \textbf{Color}. 
     Since some of the hydrodynamical simulations in CAMELS do not contain galaxy magnitudes (these properties are only available for the IllustrisTNG suite), we employ a selection based on the ``quenched'' and ``not-quenched'' galaxies.
     This definition derives from the values of the specific star formation rate (sSFR = SFR$/\mathrm{M}_{\star}$ $[\mathrm{yr}^{- 1} \mathrm{M}_{\odot}]$, where a galaxy's SFR is defined as the sum of the individual SFR of all gas cells in its subhalo), according to Reference \cite{SIMBA2019}\footnote{We have checked that this choice is similar to the color bi-modality, following Reference \cite{color_Illustris-2018} for IllustrisTNG CV boxes. For a complete correspondence on IllustrisTNG color and SFR selection, see Reference \cite{Donnari2019}.}, i.e.:
  \begin{itemize}
   \item {\bf Blue:} $sSFR < 10^{- 10.8} ~ \mathrm{yr}^{- 1} \, \mathrm{M}_{\odot}$, 
   \item {\bf Red:} $sSFR > 10^{- 10.8} ~ \mathrm{yr}^{- 1} \, \mathrm{M}_{\odot}$. 
 \end{itemize}  
     \item \textbf{Star formation rate}. The second criterion we use is based on the galaxy's SFR, where we define:
  \begin{itemize}
   \item {\bf Star forming:} $SFR > 0$
   \item {\bf Non star-forming:} $SFR = 0$
  \end{itemize}
 \end{itemize}
\end{itemize}

In all these scenarios we are modifying the CAMELS catalogs in order to include these effects. Thus, we include them in all training, validation, and testing catalogs. We emphasize that the above effects do not represent all possible systematic effects that appear in real surveys \cite{Andersen2016}.
More realistic systematics will be analyzed in future work, with mocks/simulations informed by real survey characteristics.

%%%%%%%%%%%%%%%%%%%%%%%%%%%%%%%%%%%%%%%%%%%%
\section{Methodology}
\label{sec:methodology}
%%%%%%%%%%%%%%%%%%%%%%%%%%%%%%%%%%%%%%%%%%%%

This section is devoted to describing: (1) how we built the graphs from the galaxy catalogs (Section \ref{sec:the_graph}); (2) the
details behind the architecture of the GNNs (Section \ref{sec:architecture}); (3) the use of the moment neural networks (MNN) to do the
likelihood-free inference (Section \ref{sec:free_likelihood_inference}); (4) the training procedure and optimization choices
(Section \ref{sec:train_optm}); and (5) the evaluation of the models, where we present the metrics we analyze
(Section \ref{sec:scores}).

%%%%%%%%%%%%%%%%%%%%%%%%%%%%%%%%%%%%%%%%%%%%
\subsection{Galaxy graphs}
\label{sec:the_graph}

Graphs are mathematical units composed by nodes ($\mathbf{n}_i$), edges ($\mathbf{e}_{i j}$, connecting a node $i$ to a node
$j$), and global properties ($\mathbf{g}$), each one of them characterized by a set of properties, usually called as
attributes \citep{Zhou2018, Battaglia2018, Gilmer2017}. The graphs are the input for the GNNs, not the galaxies per se.
Then, we built the graphs from galaxy catalogs, using the galaxy positions (to find the edges and the edge properties); their
peculiar velocities (only the $z$ component, and transforming it according to
$v_z \rightarrow \mathrm{sign} (v_z) \times \log_{10} \left[ 1 + \mathrm{abs} (v_z) \right]$), as node attributes;
and the logarithm of the number of galaxies in the graph: $\log_{10} ( N_g )$, as global attribute.
This was done as presented in Reference \cite{deSanti2023}, where we link galaxies if they are close
enough to be inside a radius centered on each of them, $r_{link}$. The value of the linking radius was a tuned hyperparameter, as we will describe in Section \ref{sec:train_optm}.
Also, we are following the GNNs method presented in Reference \cite{pablo-galaxies-2022} (and used in References \cite{Massara2023, helen2023, helen-halos-2022, lucas2022} for halos, and in Reference \cite{deSanti2023} for galaxies). 

The edge attributes are related to the spatial distribution of galaxies (their positions). We choose these features to make the graph invariant under rotations and translations. The edge features are:
\begin{equation}
 \mathbf{e}_{i j} = \left[ \frac{|\mathbf{d}_{i j}|}{r_{link}}, \alpha_{i j}, \beta_{i j} \right] \, ,
\end{equation}
where:
\begin{align}
 \mathbf{d}_{i j} & = \mathbf{r}_{i} - \mathbf{r}_{j} ,\\
 \pmb{\delta}_{i} & = \mathbf{r}_{i} - \mathbf{c} ,\\
 \alpha_{i j} & = \frac{ \pmb{\delta}_{i}}{ |\pmb{\delta}_{i}| } \cdot \frac{ \pmb{\delta}_{j}}{ |\pmb{\delta}_{j}| },\\   
 \beta_{i j} & = \frac{ {\pmb \delta}_{i}}{ |{\pmb\delta}_{i}| } \cdot \frac{ \mathbf{d}_{i j}}{ |\mathbf{d}_{i j}| } \, .
\end{align}
Here, $\mathbf{r}_{i}$ is the position of a galaxy $i$ and $\mathbf{c}=\sum_i^N \mathbf{r}_i/N$ being the catalog centroid.
The {\em distance} $\mathbf{d}_{i j}$ is the difference of $2$ galaxy ($i$ and $j$) positions, the {\em difference vector}
$\pmb{\delta}_{i}$ denotes the position of a galaxy $i$ with respect to the centroid, $\alpha_{i j}$
is the (cosine of) the angle between the difference vectors of $2$ galaxies, while $\beta_{i j}$ represents the angle between
the difference vector of a galaxy $i$ and its distance to another galaxy $j$. 
We account for periodic boundary conditions (PBC) when computing distances, angles, and reverse edges
\cite{deSanti2023, pablo-galaxies-2022}.

When considering errors in the line-of-sight distances (see line-of-sight distance uncertainties, in Section \ref{sec:observational_effects}), the galaxies only have $x$- and $y$- positions; thus we build $2$D graphs instead of $3$D graphs. Even removing one dimension of the galaxies we keep
all the other graph attributes, preserving the rotational and translational symmetries 
as done in the $3$D version. More details about them, and about
all the graphs in general, are presented in Appendix \ref{sec:GNN_and_graph_details}.

We normalize the graph labels ($\Omega_{\rm m}$ values), as
\begin{equation}
 \Omega_{\rm m, i} \rightarrow \frac{ \left( \Omega_{\rm m, i} - \Omega_{\rm m, min} \right) }{ \left( \Omega_{\rm m, max} - \Omega_{\rm m, min} \right) } ,   
\end{equation}
where $\Omega_{\rm m, min}=0.1$ and $\Omega_{\rm m, max}=0.5$ represent the minimum and the maximum values.

%%%%%%%%%%%%%%%%%%%%%%%%%%%%%%%%%%%%%%%%%%%%
\subsection{GNNs architecture}
\label{sec:architecture}

The architecture we employ here is the same as that implemented in Reference \cite{deSanti2023} and closely
follows the one presented in \textsc{CosmoGraphNet}\footnote{Available on \textsc{Github} repository \href{https://github.com/PabloVD/CosmoGraphNet}{https://github.com/PabloVD/CosmoGraphNet}, DOI: \href{https://doi.org/10.5281/zenodo.6485804}{10.5281/zenodo.6485804}.} \citep{pablo-galaxies-2022}. 
Primarily, the GNNs are associated with a MNN and trained to infer the value of $\Omega_{\rm m}$ for any
input graph.
This task is done by transforming the graph attributes (node $\mathbf{n}_i$ and edge $\mathbf{e}_{i j}$ properties are updated
using the information contained in the neighboring nodes $\mathbf{n}_i$, edges $\mathbf{e}_{i j}$, and global $\mathbf{g}$
characteristics), while the graph structure (edge indexes) is preserved. Then, the updated graph information is compressed
and converted by a usual multi-layer perceptron (MLP) to constrain $\Omega_{\rm m}$, employing the moment neural loss function \cite{Jeffrey2020}. Also, by construction, GNNs preserve the graph symmetries, i.e. keep the permutational invariance \citep{Bronstein2021, Battaglia2018, Gilmer2017} and the edge attributes consider translational and rotational symmetries, accounting for
PBC too.

The architecture used in the GNNs is called {\em message passing scheme}, where each layer is responsible for updating the node
and edge features, taking as input the graph and delivering as output its updated version. We can write the way that the
node and edge features at layer $\ell + 1$ are updated as:

\begin{itemize}

 \item {\bf Edge model:}
  \begin{equation}
   \mathbf{e}_{i j}^{(\ell + 1)} = \mathcal{E}^{(\ell + 1)} \left( \left[ \mathbf{n}_{i}^{(\ell)}, 
   \mathbf{n}_{j}^{(\ell)}, \mathbf{e}_{i j}^{(\ell)} \right]  \right) ,
   \label{eq:edge_model}
  \end{equation}
  being $\mathcal{E}^{(\ell + 1)}$ a MLP;
 
 \item {\bf Node model:}
  \begin{equation}
   \mathbf{n}_{i}^{(\ell + 1)} = \mathcal{N}^{(\ell + 1)} \left( \left[ \mathbf{n}_{i}^{(\ell)}, 
   \bigoplus_{j \in \mathfrak{N}_i} \mathbf{e}_{i j}^{(\ell + 1)}, \mathbf{g} \right]  \right) ,
   \label{eq:node_model}
  \end{equation}
  with $\mathfrak{N}_i$ representing all neighbors of node $i$, $\mathcal{N}^{(\ell + 1)}$ a MLP, and $\oplus$ a
  multi-pooling operation \citep{Corso2020} responsible to concatenate several permutation invariant operations:
  \begin{equation}
   \bigoplus_{j \in \mathfrak{N}_i} \mathbf{e}_{i j}^{(\ell + 1)} = \left[ 
   \max_{j \in \mathfrak{N}_i} \mathbf{e}_{i j}^{(\ell + 1)},
   \sum_{j \in \mathfrak{N}_i} \mathbf{e}_{i j}^{(\ell + 1)}, 
    \frac{ \sum_{j \in \mathfrak{N}_i} \mathbf{e}_{i j}^{(\ell + 1)} }{ \sum_{j \in \mathfrak{N}_i} }
   \right] .  \label{eq:multi-pooling}
  \end{equation}
\end{itemize}
Note that the number of layers to perform this update is a hyperparameter to be chosen in the optimization scheme. 
We also made use of residual layers in the intermediate layers \citep{deSanti2023, pablo-galaxies-2022, Li2017}.

The updated version of the initial graph, after $N$ message passing layers, is collapsed it into a $1$D feature vector
according to
\begin{equation}
  \mathbf{y} = \mathcal{F} \left( \left[ \bigoplus_{i \in \mathfrak{F}} \mathbf{n}_i^N, \mathbf{g}
  \right] \right) , \label{eq:last_layer}
\end{equation}
where $\mathcal{F}$ is the last MLP, $\oplus_{i \in \mathfrak{F}}$ the last multi-pooling operation (operating over all nodes in
the graph $\mathfrak{F}$), and $\mathbf{y}$ the target of the GNNs (i.e. $\Omega_{\rm m}$).

The MLPs used in each update block are composed of a series of fully connected layers, using ReLU activation functions (except for the last layer). The number of layers in the MLPs, the number of neurons per layer, the weight decay, and the
learning rate were considered as hyperparameters. All the architectures presented in this work use
\textsc{PyTorch Geometric} \cite{pytorch-geometric}.

%%%%%%%%%%%%%%%%%%%%%%%%%%%%%%%%%%%%%%%%%%%%
\subsection{Likelihood-free inference and the loss function}
\label{sec:free_likelihood_inference}

The final product of the ML models presented is the inference of $\Omega_{\rm m}$, making use of moment neural networks (MNN) by predicting the marginal
posterior mean $\mu_i$ and standard deviation $\sigma_i$:
\begin{equation}
 \mathbf{y}_i(\mathcal{G}) = [\mu_i(\mathcal{G}), \sigma_i(\mathcal{G})] ,
\end{equation}
with
\begin{align}
 \mu_i ( \mathcal{G} ) & = \int_{ \Omega_{\rm m}^i } d \Omega_{\rm m}^i ~ \Omega_{\rm m}^i ~ p (\Omega_{\rm m}^i | \mathcal{G} ) \\
 \sigma^2_i ( \mathcal{G} ) & = \int_{ \Omega_{\rm m}^i } d \Omega_{\rm m}^i ~ ( \Omega_{\rm m}^i - \mu_i)^2 ~ p (\Omega_{\rm m}^i | \mathcal{G} ) \, .
\end{align}
In this notation, $\mathcal{G}$ represents the input graph and $p (\Omega_{\rm m}^i | \mathcal{G} )$ the marginal posterior, taken
according to
\begin{equation}
 p (\Omega_{\rm m}^i | \mathcal{G} ) = \int_{ \Omega_{\rm m}^i } \prod_{i = 1}^{n} d \Omega_{\rm m}^i ~ p (\Omega_{\rm m}^1, \Omega_{\rm m}^2, \dots, \Omega_{\rm m}^n | \mathcal{G} ) .
\end{equation}
These predictions are done without making any assumption about the form of the posterior, using a specific loss function
according to \cite{Jeffrey2020}, designed to infer the first $2$ moments of the posterior:
\begin{equation}
 \mathcal{L} = 
 \log \left[ \sum_{ i \in \mathrm{batch} } \left( \Omega_{\rm m}^{i} - \mu_{i} \right)^2 \right] +  \log \left\{ \sum_{ i \in \mathrm{batch} } \left[ 
 \left( \Omega_{\rm m}^{i} - \mu_{i} \right)^2 - \sigma_{i}^2 \right]^2 \right\} , \label{eq:loss}
\end{equation}
where $i$ represents the samples in a given batch. 
We take the logarithms of the differences according to Reference \cite{Paco2022}.
We stress that the best evaluation of the method makes use of the true values of $\Omega_{\rm m}$, in order to access the quality of the predictions for both $\mu$ and $\sigma$.

The error of the model is $\sigma_i$. This error only represents the aleatoric error, and
does not include the epistemic one (i.e.\ the error intrinsically related to the ML model). We do not report the magnitude
of the epistemic errors because they correspond to $\sim 10 \%$ of the aleatoric one.

%%%%%%%%%%%%%%%%%%%%%%%%%%%%%%%%%%%%%%%%%%%%
\subsection{Training procedure and optimization}
\label{sec:train_optm}

Authors from References \cite{deSanti2023} and \cite{Yueying2023} found
that the ML models trained on Astrid were precise and robust because the Astrid simulations encompass a diverse range in some galaxy properties and galaxy number density.
For this reason, all the models presented in this paper were trained on graphs from the LH set 
of the Astrid simulations.

The division of the $1000$ LH simulations was: $850$ for training, $100$ boxes for validation, and $50$ boxes for testing (all of these subsets were built considering the observational effects). 
Besides, we have employed a data augmentation in the way we have copied each simulation box into $10$ different catalogs with
galaxies constrained by a stellar masses cut of $1.3 R \times 10^8~M_\odot/h$, where $R$ is a random number uniformly
distributed between $1$ and $2$. This trick allows us to marginalize over different minimum threshold values for stellar
masses, and it was also used in Reference \cite{deSanti2023} and a similar strategy was taken by References 
\cite{helen2023, helen-halos-2022}. We also tried to marginalize over the sSFR cut, to select the galaxies by color, but the
results were similarly accurate and precise as the ones only considering the cut on stellar mass.

Other details related to the implementation are: the number of epochs ($300$), the Adam optimizer \cite{Adam}, and the batch size of $25$. The optimization of the hyperparameters (learning rate, weight decay, linking radius,
number of message passing layers, and number of hidden channels per layer of the MLPs) was done using \textsc{Optuna}
\cite{optuna_2019}. We have performed a Bayesian optimization with Tree Parzen Estimator (TPE) \cite{Bergstra2011}, in order to find the best set of values. 
We also used at least $100$ trials to sample the hyperparameter space, minimizing the validation loss, computed using an early-stopping scheme. Then, we save only the model with the minimum validation error and use it for robustness tests.

Note that the value for the linking radius was found, most of the time, equal
to $r_{link} \sim 1.25$ $h^{- 1}$ Mpc, as in the previous work \citep{deSanti2023}. We present
different values found for some models' linking radii in Table 
\ref{tab:linking_radius}, see Appendix \ref{sec:GNN_and_graph_details}.

%%%%%%%%%%%%%%%%%%%%%%%%%%%%%%%%%%%%%%%%%%%%
\subsection{Performance Metrics}
\label{sec:scores}

The scores used to validate each model are presented in this section. They quantify the accuracy and precision of the network's predictions. For each graph, $i$, the models output $\mu_i$ and $\sigma_i$, the mean and standard deviation of the posterior for $\Omega_{\rm m}$ where the true value is $\Omega_{\rm m}^i$:

\begin{itemize}

 \item {\bf Root Mean Squared Error (RMSE):}
  \begin{equation}
   \text{RMSE} = \sqrt{ \frac{1}{N} \sum_{i = 1}^N 
   \left( \Omega_{\rm m}^i - \mu_i \right)^2 } . \label{eq:RMSE}
  \end{equation}
 Low values are preferable.

 \item {\bf Coefficient of determination:}
  \begin{equation}
    R^2 = 1 - \frac{ \sum_{i = 1}^N \left( \Omega_{\rm m}^i - \mu_i \right)^2 }{ 
    \sum_{i = 1}^N \left( \Omega_{\rm m}^i - \bar{\Omega}_{\rm m}^i \right)^2 } , \label{eq:R2} 
  \end{equation}
  with $\bar{\Omega}_{\rm m}^i = \frac{1}{N} \sum_{i = 1}^N \Omega_{\rm m}^i$. Values closer to $1$ indicate a better model.

 \item {\bf Pearson Correlation Coefficient (PCC):}
  \begin{equation}
   \label{eq:PCC}
   \rm{PCC} = \frac{\rm{cov} \left( \Omega_{\rm m}, \mu \right)}
       {\sigma_{\Omega_{\rm m}} \sigma_\mu} .
  \end{equation}
  Measures the positive/negative linear relationship between truth values and inferences. Well correlated predictions are close to $\pm 1$, and poor correlations are close to $0$.

 \item {\bf Bias:}
  \begin{equation}
   b = \frac{1}{N} \sum_{i = 1}^N ( \Omega_{\rm m}^i - \mu_i ) .
  \end{equation}
  This number quantifies how biased the predictions are with respect to the truth. Values close to $0$ indicate a less biased model.
 
 \item {\bf Mean relative error:}
  \begin{equation}
  \label{Eq:eps}
   \epsilon = \frac{1}{N} \sum_{i = 1}^N \frac{ | \Omega_{\rm m}^i - \mu_i | }{ \mu_i } .
  \end{equation}
  Low values indicate the model is precise.
  
 \item {\bf Reduced chi squared:}
  \begin{equation}
   \chi^2 = \frac{1}{N} \sum_{i = 1}^N \left( \frac{\Omega_{\rm m}^i - \mu_i }{\sigma_i} \right)^2  .  \label{eq:reduced}
  \end{equation}
  This statistic quantifies the accuracy of the estimated errors.
  Values of $\chi^2$ close to $1$ indicate that the errors are accurately predicted (the second term of Equation \ref{eq:loss} goes to zero), while values larger/smaller than $1$ indicate
  the model is under/over predicting the errors.
  
\end{itemize}

In the following plots (see Section \ref{sec:results}), we mainly present the $\chi^2$ statistics. This choice was based on three reasons: 
(1) it is the only metric able to assess the quality of both the predicted $\mu$ and $\sigma$; 
(2) it the statistic we use to detect and remove outlier catalogs (catalogs with $\chi^2 > 10$ -- see more details in Section \ref{sec:results}); 
(3) it is the statistic we use to see if each model is robust against the systematic in question.
Although there is no universally accepted measure of robustness, a good value of $\chi^2$ is always correlated with a good performance in the other metrics as well. We present the complete set of scores, for each model, in the Appendix \ref{sec:complete_set}, also showcasing them in the results section.

\begin{figure*}[!ht]
 \centering
 \includegraphics[scale=0.29]{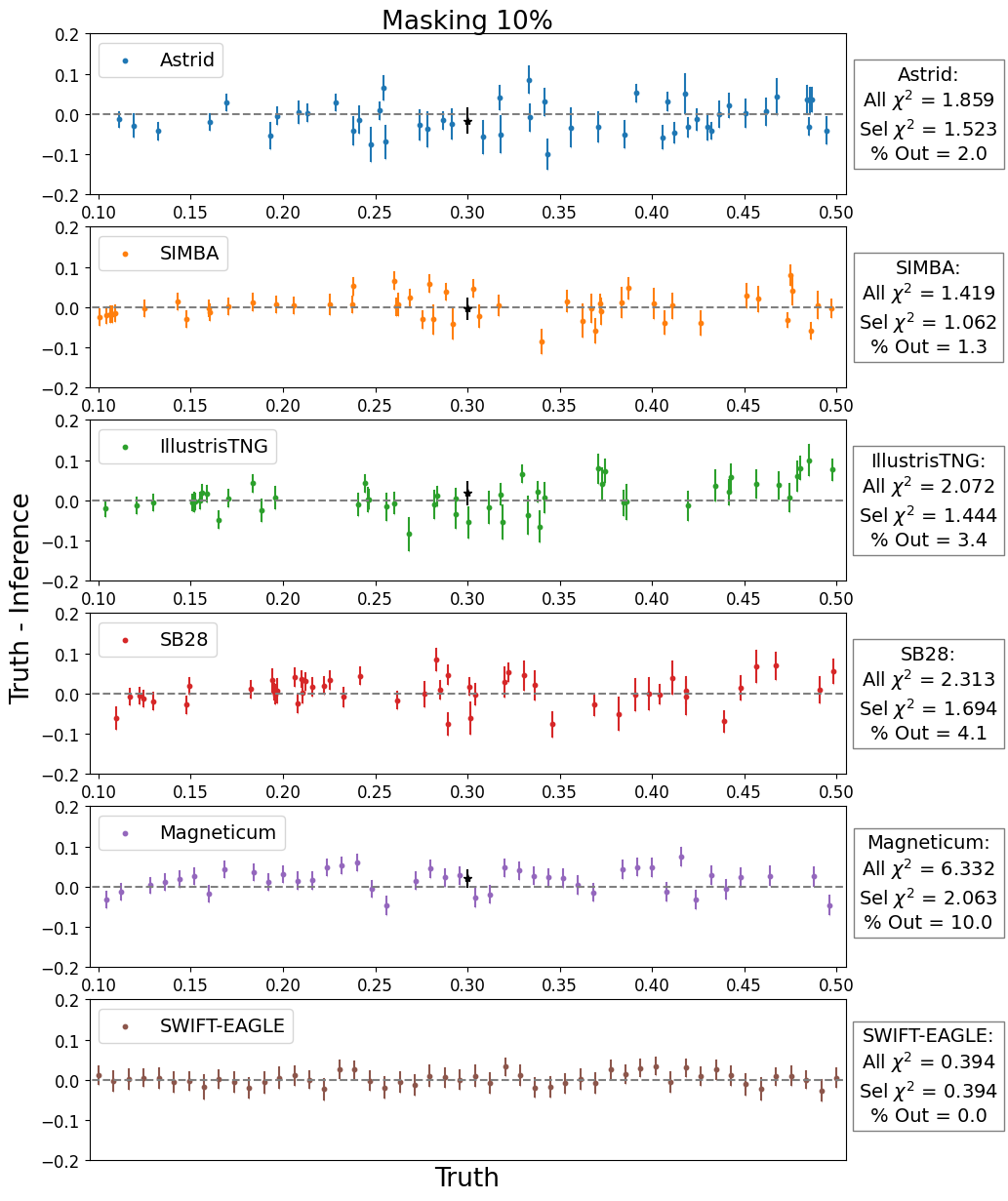}
 \includegraphics[scale=0.29]{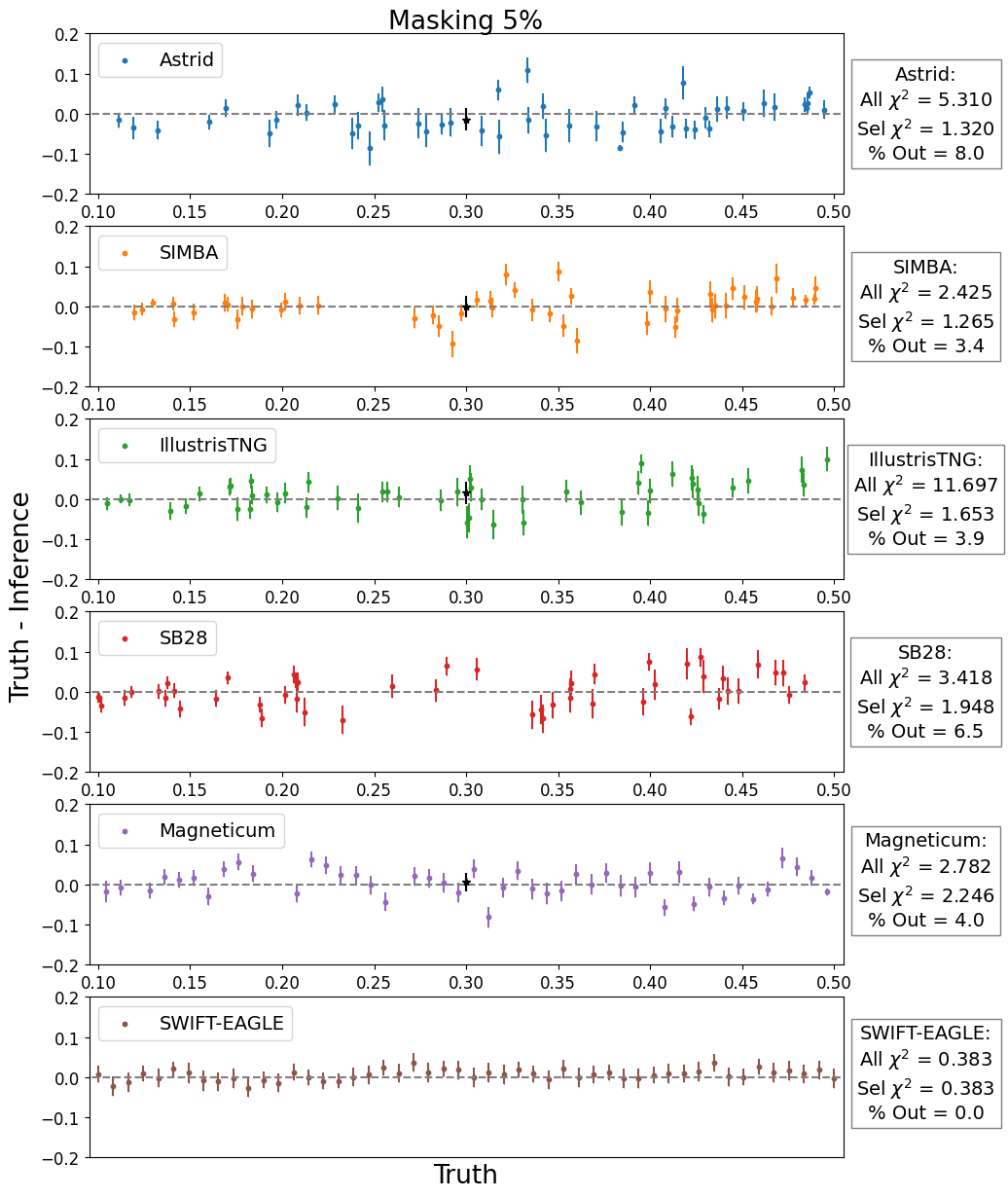}
 \caption{{\bf Truth - Inference of $\Omega_{\rm m}$ -- Masking:} removing $10 \%$ and $5 \%$ of the galaxies, respectively on the left and on the right panels. We present the predictions for galaxy catalogs from Astrid, SIMBA, IllustrisTNG, SB28, Magneticum, and SWIFT-EAGLE. For each simulation suite, we indicate the average $\chi^2$ value across all galaxy catalogs in the test set. We also list the $\chi^2$ values after removing outliers, which are selected as catalogs whose predictions exhibit $\chi^2 > 10$ and present the percentage of outliers (percentage of catalogs removed after this selection).}
 \label{fig:mask_results}
\end{figure*}

%%%%%%%%%%%%%%%%%%%%%%%%%%%%%%%%%%%%%%%%%%%%
\section{Results} 
\label{sec:results}
%%%%%%%%%%%%%%%%%%%%%%%%%%%%%%%%%%%%%%%%%%%%

In this section we present the results of training and testing our models with the different considered systematic effects.
The scores are taken on galaxy catalogs with different cosmologies, astrophysical parameters, and subgrid physic models, in order to
test the coverage of the predictions and the robustness of the different ML models.
We account for models related to the following observational effects: masking (Section \ref{sec:mask_results}),
velocity errors (Section \ref{sec:vel_results}), masking and velocity errors (Section
\ref{sec:mask_and_vel_results}), errors in the line-of-sight positions (Section \ref{sec:2Dpos_1Dvel}), and galaxy selection
(Sections \ref{sec:color_results} and \ref{sec:star_forming_results}).

In all the analyses, we are presenting the metrics measured in the complete data set (unless for Astrid, because it was used to
train the models and, then, the test set contains only $50$ simulations), naming them as ``all''.
In the majority of the cases, our results are highly affected by a outliers, which can be seen as an indicative of the failure of the model.
To avoid contamination from them, we also report the scores without them (for {\em selected} predictions with $\chi^2 < 10$).
Note that the fraction of outliers is something $\sim 10 \%$, as can be seen by the percentage of outliers, also reported in the main plots.
This small fraction indicates that we can analyze the predictions without them, specially because we do not have a definitive explanation for these outliers.

In our previous work \cite{deSanti2023}, we excluded outlying predictions considering $\chi^2 > 14$. These predictions corresponded to physical models with extreme AGN and SN feedback, which were few enough across the entire data set that excluding them did not affect the robustness and results of our previous GNNs. However, in this work, our outliers do not always have extreme values for the astrophysical parameters, but they suffer from the observational effects implemented, due to a more complex data set. 

Here, we describe the several methods we explored for detecting and analyzing outliers in the model predictions. First, we analyzed the latent space of the last aggregation layer (see Equation \ref{eq:last_layer}), reducing it to a low component space ($2$, $3$, $6$ dimensions) with three different dimensional reduction techniques: Uniform Manifold Approximation and Projection (UMAP) \cite{UMAP}, T-distributed Stochastic Neighbor Embedding (TSNE) \cite{TSNE}, and Principal Component Analysis (PCA) \cite{PCA}. We then looked for outliers on this reduced dimensional space by making use of traditional outlier methods of detection such as: $\chi^2$ selection -- to identify outliers based on their $\chi^2$ values and a threshold; Local Outlier Factor (LOF) -- to identify outliers based on the local density deviation of a given data point with respect to its neighbors \cite{LOF}; One-Class Support Vector Machine (One-Class SVM) -- to identify outliers by finding a boundary that separates the normal data points from the outlier \cite{SVM}; and Normalizing Flows (NF) -- to identify outliers based on the training latent distribution \cite{NF}. In none of these methods the detected outliers by one method matched with the others. Additionally, apart from the $\chi^2$ selection, none of the outliers detection algorithms were able to produce good scores to all the metrics analyzed with less than $10\%$ of outliers. This is indicative that none of these methods worked well and of the of low correlations between good/bad predictions and the latent space. The underlying reasons for these outliers will be examined in future investigations. While we search for better ways of detecting outliers, we keep the $\chi^2$ selection for discarding them when necessary.

We reinforce that the points in the plots only contain $50$ predictions for ease of reading\footnote{In order to avoid overcrowding the plot.}, but the scores were computed for the entire test data set into consideration. Moreover, the points corresponding
to the black dots are computed as an average over the mean and standard deviations for all the points in the CV set of each
simulation.

%%%%%%%%%%%%%%%%%%%%%%%%%%%%%%%%%%%
\subsection{Masking} 
\label{sec:mask_results}

\begin{figure*}[!ht]
 \centering
 \includegraphics[scale=0.29]{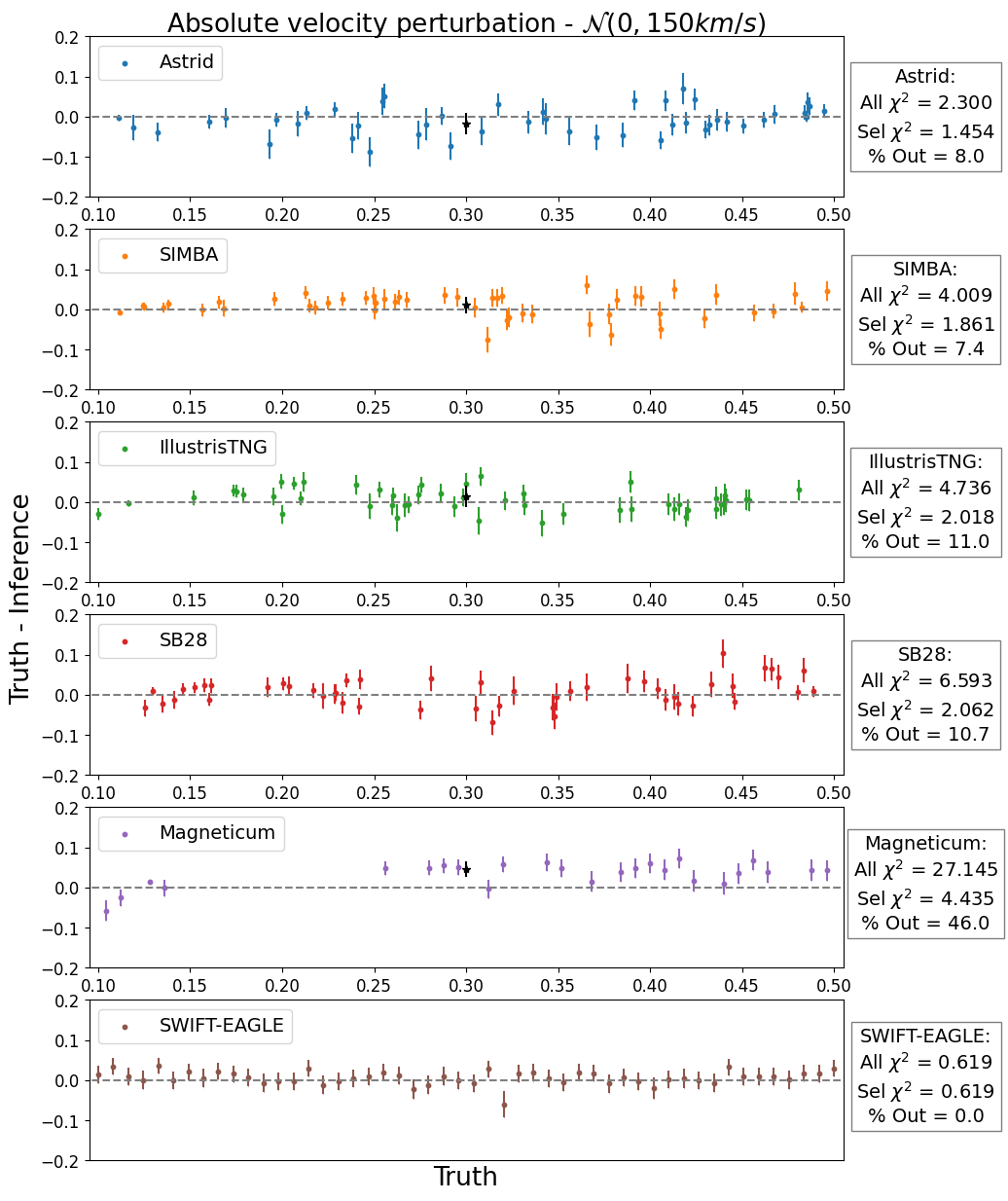}
 \includegraphics[scale=0.29]{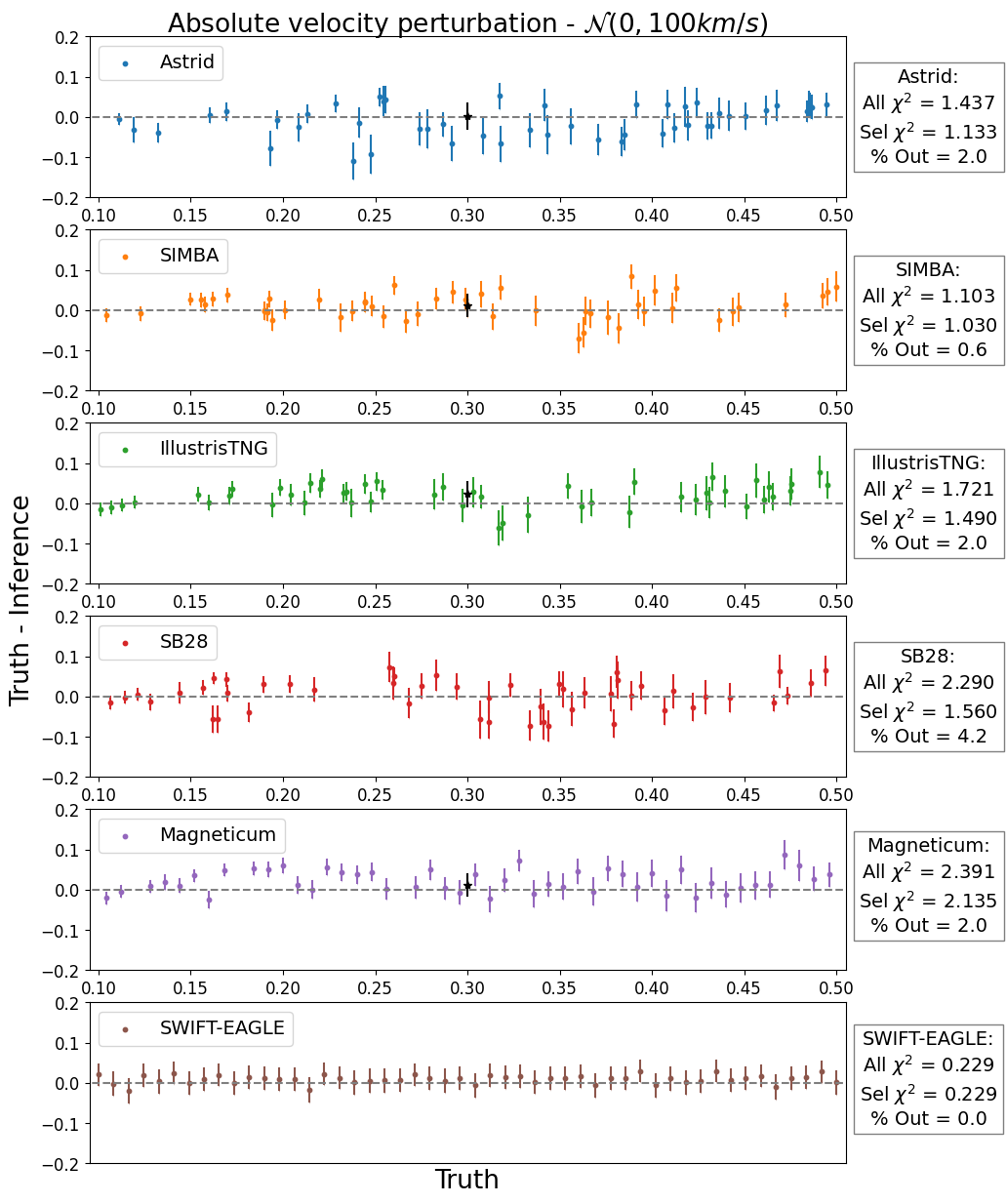}
 \caption{{\bf Truth - Inference of $\Omega_{\rm m}$ -- Peculiar velocity uncertainties: absolute error.} For $V = 150$ km/s and $100$ km/s for each galaxy velocity, respectively on
 the left and the right panels. We present the predictions for galaxy catalogs from Astrid, SIMBA, IllustrisTNG, SB28, Magneticum, and SWIFT-EAGLE. For each simulation suite, we indicate the average $\chi^2$ value across all galaxy catalogs in the test set. We also list the $\chi^2$ values after removing outliers, which are selected as catalogs whose predictions exhibit $\chi^2 > 10$ and present the percentage of outliers (percentage of catalogs removed after this selection).}
 \label{fig:additive}
\end{figure*}

We start by showing the results of training the GNNs on catalogs that contain only the $3$D positions and $1$D velocity
considering a random mask applied to them, i.e. we randomly remove (mask out) $5 \%$ and $10 \%$ of the galaxies per catalog.
This means that we are removing the same fraction of galaxies from all the different boxes for the different hydrodynamic
simulations considered. This analysis considers the power of the GNNs to constrain $\Omega_{\rm m}$ from catalogs of reduced
information (less galaxies).

The results for the LH and CV sets are presented in Figure \ref{fig:mask_results} and in Appendix \ref{sec:complete_set}.
Overall (independent of whether outliers are discarded or not), in both scenarios, the network is behaving robustly across nearly all
hydrodynamic suites (which is confirmed with almost all $\chi^2$ values close to
$1$).
The only exception is Magneticum, for which the final value is still poor ($\chi^2 \in [2.06, 2.25]$, respectively for masking
$10 \%$ and $5 \%$), even after removing the high $\chi^2$ outliers. 
The models trained on Astrid and tested on Astrid already have higher $\chi^2$ values before removing the outliers, $\chi^2 \in [1.86, 5.31]$ (again, respectively for masking $10 \%$ and $5 \%$), which correspond to $\sim 8 \%$ of removed catalogs, specifically for masking $5 \%$ of the galaxies. 
After removing the outliers, the results are comparable to the ones found in Reference \cite{deSanti2023} (see Appendix \ref{sec:best_model}), with $RMSE \sim 0.04$, $R^2 \sim 0.8$,
$PCC \sim 0.9$, $\epsilon \sim 11 \%$, and $\chi^2 \sim 1$ for both models.

The results for SIMBA, IllustrisTNG, and SB28 are all well constrained, especially after removing the outliers.
The fraction of outliers in these data sets are $\in [3, 7] \%$, which means that the bad predictions derive from a small number of catalogs. The model also performs well for SB28, which contains catalogs with broader
cosmological and astrophysical parameter variations. Results are, however, worse for Magneticum, where the model yields higher $\chi^2$, before removing outliers: we found $\chi^2$ values of $\sim 6$ to $\sim 3$, respectively for masking $10 \%$ and $5 \%$ of the galaxies.

Overall, we find that our models are robust to masking effects, and as expected, the smaller the effect, the better the results.

\begin{figure*}[!ht]
 \centering
 \includegraphics[scale=0.29]{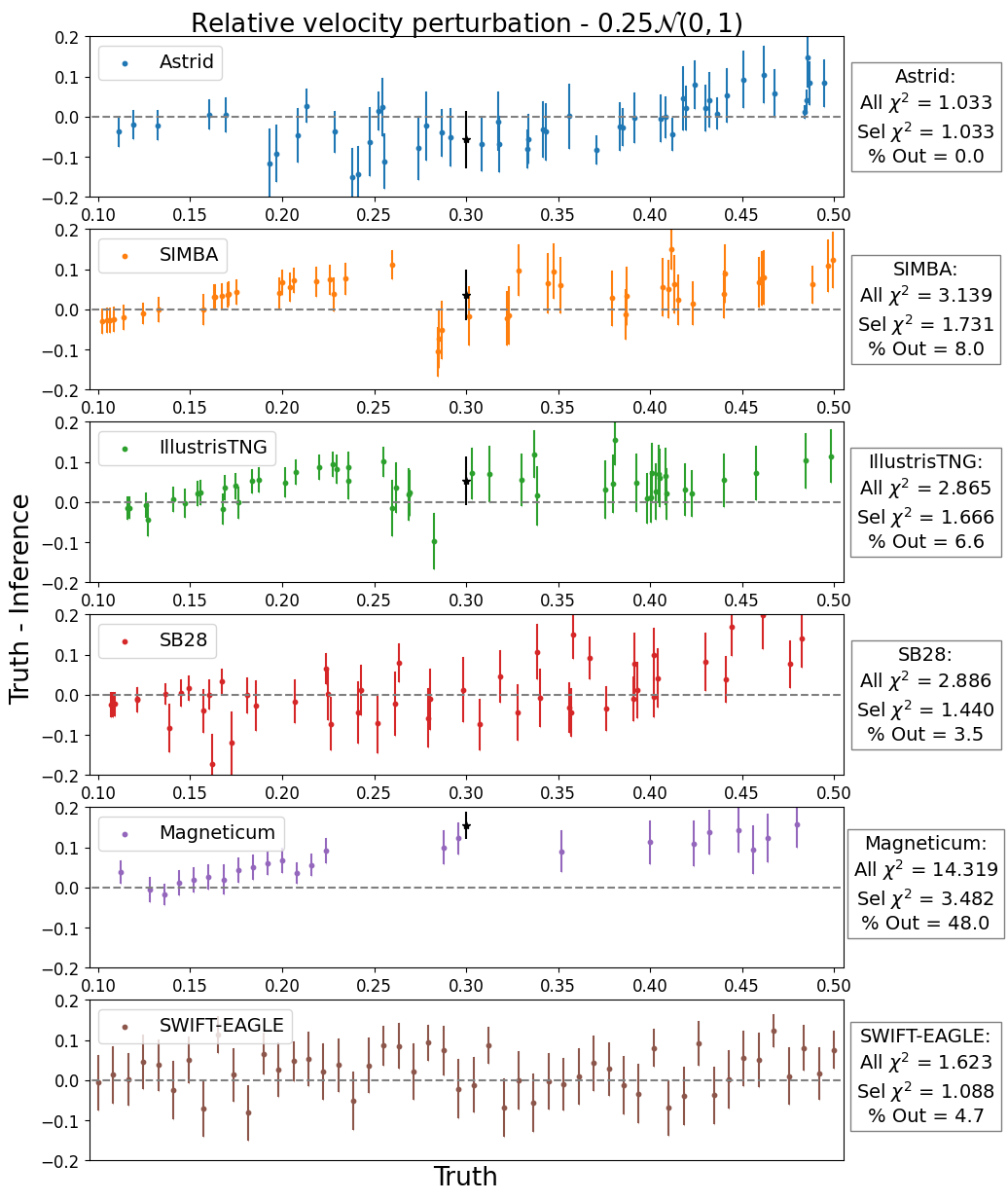}
 \includegraphics[scale=0.29]{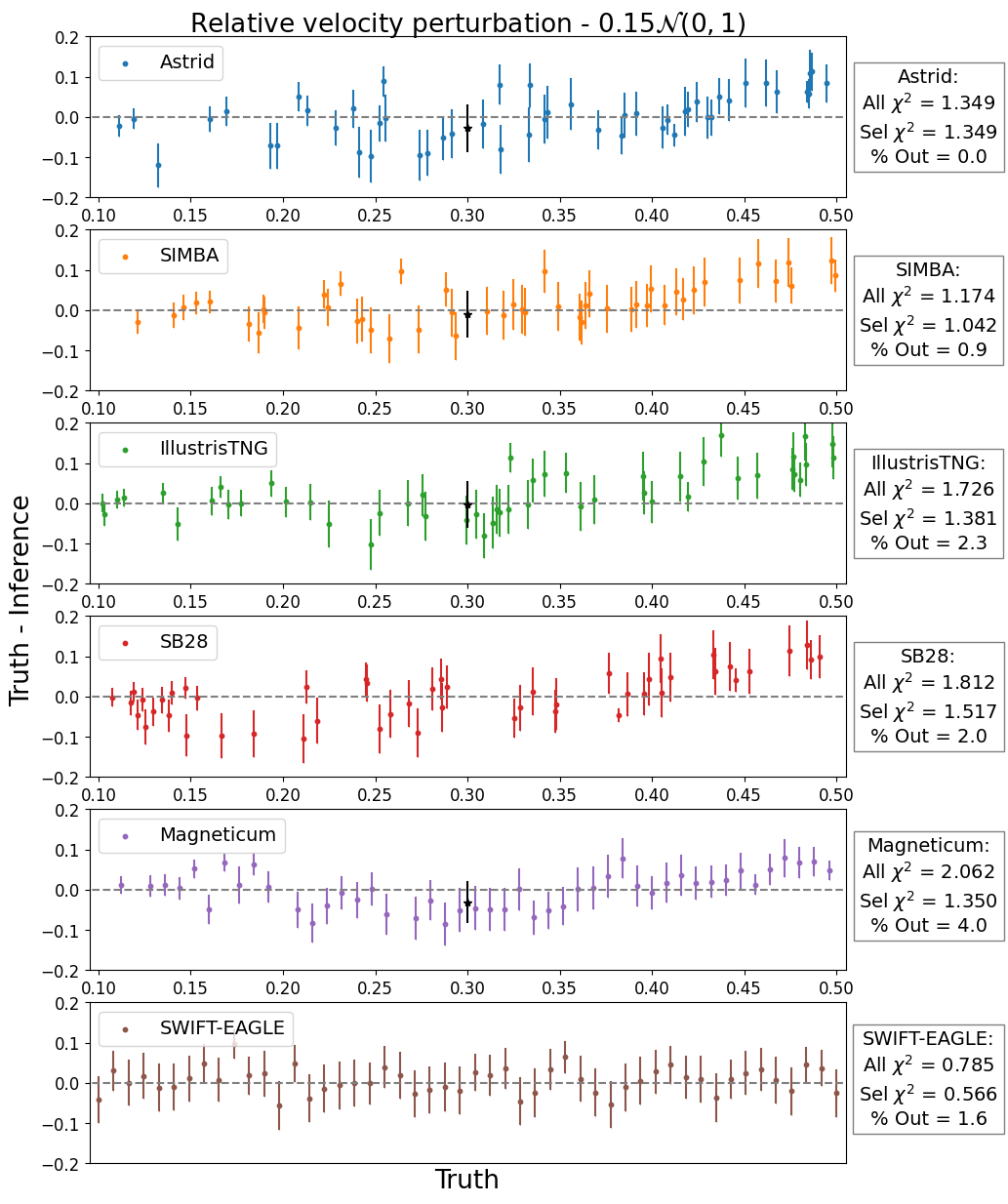}
 \caption{{\bf Truth - Inference of $\Omega_{\rm m}$ -- Peculiar velocity uncertainties: relative error.} For $P = 25 \%$ and $15 \%$ of the galaxy velocities, respectively on
 the left and on the right panels. We present the predictions for galaxy catalogs from Astrid, SIMBA, IllustrisTNG, SB28, Magneticum, and SWIFT-EAGLE. For each simulation suite, we indicate the average $\chi^2$ value across all galaxy catalogs in the test set. We also list the $\chi^2$ values after removing outliers, which are selected as catalogs whose predictions exhibit $\chi^2 > 10$ and present the percentage of outliers (percentage of catalogs removed after this selection).}
 \label{fig:multiplicative}
\end{figure*}

%%%%%%%%%%%%%%%%%%%%%%%%%%%%%%%%%%%%%%%
\subsection{Peculiar velocity uncertainties} 
\label{sec:vel_results}

We now discuss the results related to errors in the galaxy's peculiar velocities. As explained in Section \ref{sec:observational_effects}, we are considering {\em absolute} and {\em relative} errors, and
we present the results in Figures \ref{fig:additive} and \ref{fig:multiplicative} and in Appendix \ref{sec:complete_set}. 
A rough comparison of both indicates that the predictions achieved with the absolute uncertainties are worse than the relative ones.

The model trained on Astrid and tested on Astrid performs slightly worse than when considering absolute uncertainties ($RMSE \sim [0.04, 0.06]$, $R^2 \sim [0.5, 0.9]$, $PCC \sim [0.8, 0.9]$, $\epsilon \sim [10, 17] \%$, and $\chi^2 \sim [1.0, 2.3]$, respectively for the highest value of relative and absolute uncertainties), and it is biased for the CV set in both considerations (relative and absolute). 
Additionally, the model which considers the absolute error (with $V = 150$ km/s)
achieves a $\chi^2 \sim 2.3$, which indicates that it is more challenging for the GNNs to perform their predictions, as expected.

\begin{figure*}[!ht]
 \centering
 \includegraphics[scale=0.29]{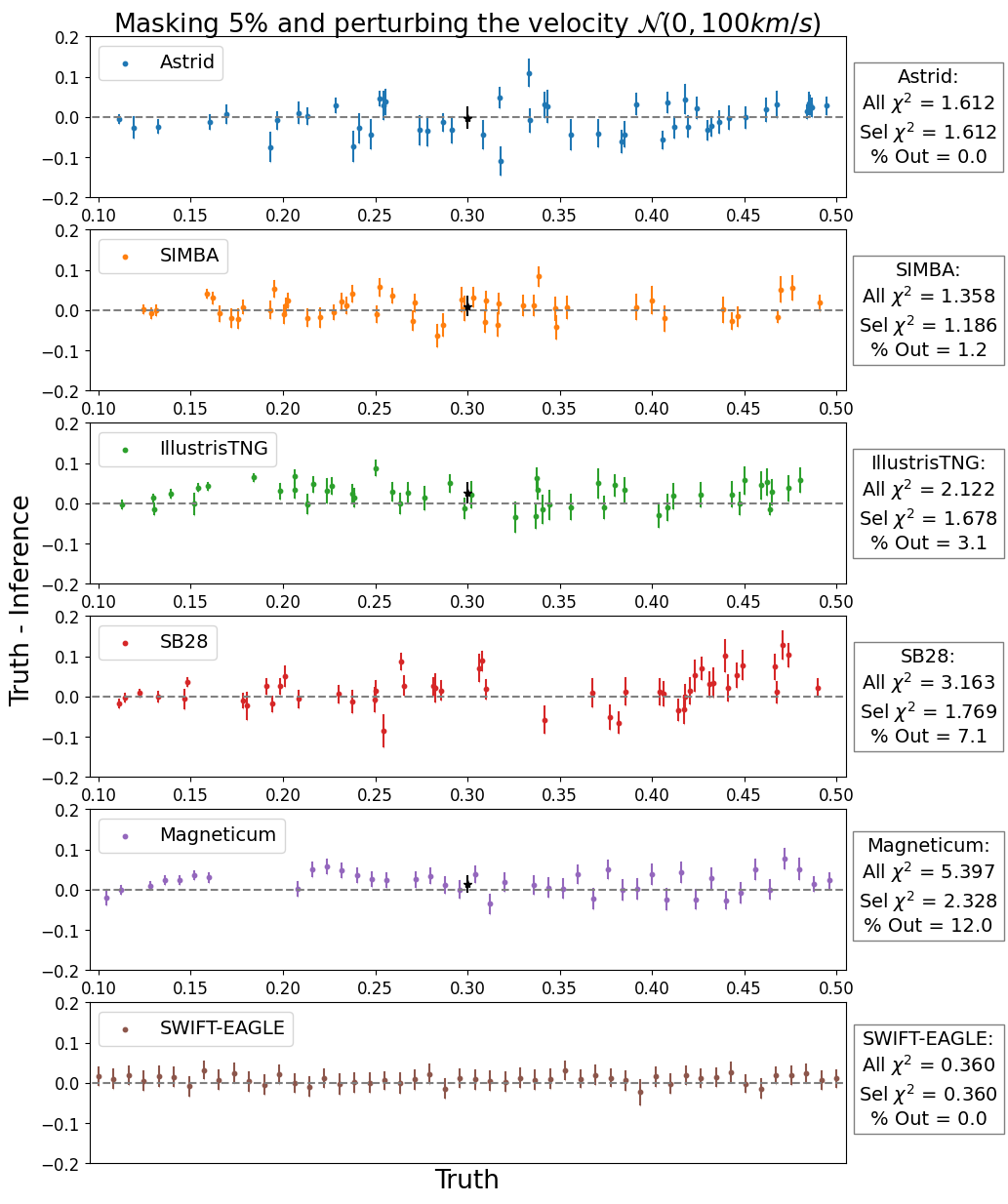}
 \includegraphics[scale=0.29]{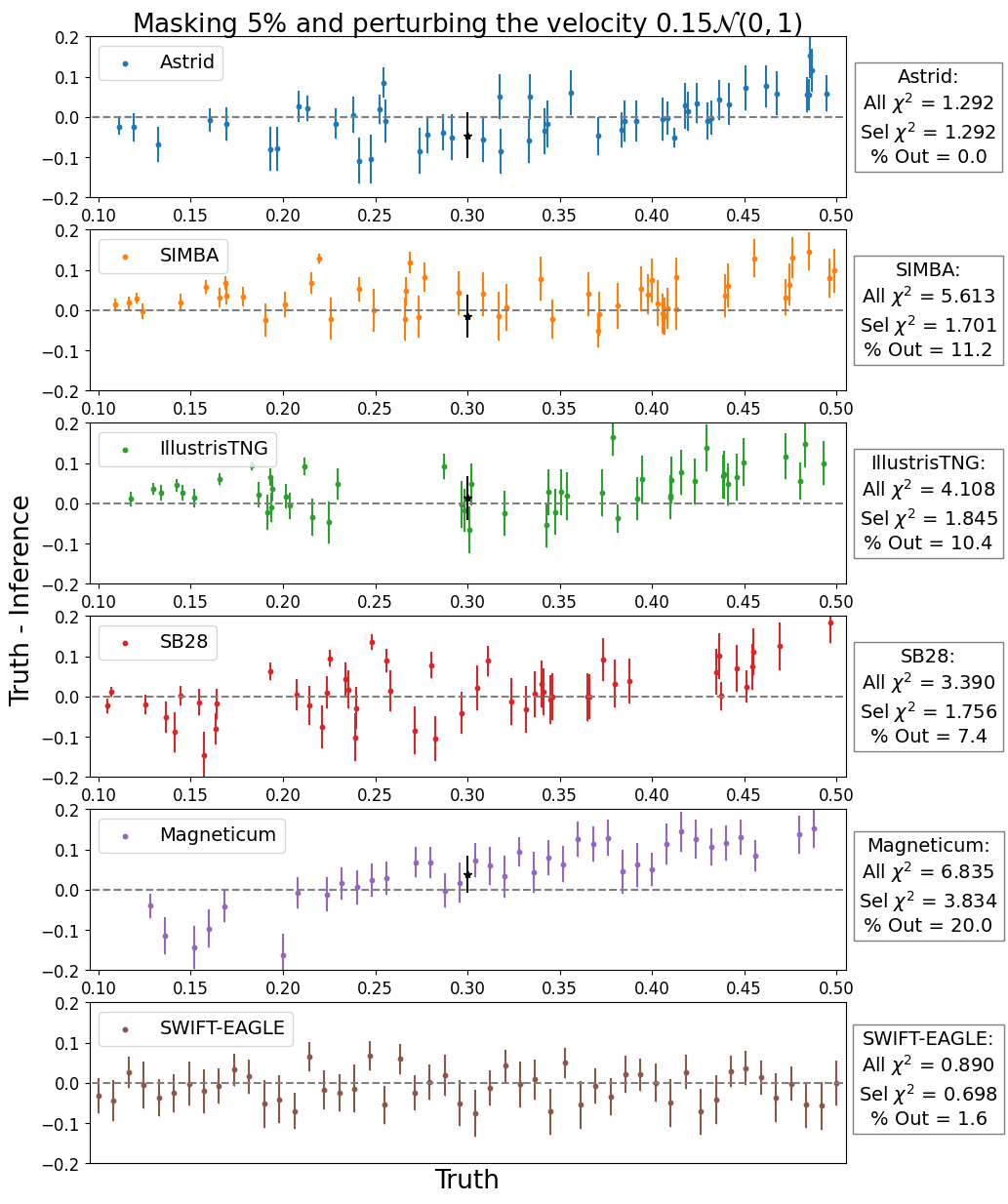}
 \caption{{\bf Truth - Inference of $\Omega_{\rm m}$ -- Masking and perturbing the galaxy velocities.} In this figure we are masking the galaxies in $5 \%$ and considering galaxy velocity uncertainties absolutely and relatively, respectively on the left and on the right panels.
 We present the predictions for galaxy catalogs from Astrid, SIMBA, IllustrisTNG, SB28, Magneticum, and SWIFT-EAGLE. For each simulation suite, we indicate the average $\chi^2$ value across all galaxy catalogs in the test set. We also list the $\chi^2$ values after removing outliers, which are selected as catalogs whose predictions exhibit $\chi^2 > 10$ and present the percentage of outliers (percentage of catalogs removed after this selection).}
 \label{fig:masking_and_vel_perturbing}
\end{figure*}

However, we do find a robust model for SIMBA, IllustrisTNG, SB28, and SWIFT-EAGLE, especially after selecting the points by
$\chi^2$ values (see Figures \ref{fig:additive} and \ref{fig:multiplicative}, with respective $\chi^2$ values close to $1$, after removing the ``outliers'').
As expected, increasing the velocity errors yields worse performance, indicating a possible 
limit of the prediction power of the GNNs. 
This is reflected in the number of outliers ($\chi^2 > 10$), which ranges from $[0.9, 8.0] \%$
for the relative errors, to $[0.6, 11] \%$ for the absolute error, which is, sometimes, higher than $10 \%$ of the datasets. 
In the case of Magneticum, the need to remove bad predictions achieve $\sim 48 \%$ of the whole test set, evidencing that the 
number of galaxies is not the most important property for the galaxy graphs, and, thus, that the GNNs fail to extrapolate their predictions to Magneticum in this specific case (see the discussion in the Appendix \ref{sec:GNN_and_graph_details} and Reference \cite{deSanti2023}). 

One interesting point is that the linking radii of these models (where we used the relative uncertainty in the velocities) were lower than when not perturbing the velocities (for more details see Table \ref{tab:linking_radius}).

Overall, the precision of the networks is worse when errors are modeled as absolute instead of relative. Removing the outliers yields similar performance across the $2$ systematic models.
We thus conclude that our GNNs are able to deal with small errors on the peculiar velocities, at the expense of accuracy.
One of the conclusions that we can draw from these results is that, depending on the survey considered in future analysis, it may beneficial to select galaxies with smaller redshift errors.

%%%%%%%%%%%%%%%%%%%%%%%%%%%%%%%%%%%%%%%%%%%

\subsection{Masking and peculiar velocity uncertainties} 
\label{sec:mask_and_vel_results}

We have studied the effects of masking and velocity errors separately in Sections \ref{sec:mask_results} and
\ref{sec:vel_results}, showing that the GNNs perform better with smaller masks, and when velocity errors are absolute with an error of 100 km/s or when they are modeled as relative with a 15\% amplitude. We now investigate how our model can deal with both effects simultaneously and show the results in Figure \ref{fig:masking_and_vel_perturbing} and Appendix \ref{sec:complete_set}.

The overall analysis indicates that the model accounting for the absolute velocity errors 
performs better than the relative one, while considering the mask effects together. 
The score values without removing any outliers for the models tested on SIMBA, IllustrisTNG, SB28, and SWIFT-EAGLE are in the ranges:
$RMSE \in [0.041, 0.136]$ (relative uncertainty) versus $RMSE \in [0.015, 0.046]$ (absolute uncertainty); 
$R^2 \in [- 2.588, 0.727]$ (relative uncertainty) versus $R^2 \in [0.833, 0.937]$ (absolute uncertainty); 
$PCC \in [- 0.004, 0.911]$ (relative uncertainty) versus $PCC \in [0.925, 0.973]$ (absolute uncertainty); 
$b \in [- 0.008, 0.037]$ (relative uncertainty) versus $b \in [- 0.005, 0.025]$ (absolute uncertainty); 
$\epsilon \in [11, 49] \%$ (relative uncertainty) versus $\epsilon \in [4.0, 12.3] \%$ (absolute uncertainty); 
and $\chi^2 \in [0.89, 6.80]$ (relative uncertainty) versus $\chi^2 \in [0.36, 5.40]$ (absolute uncertainty). 
The percentage of catalogs discarded due to their $\chi^2$ values are $\sim 20 \%$ versus $12 \%$ at maximum, respectively for relative and absolute uncertainties. 
One explanation for this result is that the relative error ends up with larger error bars,
which increases the level of the difficulty to do the predictions associated with the mask
effects. 
After selecting the good predictions both models are comparable to the model considering all
galaxies, apart for their predictions for Magneticum, in the case of the relative velocity
uncertainties, that achieves $\epsilon \sim 26 \%$, and $\chi^2 \sim 3.8$. 

Therefore, the main conclusion is that the GNNs can handle masking and peculiar velocity errors together
but at the cost of a significant loss of robustness in some cases. 

%%%%%%%%%%%%%%%%%%%%%%%%%%%%%%%%%%%%%%%%%%%%%%%%%%%%%%%
\subsection{Line-of-sight distance uncertainties} 
\label{sec:2Dpos_1Dvel}

\begin{figure}[!ht]
 \centering
 \includegraphics[scale=0.31]{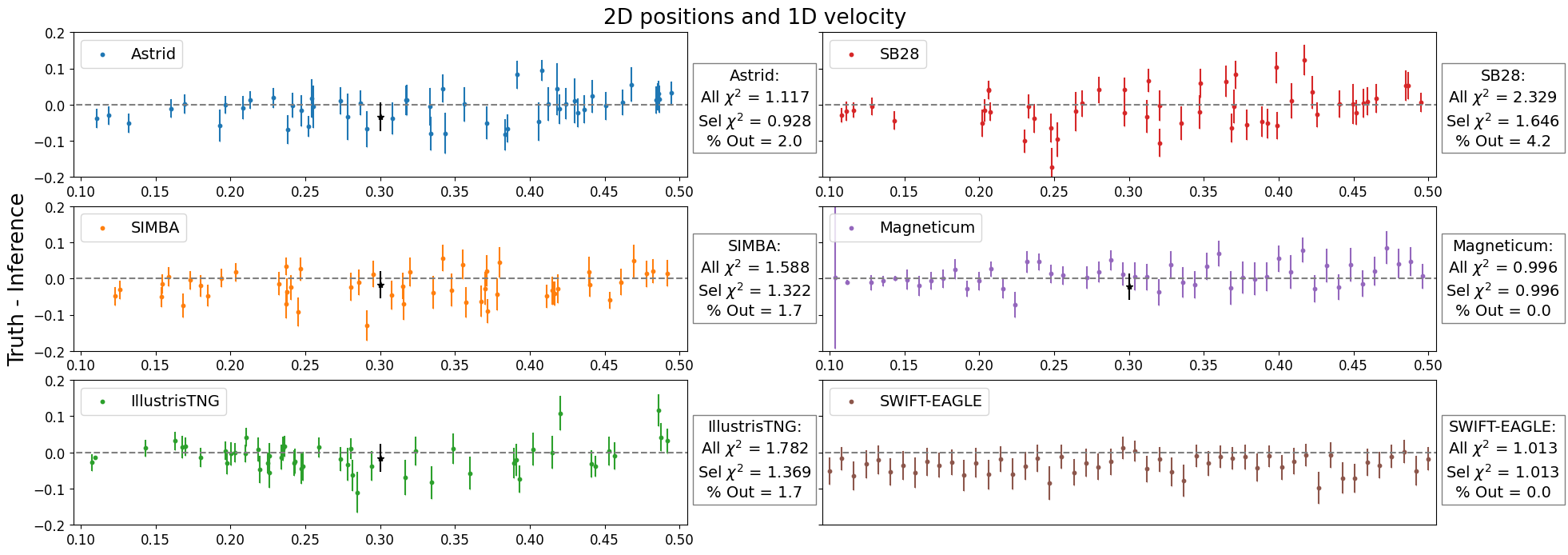}
 \caption{{\bf Truth - Inference of $\Omega_{\rm m}$ -- Line-of-sight distance uncertainties.} Considering only the $x$ and $y$
 positions and $v_z$ velocity of the galaxies. We present the predictions for galaxy catalogs from Astrid, SIMBA, IllustrisTNG, SB28, Magneticum, and SWIFT-EAGLE. For each simulation suite, we indicate the average $\chi^2$ value across all galaxy catalogs in the test set. We also list the $\chi^2$ values after removing outliers, which are selected as catalogs whose predictions exhibit $\chi^2 > 10$ and present the percentage of outliers (percentage of catalogs removed after this selection).}
 \label{fig:2Dpos_1Dvel}
\end{figure}

Finally, we consider the case where the radial distances are affected by errors. 
As explained above, we simply remove the line-of-sight component of the position and project 
all galaxies into a $2$D plane perpendicular to the line-of-sight. The main reason for doing this 
was to see the ability of the GNNs to constrain information from a reduced space dimension, usually available in photometric surveys, in contrast to the complete information 
(spatial $3$D) from spectroscopic surveys. 
Given the box size of the CAMELS simulations, we are in practice making the assumption that 
positions are known within a $25 h^{- 1}$ Mpc.

The results, shown in Figure \ref{fig:2Dpos_1Dvel} and Appendix \ref{sec:complete_set}, evidence that the model found scores 
comparable to the ones found using the entirety of galaxy position information 
\citep{deSanti2023}, even considering the predictions without removing outliers (see also Figure \ref{fig:2Dpos_1Dvel}, where almost all $\chi^2$ values are close to $1$, apart for SB28).
For instance, $RMSE \sim 0.04$, $R^2 \sim 0.8$, $PCC \sim 0.9$, $b \sim 0.01$, $\epsilon \sim 10 \%$ (even
better than $\epsilon \sim 12 \%$), and $\chi^2 \sim 1.1$ (again better than 
$\chi^2 \sim 1.6$), for the model trained on Astrid and tested on Astrid (see Appendix
\ref{sec:best_model})\footnote{We have to keep in mind that the scores can vary a bit
according to the model chosen by \textsc{optuna}, therefore, this is does not mean that the 
model found now is better, but compatible to the best model found in Reference \cite{deSanti2023}.}. 
We found that the predictions after the selection of $\chi^2 > 10$ are good, ending up in a
robust model which corresponds to removing always less than $4 \%$ of the catalogs. 
The only caveat of this current model is that it presents a small bias, especially for the CV
sets. 
This is evidence for the predictions on Astrid ($b \sim - 0.034$ versus $b \sim - 0.016$,
before). 
Another bad performance follows for SWIFT-EAGLE, with $\epsilon \sim 11.3 \%$ versus $\epsilon \sim 4.0 \%$, before.
But improvement was found for Magneticum, where now $\chi^2 \sim 1.0$ and before $\chi^2 \sim 2.21$.

\begin{figure*}[!ht]
 \centering
 \includegraphics[scale=0.29]{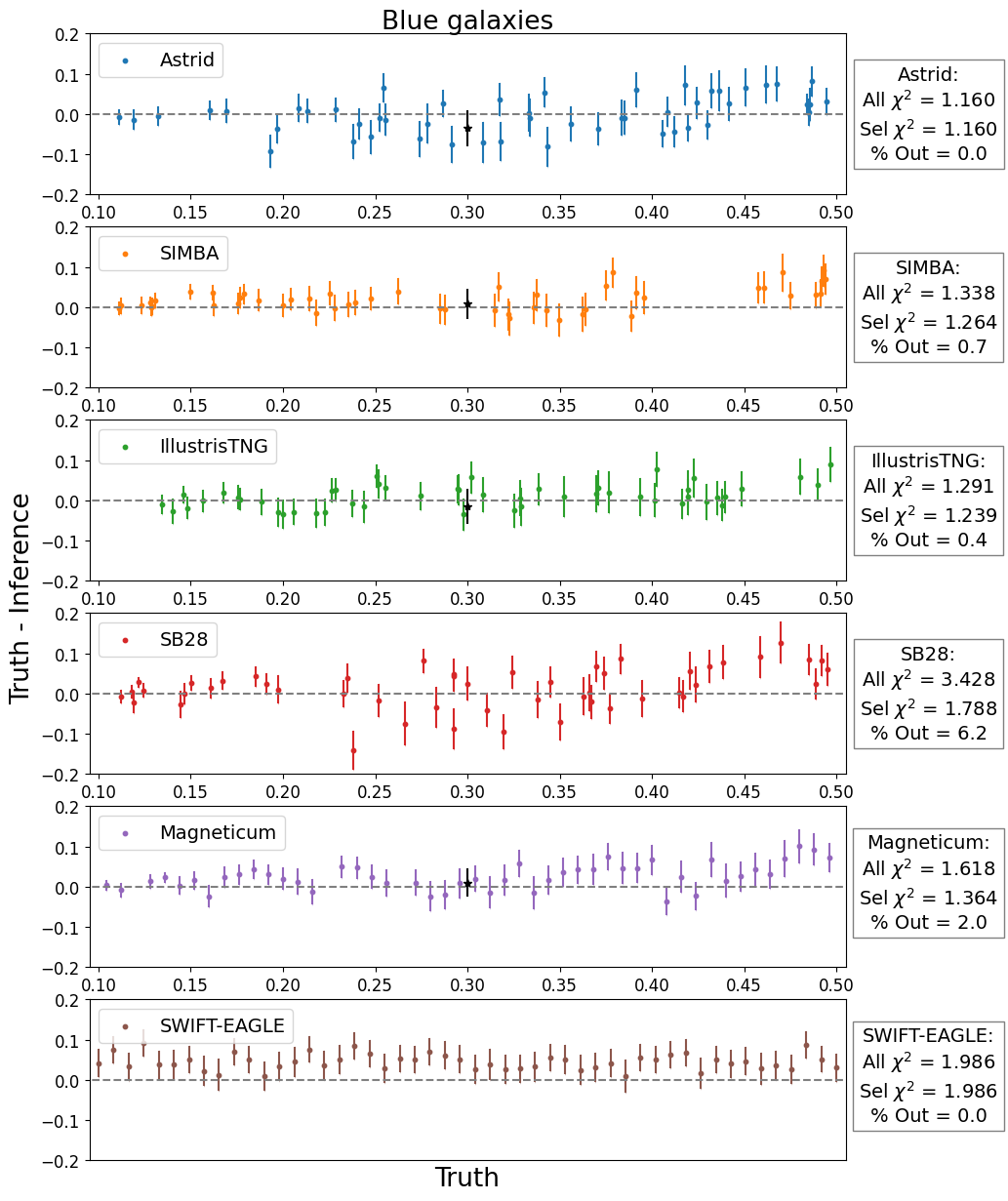}
 \includegraphics[scale=0.29]{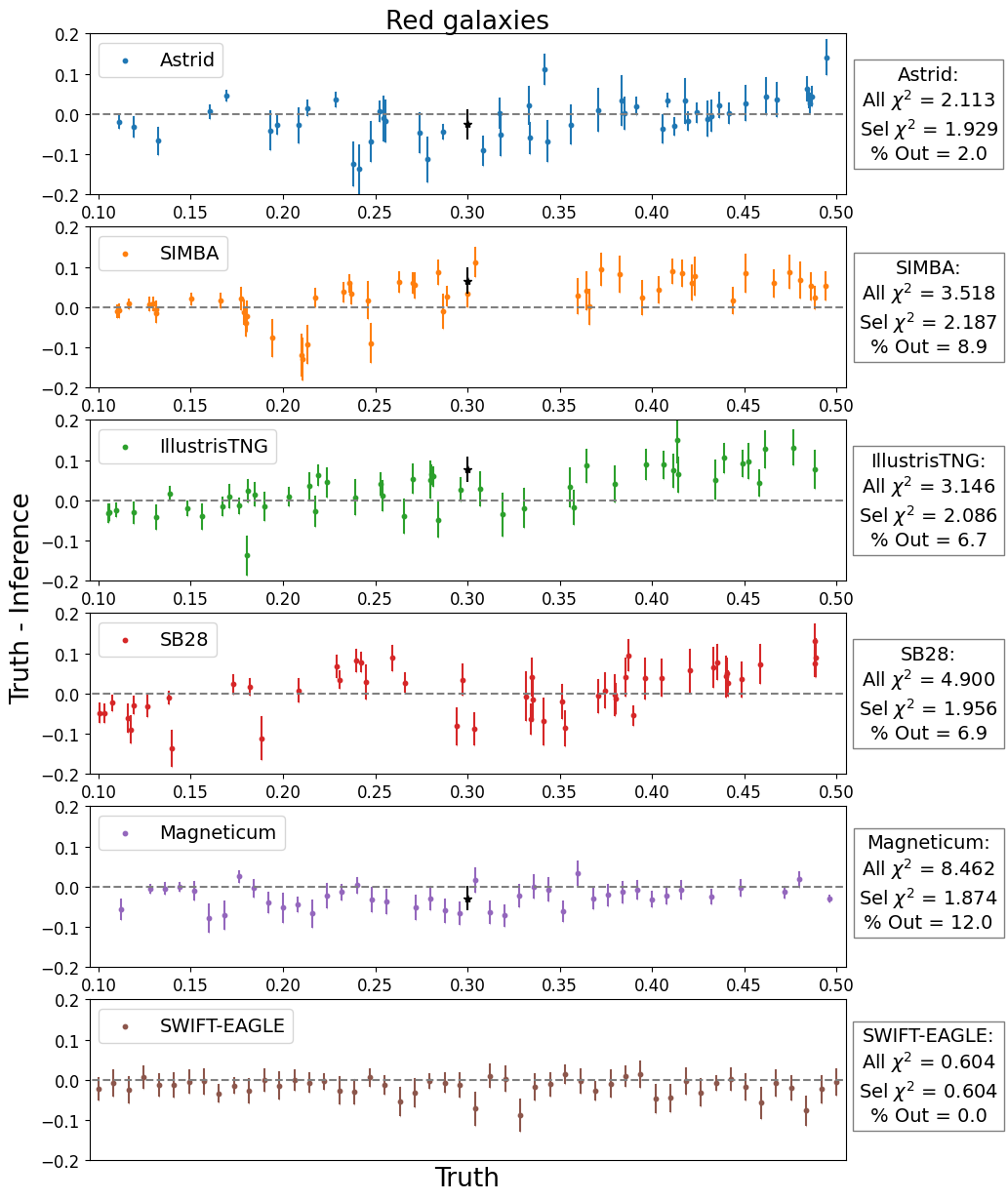}
 \caption{{\bf Truth - Inference of $\Omega_{\rm m}$ -- Galaxy selection: color.} Blue and red galaxies, respectively on the left and on the right panels. We present the predictions for galaxy catalogs from Astrid, SIMBA, IllustrisTNG, SB28, Magneticum, and SWIFT-EAGLE. For each simulation suite, we indicate the average $\chi^2$ value across all galaxy catalogs in the test set. We also list the $\chi^2$ values after removing outliers, which are selected as catalogs whose predictions exhibit $\chi^2 > 10$ and present the percentage of outliers (percentage of catalogs removed after this selection).}
 \label{fig:color_selection}
\end{figure*}

The main reason for this slightly worse performance now follows the fact that in $2$ dimensions
we have less information, from the galaxy positions, especially encompassing the loss in
information for large scales. 
Even though, with only $25 h^{- 1}$ Mpc boxes, from CAMELS, this loss in information can not be so
evident. 
In any case, we already know that GNNs can constrain galaxy information for reduced dimensions 
in graphs, as seen in Reference \cite{Wu2023}. 
Note that the linking radius for this model was larger than all the other found values:
$r_{link} \sim 2 h^{- 1}$ Mpc (see Table \ref{tab:linking_radius}). 
Also, we tested to use other velocity components to train a new model and the results are comparable, what proves
that we keep the homogeneity and isotropy in this analysis.

%%%%%%%%%%%%%%%%%%%%%%%%%%%%%%%%%%%%%%%%%%%%
\subsection{Galaxy selection: color} 
\label{sec:color_results}

In Figure \ref{fig:color_selection} and Appendix \ref{sec:complete_set} we show the impact of selecting galaxies based on their color. 
As explained in Section \ref{sec:observational_effects} we classify galaxies into blue and red
according to their sSFR values. 
We find that the GNNs perform well for both types,
though the model is slightly better and more robust for the blue galaxies.
Even after discarding the outliers, the predictions for the red galaxies are not as good as for 
the blue galaxies (what can be confirmed with the $\chi^2$ values in Figure \ref{fig:color_selection}, with almost
all the $\chi^2$ values for the red galaxies around $2$):
$RMSE \in [0.028, 0.053]$ (for red galaxies) $\times$ 
$RMSE \in [0.040, 0.052]$ (for blue galaxies), 
$R^2 \in [0.689, 0.870]$ (for red galaxies) $\times$, 
$R^2 \in [0.736, 0.860]$ (for blue galaxies),
$PCC \in [0.874, 0.965]$ (for red galaxies) $\times$, 
$PCC \in [0.902, 0.964]$ (for blue galaxies),
$b \in [- 0.007, - 0.025]$ (for red galaxies) $\times$, 
$b \in [- 0.001, 0.045]$ (for blue galaxies), 
and $\chi^2 \in [0.62, 2.19]$ (for red galaxies) $\times$
$\chi^2 \in [1.16, 2.04]$ (for blue galaxies). 
Also, the percentage of outliers is always large for the predictions in the red catalogs
($6 \% \rightarrow 12 \%$, from blue to red galaxies). 
The same effect is noticeable for the predictions in the CV set, where we always find a bias, or a
large bias, for the red ones.

A reason for this behavior may be the low number density and higher clustering associated to the
red galaxies. 
A possible solution to this selection may be to incorporate the green valley to the red galaxies, 
in the way to increase the number of galaxies per catalog, in different regions of the boxes. 
Although, in both cases we still have a robust model, after selecting the best predictions, which
correspond to removing less than $\sim 12 \%$ of the catalogs in the different data sets, getting
scores slightly worse than the GNNs trained with all the galaxies. 
Nevertheless, this analysis shows that the GNNs account for clustering effects, since blue and red
galaxies trace differently the underlying matter content and are disposed in different regions of
the halos (while blue galaxies are mostly in the neighboring regions, the red ones are mostly at 
the centers \citep{Sales2015, Knobel2013}). 
Also, different values for the linking radius were found for both models (considering only
red or blue galaxies), as shown in Table \ref{tab:linking_radius}. 

%%%%%%%%%%%%%%%%%%%%%%%%%%%%%%%%%%%%%%%
\subsection{Galaxy selection: star formation rate} 
\label{sec:star_forming_results}

\begin{figure*}[!ht]
 \centering
 \includegraphics[scale=0.29]{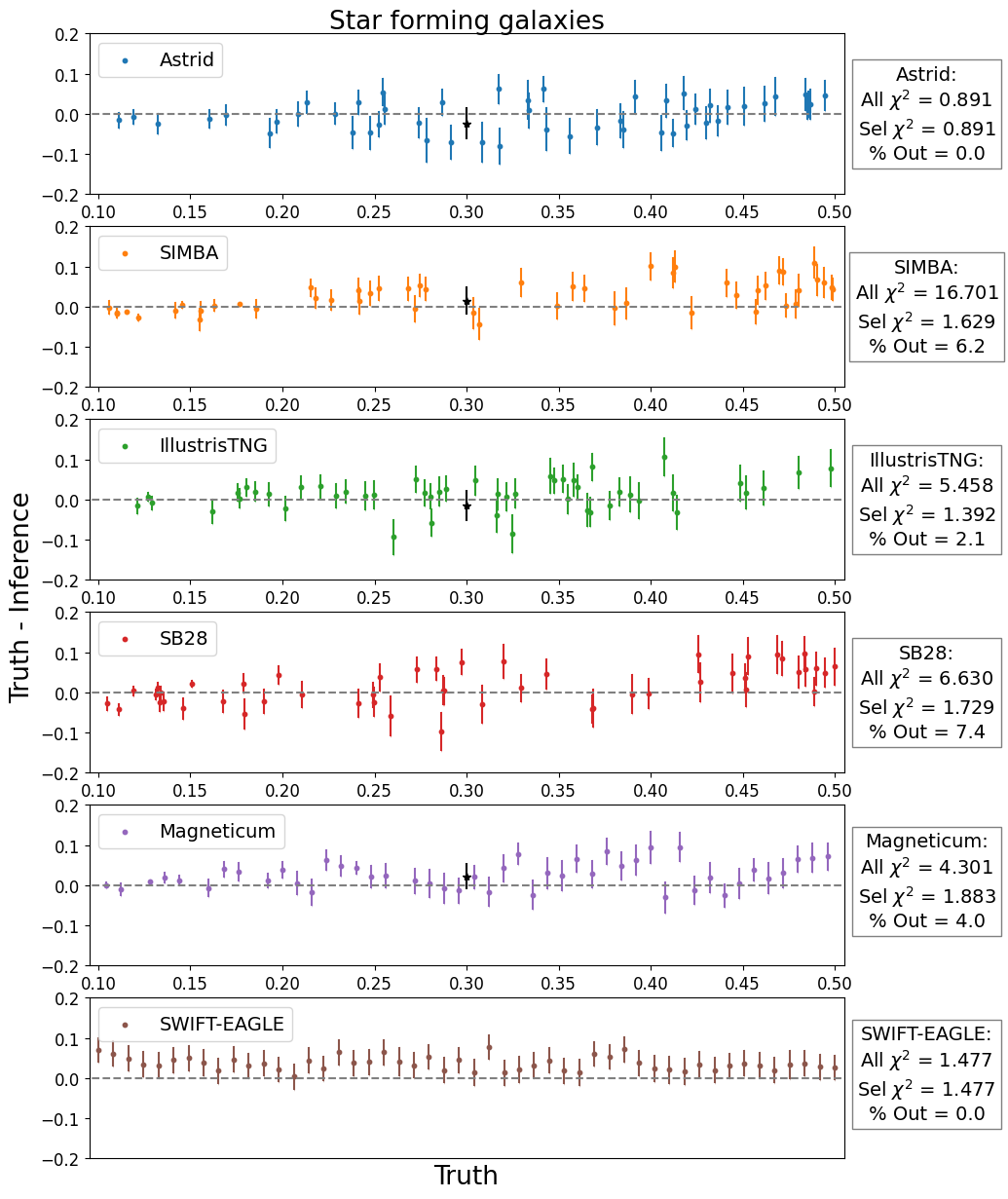}
 \includegraphics[scale=0.29]{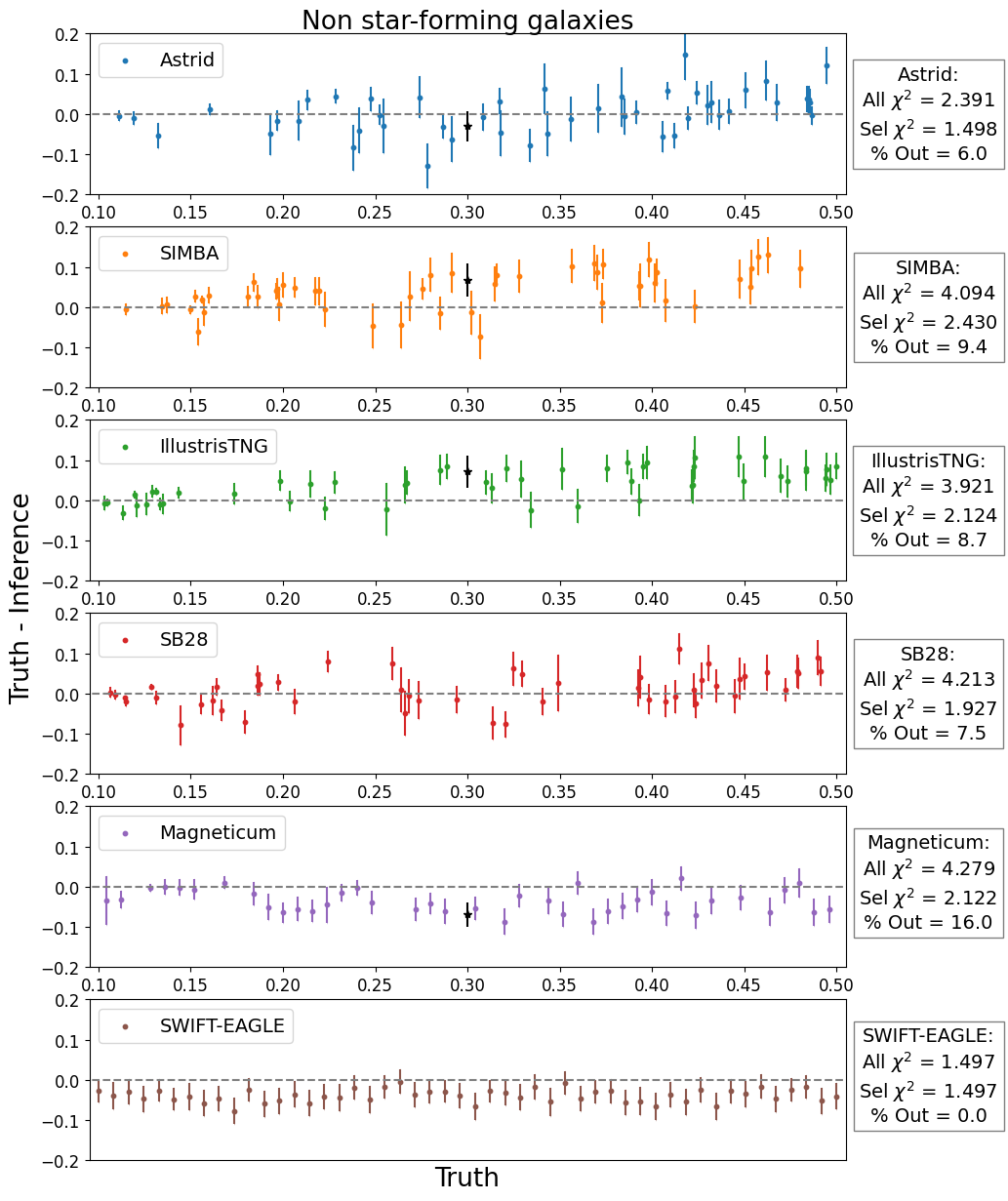}
 \caption{{\bf Truth - Inference of $\Omega_{\rm m}$ -- Galaxy selection: star formation rate.} Forming and non star-forming, respectively on the left and on the right panels. We present the predictions for galaxy catalogs from Astrid, SIMBA, IllustrisTNG, SB28, Magneticum, and SWIFT-EAGLE. For each simulation suite, we indicate the average $\chi^2$ value across all galaxy catalogs in the test set. We also list the $\chi^2$ values after removing outliers, which are selected as catalogs whose predictions exhibit $\chi^2 > 10$ and present the percentage of outliers (percentage of catalogs removed after this selection).}
 \label{fig:star_forming_selection}
\end{figure*}

We now study the results when galaxies are selected according to their star formation rate. 
We show the results in Figure \ref{fig:star_forming_selection} and 
Appendix \ref{sec:complete_set}. 
We note that in our previous work \cite{deSanti2023} and in all the different analyses in 
the present paper, we have considered galaxies with cuts only on stellar mass. 
In this way, the fraction of non star-forming galaxies was always there.

In both scenarios, the scores without removing any outliers evidence a poor performance (also confirmed by the
$\chi^2$ values presented in Figure \ref{fig:star_forming_selection}, being always around $2$):
$\epsilon \sim [15, 17] \%$, respectively for star forming and non star-forming for SB28; 
$\chi^2 \sim [16.7, 4.1]$, for SIMBA, in the case of star forming and non star-forming
galaxies; 
and $\chi^2 \sim 4.3$, for Magneticum, while considering both cases. 
We find that when outliers are removed, the model behaves better and are robusts: 
$\epsilon \lesssim [13, 16] \%$ and $\chi^2 \lesssim [1.9, 2.4]$ respectively for
star forming and non star-forming galaxies. 
On the other hand, the results for non star-forming galaxies indicate that the GNNs are not
able to extrapolate its accuracy attained for Astrid to other simulations (because
almost all the $\chi^2 \sim 2$).

There is a clear difference between the results for star-forming and non star-forming 
galaxies, which is observed in the best predictions for the former case, for a couple of
reasons. 
The first explanation is that catalogs selected by their non star-forming galaxies
have low number densities. 
Secondly, using the criteria we employed in this work (SFR $= 0$ galaxies), we can select
galaxies with very low masses that are likely numerical artifacts and, therefore cannot be 
trusted \citep{Donnari2019}.
Moreover, this is different depending on the different setups for the different 
combinations of each point in the LH CAMELS. 
We also found different (and lower than $r_{link} \sim 1.25 h^{- 1}$ Mpc) values for the
linking radius of these selected models (see Table \ref{tab:linking_radius}).

%%%%%%%%%%%%%%%%%%%%%%%%%%%%%%%%%%%%%%%%%%%%
\section{Discussion and conclusions} 
\label{sec:disc_and_conc}
%%%%%%%%%%%%%%%%%%%%%%%%%%%%%%%%%%%%%%%%%%%%

Efforts to constrain cosmological parameters from galaxy redshift surveys have taken many different approaches
\citep{deSanti2022JCAP, CARPool2022, Gualdi2021, Banerjee2021, Hahn2020, Uhlemann2020, Alan2018, Taylor2013}. 
More specifically, the use of galaxy positions and peculiar velocities to predict parameters such as $\Omega_{\rm m}$ has a long and
highly successful history \citep{Howlett2017, Cen1994, kaiser}. 
This has stimulated investment in galaxy surveys that allow us to measure peculiar velocities
\citep{SLOAN2022, Kourkchi2020, Howlett2017}, but up to now we could only count with traditional methods to analyze those data sets
\citep{Lai2023, Howlett2019}.
ML techniques are emerging as a promising toolkit that can tackle the problem of inferring cosmological parameters from large-scale
structures \cite{lucia2022, Paco2021}, and, recently, the use of
GNNs have seen as a promising alternative in the field \citep{Massara2023, deSanti2023, pablo-galaxies-2022, helen-halos-2022, lucas2022, Anagnostidis_2022}.

While traditional methods still need a theory template, as well as a summary statistics
\citep{MCMC2010, VI2022, NS2006}, ML methods can often connect data directly with the predictions,
especially with the advent of field level likelihood-free inference \citep{Lemos2023, Wang2023, Cranmer2020}.
In this scenario, GNNs are instrumental as a method that can analyze galaxy redshift surveys 
in a way that accounts for the positions and velocities of those objects relative to each other.
GNNs are designed to work with sparse and irregular data \citep{Bronstein2021, Battaglia2018, Gilmer2017}, they can handle physical
symmetries \citep{pablo-galaxies-2022}, and they do not impose a cutoff on scale to extract information.

GNNs have already been shown to lead to fairly accurate predictions for $\Omega_{\rm m}$ while using phase-space information together with other  properties, for both galaxies \citep{pablo-galaxies-2022} and halos  \citep{helen-halos-2022}. 
An important caveat related to these works was that, when using galaxies the model was not robust across different subgrid physical models, whereas when halos were employed, the methods had to rely on unobservable halo properties.
However, in Reference \cite{deSanti2023} we created a GNN that was designed to take only the galaxy positions and velocities, obtaining a model able to extrapolate the predictions even when changes are made to the astrophysics modelling, subgrid physics, and subhalo/galaxy finder.
The next step now is to  consider real-world and systematic effects in these galaxy catalogs, in order to show that GNNs are viable tools that can handle observational data.

For the reasons above, in this work we have trained different GNNs to be tested on thousands of galaxy catalogs from CAMELS in order to infer
$\Omega_{\rm m}$ at the field level, using a likelihood-free approach and considering different observational effects, such as masking, velocity errors, distance errors, and galaxy selection. 
In our analyses we sought models that are capable of making robust and accurate predictions.
The main results of those tests are as follows:
\begin{itemize}

 \item All the models presented here were trained on Astrid catalogs, using only galaxy phase-space information (positions and velocities). 
 When tested on Astrid
 catalogs, they were able to predict $\Omega_{\rm m}$ with scores (mean relative error and reduced chi equared -- see Equations \ref{Eq:eps} and \ref{eq:reduced}) up to $\epsilon \sim 17 \%$, and $\chi^2 \sim 1.9$, even when including systematic effects, as well as different cosmologies and  astrophysical parameters.
 
 \item The robustness of those models were put to the test on SIMBA, IllustrisTNG, SB28, Magneticum, and SWIFT-EAGLE, showing that, indeed, we still achieved models that are able to extrapolate their predictions despite
 the inclusion of systematics and different subgrid physical methods. 
 These robust models achieved scores up to
 $\epsilon \sim 26 \%$ and $\chi^2 \sim 4$.
 
 \item For each of the tests we have presented the scores corresponding to the entire test set, as well those which result from selecting the predictions with
 $\chi^2 < 10$ (i.e., removing the outliers). 
 The typical percentage of outliers correspond to $\sim 10 \%$ of the samples -- except in the case of Magneticum, in particular when considering velocity uncertainties. This showcases the fact that the number of galaxies per catalog is not the most important feature leading to a robust model in this case (since Astrid retains the broadest variation, encompassing the higher number of galaxies of Magneticum -- see Reference \cite{deSanti2023} and the discussion in the Appendix \ref{sec:GNN_and_graph_details}), evidencing the importance of galaxy velocities from the phase space.
 
 \item Our models can handle all the observational effects considered in the present work (masking effects, peculiar velocity errors, radial distance errors, galaxy selection effects), with different impacts on the accuracies of the predictions. 
 The effect with the least impact on the results, compared to the model from our previous paper \cite{deSanti2023} and all the other observational effects presented here, was 
 masking (assuming a 5\% loss in area): $\epsilon = 10.92 \%$ and $\chi^2 = 1.32$, for testing on Astrid and removing only $\in [3, 7] \%$ of outliers in the robustness tests (i.e., while tested on SIMBA, IllustrisTNG, SB28, Magneticum, and SWIFT-EAGLE).
 On the other hand, velocity uncertainties led to the worst scores, especially due to the biased CV predictions. 
 This reinforces our interpretation that the networks are effectively using phase-space information in order to draw their predictions.
 We also showed that we obtain fair predictions when considering both a mask and velocity uncertainties: $\chi^2 = [1.61, 1.29]$ (respectively for absolute and relative perturbation with $5 \%$ of masking) for the model tested on Astrid.
 
 \item We have shown that our methods are able to infer the value of $\Omega_{\rm m}$ from catalogs even after taking into account the line-of-sight redshift-space effects, i.e., with information
 which is limited to the $2$D spatial positions and $1$D velocities. 
 On the basis of these realistic data sets we obtain
 $\epsilon \sim 10 \%$, and $\chi^2 \sim 1.1$ for tests on Astrid. 
 These scores are almost as accurate as those for the catalogs with full $3$D positions and velocities (see Appendix \ref{sec:best_model}). 
 
\end{itemize}

\begin{figure}[h!]
    \centering
    \includegraphics[scale=0.38]{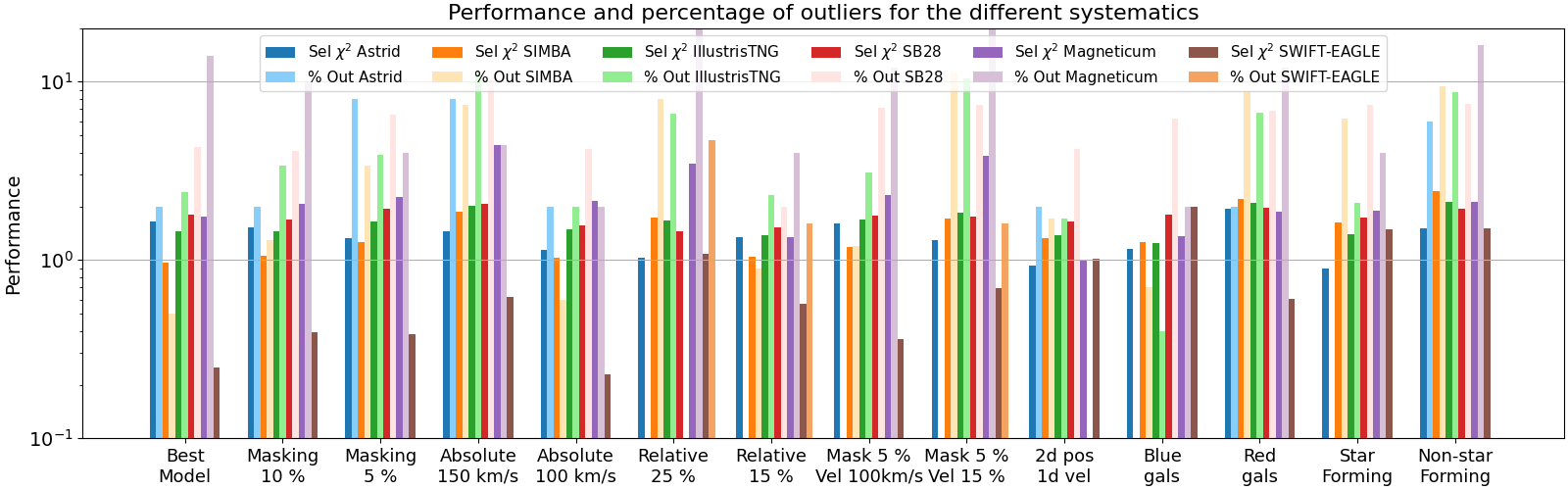}
    \caption{\textbf{$\chi^2$ performance of selected samples and percentage of outliers.} We present the results for all the systematics considered in the work (masking $10$\% and $5$\% of the galaxies, absolute velocity perturbation of $150$km/s and $100$km/s, relative velocity perturbation of $25$\% and $15$\%, masking $5$\% of the galaxies and absolute velocity perturbation of $100$km/s, masking $5$\% of the galaxies and relative velocity perturbation of $15$\%, 2D positions and 1D galaxy velocities, blue and red galaxies, star and non-star forming galaxies), compared with the best model, from Reference \cite{deSanti2023}.}
    \label{fig:summary}
\end{figure}

Taking into account all the tests that were presented, one of the foremost applications of GNNs would be to analyze large-area surveys with low-redshift, blue galaxies with high SFR.

Some future points that should be addressed are the following:
\begin{itemize}

 \item A more thorough understanding of the effects of super-sample covariance, similar to what was done in Reference \cite{deSanti2023}. 
 For that we need larger boxes to retrain the GNNs, and to study these effects in larger simulations with different resolutions. 
 We plan to address this problem with the next generation of simulations of CAMELS.
 This would help taking into account another limitation of the GNNs up to now: constraining more than only $\Omega_{\rm m}$, such as $\sigma_8$ and other
 cosmological parameters \cite{deSanti2023, Ni2023}.

 \item A deeper investigation about the reasons for the poor predictions after introducing the systematics (outliers detection). We can explore this by assessing the full posterior predictions for $\Omega_{\rm m}$ (e.g., using Normalizing Flows to replace the MNNs) or by predicting its full likelihood \cite{biwei2023} and re-training the models on mock galaxy catalogs designed for a given survey.
 
 \item Comparing the extrapolation power of the GNNs when trained on galaxy catalogs generated with faster techniques, such as the ones produced by semi-analytical models \cite{lucia2022, summarizeSAM-2015, Somerville1999}, abundance matching \cite{Contreras2021, Klypin2015}, halo-occupation distribution catalogs \cite{Berlind2003}, and others.
 
 \item Understanding the challenges ahead as we make progress towards our ultimate goal, which is to train the models presented here on simulated or mock galaxy catalogs and then to extract predictions from real data.
 
\end{itemize}

In this work we considered several systematic effects as if they were independent. 
However, in real surveys all these effects (and more) appear simultaneously. 
A more realistic test of the GNNs would be with catalogs that include the main effects all at once, such as with, e.g., the $2$D positions of observed red galaxies with approximated peculiar velocity errors, including selection effects and with bright foreground stars masked out. Presently, this is challenging due to the limited volume of the CAMELS hydrodynamic suites.
However, with future CAMELS-like simulations covering larger volumes, such tests will be carried out. 

Figure \ref{fig:summary} shows a summary of {the model performance for each method, indicating that after removing outlier predictions, which constitute only $\sim 10 \%$ of each sample, the $\chi^2$ remain close to $1$ for all analyzed systematics.} 
For some of these observational effects, the results still demonstrate robust performance (for instance, masking, reduced values for absolute and relative velocity perturbations, the joint of these effects, utilizing only 2D positions and 1D velocities, and selecting only blue and star-forming galaxies). On the other hand, other systematics pose a greater challenge for the model, resulting in a larger impact on the accuracy of the inferred parameters (for example, larger absolute and relative velocity perturbations or selecting only red and non-star forming galaxies. This is especially evident when using galaxy catalogs from Magneticum).

In conclusion, we have shown that the method proposed by
\citet{deSanti2023} to recover cosmological parameters from galaxies, and further developed here, is relatively robust to observational effects. We acknowledge that future steps for improving the GNN performance on observational effects can be taken. For instance, training on an even wider parameter space that includes not only cosmology and astrophysics but also systematic effects. Moreover, we can also design models which are more accurate within a given range of scales and with specific selection criteria. Therefore, this paper represents an important first step towards applying these methods to real galaxy catalogs.

\begin{table*}[!ht]
 \caption{\label{tab:linking_radius} Values of the linking radius found by \textsc{optuna} for the selected models.}
 \begin{center}
  \begin{tabular}{cc}
   \hline\hline
   \textbf{Model}             & $\mathbf{r}_{\mathbf{link}}$  \\
   \hline\hline
 %   Best model                           & $1.205775264930918$ Mpc$/h$ \\
 %   \hline
 %   Masking $10 \%$                      & $1.246626035737614$ Mpc$/h$ \\
 %   Masking $5 \%$                       & $1.238223043626219$ Mpc$/h$ \\
 %   \hline
    Relative error: $P = 25 \%$ & $0.428 h^{- 1}$ Mpc \\
    Relative error: $P = 15 \%$ & $0.913 h^{- 1}$ Mpc \\    
 %   Aditive perturbation: $150 km/s$     & $1.239282787421417$ Mpc$/h$ \\
 %   Aditive perturbation: $100 km/s$     & $1.249512586325772$ Mpc$/h$ \\    
    \hline
%    Masking $5 \%$ and additive perturbation: $100 km/s$       & $1.247148267605575$ Mpc$/h$ \\    
%    Masking $5 \%$ and multiplicaitive perturbation: $15 km/s$ & $1.249512586325772$ Mpc$/h$ \\    
%    \hline
    Color: Blues        & $1.065 h^{- 1}$ Mpc \\    
    Color: Reds         & $1.025 h^{- 1}$ Mpc \\    
    \hline
    Non star-forming        & $1.160 h^{- 1}$ Mpc \\    
    Forming stars            & $1.231 h^{- 1}$ Mpc \\    
    \hline
    $2$D positions and $1$D velocity & $1.938 h^{- 1}$ Mpc \\    
   \cline{1-2}%
   \end{tabular}
  \end{center}
\end{table*}

\begin{figure}[!ht]
 \centering
 \includegraphics[scale=0.5]{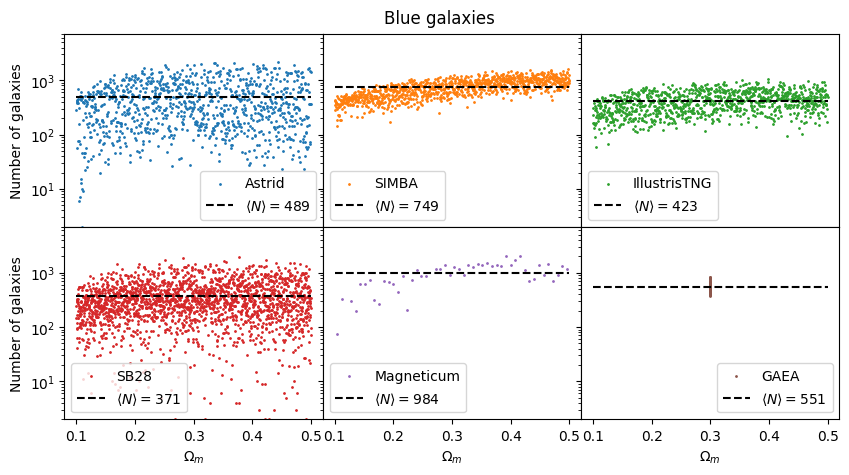}
 \includegraphics[scale=0.5]{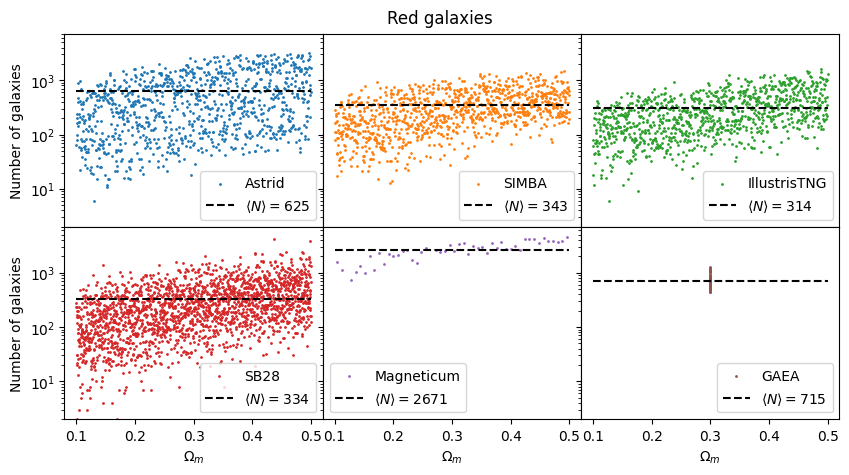}
 \caption{{\bf Galaxy selection: color.} Blue (top panel) and red (bottom panel) galaxies for the comparison of the number of galaxies versus $\Omega_{\rm m}$ values, per catalog in CAMELS simulations for Astrid (top left), SIMBA (top middle), IllustrisTNG (top right), SB28 (bottom left), Magneticum (bottom middle), and SWIFT-EAGLE (bottom right). The horizontal lines correspond to the mean number of galaxies per subgrid physical model.}
 \label{fig:color_selection-NxOmegaM}
\end{figure}

%%%%%%%%%%%%%%%%%%%%%%%%%%%%%%%%%%%%%%%%%%%%%%%%%%
\appendix

\section{GNNs and graph details}
\label{sec:GNN_and_graph_details}

\begin{figure}[!ht]
 \centering
 \includegraphics[scale=0.5]{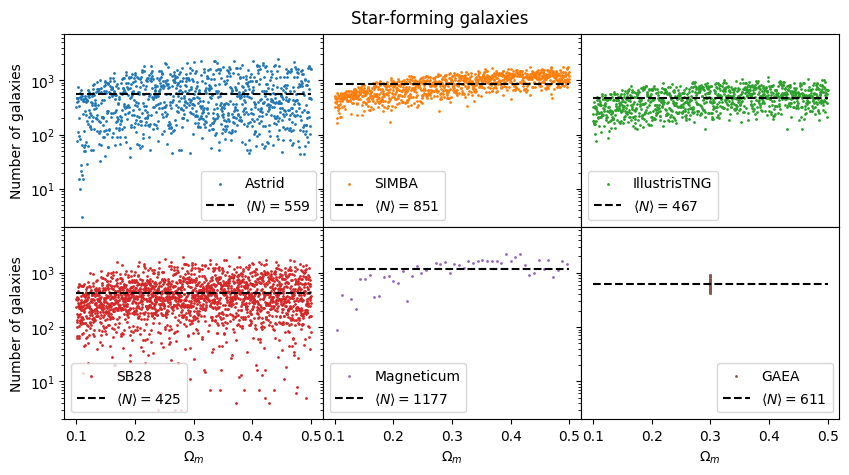}
 \includegraphics[scale=0.5]{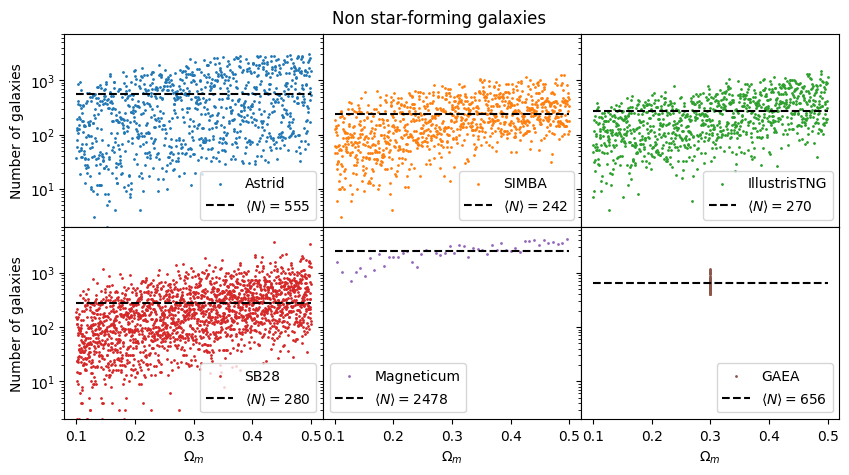}
 \caption{{\bf Galaxy selection: star formation rate.} Star forming (top panel) and non star-forming (bottom panel) galaxies for the comparison of the number of galaxies versus $\Omega_{\rm m}$ values, per catalog in CAMELS simulations for Astrid (top left), SIMBA (top middle), IllustrisTNG (top right), SB28 (bottom left), Magneticum (bottom middle), and SWIFT-EAGLE (bottom right). The horizontal lines correspond to the mean number of galaxies per subgrid physical model.}
 \label{fig:star_forming_selection-NxOmegaM}
\end{figure}

All of the graphs built in this work follow the prescription of Section 
\ref{sec:the_graph}. 
The free parameter found to build them, using \textsc{optuna}, was the link radius,
which was always a hyperparameter found to be as $\sim 1.25 h^{- 1}$ Mpc (and in agreement with Reference \cite{deSanti2023}), apart for the models presented in Table
\ref{tab:linking_radius}. 
We can say that, in the case of the relative velocity uncertainties, the impact of the random changes in the velocities lead to a smaller value to connect the galaxies. 
At the same time, clustering differences, due to the presence of more galaxies in
the center of the halos (what is the case of the red galaxies or the 
non star-forming galaxies \citep{Sales2015, Knobel2013}) led to lowest values being preferred too. On the opposite way, the model where we have
reduced the space component to only $2$ dimensions, indicates that the GNNs preferred higher values ($r_{link} \sim 2 h^{- 1}$ Mpc) to
propagate the information along the network. 
Beyond that, analyses regarding the distance among the galaxies in the
catalogs, their similar structure on their distribution, and the average number of edges keeps the same as the model presented
before \citep{deSanti2023}. They are different only for the model of Section \ref{sec:2Dpos_1Dvel}.

\begin{figure}[!ht]
 \centering
 \includegraphics[width=5.0cm, height=4.4cm]{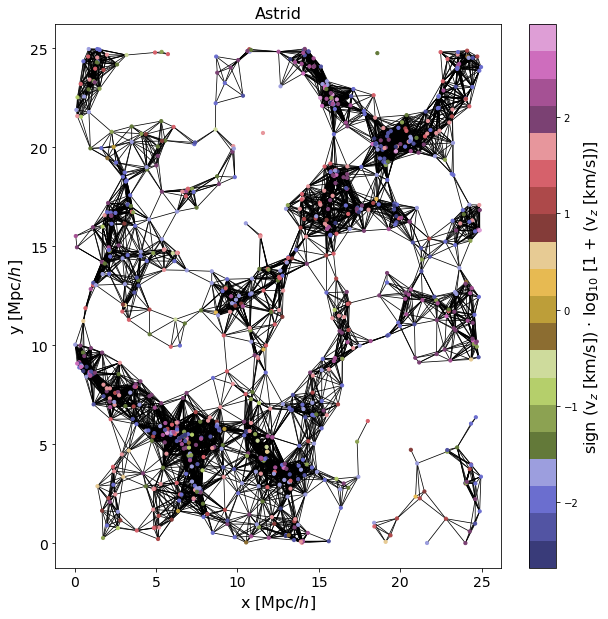}
 \includegraphics[width=5.0cm, height=4.4cm]{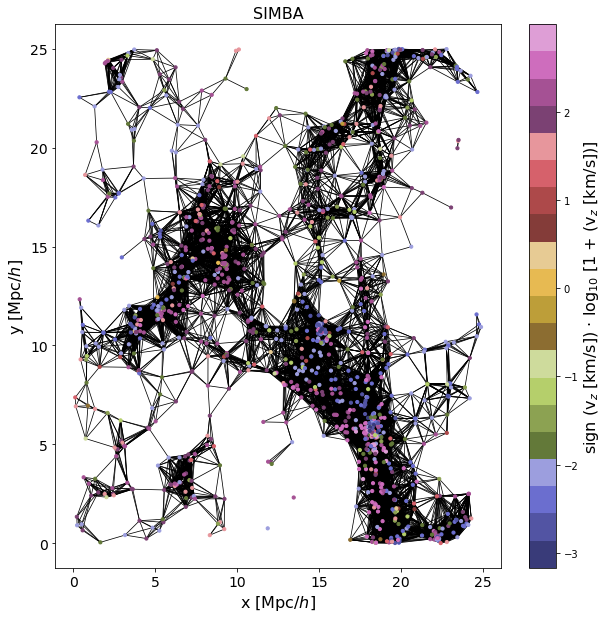}
 \includegraphics[width=5.0cm, height=4.4cm]{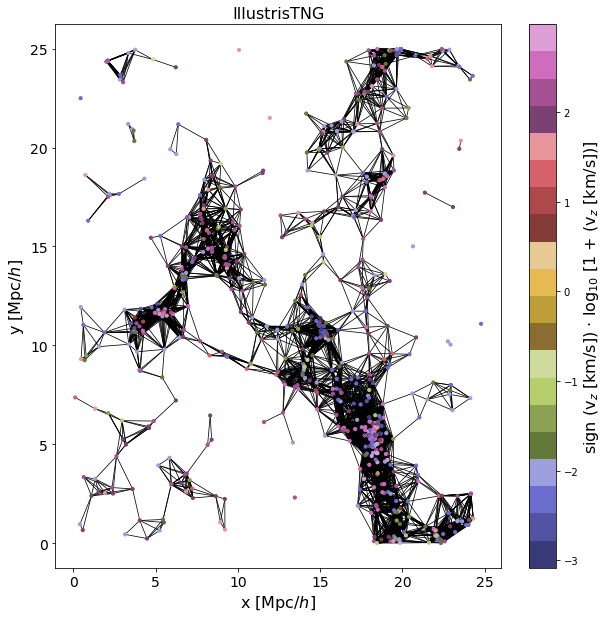}
 \includegraphics[width=5.0cm, height=4.4cm]{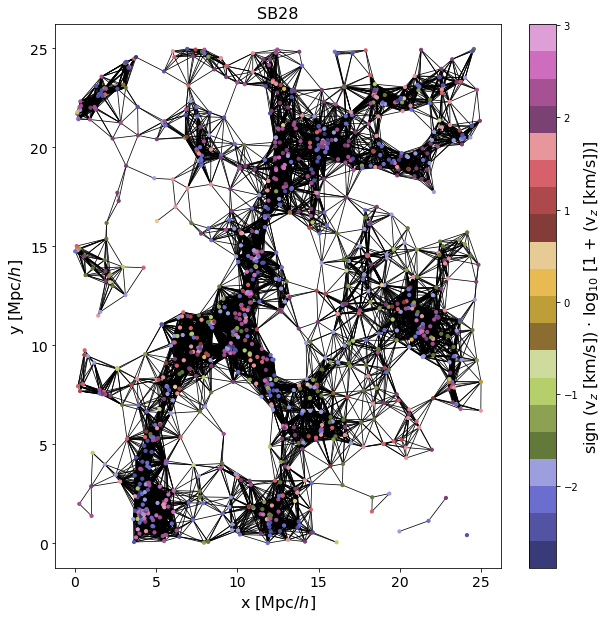}
 \includegraphics[width=5.0cm, height=4.4cm]{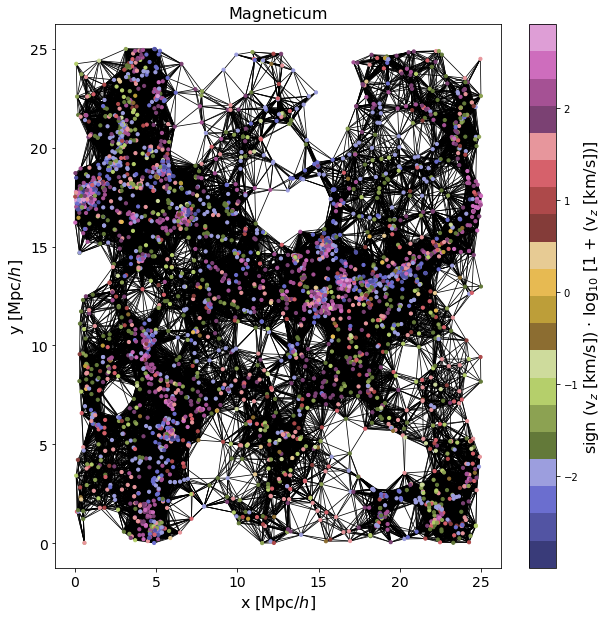}
 \includegraphics[width=5.0cm, height=4.4cm]{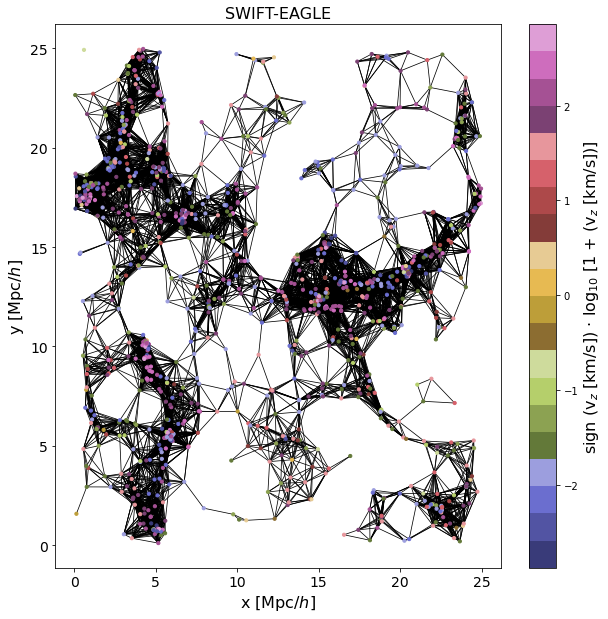}
 \caption{Examples of $2$D graphs (used in Section \ref{sec:2Dpos_1Dvel}) built from
 galaxy catalogs from different CAMELS simulations: Astrid, SIMBA, IllustrisTNG,
 Magneticum, SB28, and SWIFT-EAGLE. 
 The nodes represent the galaxies and their colors correspond to the normalized $z$
 component of their peculiar velocity. 
 Galaxies are connected by edges (shown as black lines) if their distance is smaller
 than the linking radius. 
 We stress that, in this pictorical representation, there are no galaxies which are
 linked due to PBC.}
 \label{fig:graph_details}
\end{figure}

One of the explanations for the success of our previous paper \citep{deSanti2023} was the correlation between the number of
galaxies per catalog versus the $\Omega_{\rm m}$ values. At that time we claimed that the model trained on Astrid contained
catalogs covering a huge range of number of galaxies per value of $\Omega_{\rm m}$, which contains all the other different
simulations.
In the case of the models presented here, the ones which have a impact in the number 
density are the color and star formation rate selections, respectively presented in
Sections \ref{sec:color_results} and \ref{sec:star_forming_results}.
We performed the same analysis for them in Figures 
\ref{fig:color_selection-NxOmegaM} and \ref{fig:star_forming_selection-NxOmegaM}.
They evidence that this correlation, for both scenarios, is not anymore the most important cause for the success of current the predictions. 
This is because, even having the models trained on Astrid and still having this
broader range, while looking at the tests on all the other simulations we do not 
exhibit the same level of success.
An interesting point is related to the percentage of the number of the different $2$ 
sub-populations of red and blue galaxies or even star-forming and non star-forming
galaxies from simulation to simulation. Astrid, Magneticum, and SWIFT-EAGLE, in
average, contain more red than blue galaxies; Astrid contain almost the same number 
of forming and non star-forming galaxies; SIMBA, IllustrisTNG, SB28, and Magneticum
contain more star-forming galaxies. 
Thus, the number of galaxies per catalog is not consistent between the different simulations under the observational effects.
Moreover, this proves that usual characterization of color for galaxies suffer from
the effects of the changes in the astrophysical parameters differently from
simulation to simulation in the LH/SB sequences.

The visual aspects of the $2$D graphs, used in the model of Section 
\ref{sec:2Dpos_1Dvel}, can be found in Figure \ref{fig:graph_details}. 
We present one graph for $1$ CV box of Astrid, SIMBA, IllustrisTNG, and Magneticum 
and $1$ SB/LH box of SB28 and SWIFT-EAGLE. 
Following the same prescription of \cite{deSanti2023}, we are representing the 
galaxies by points, colored according to theirs values of the transformed
velocities, and by black lines, the edge connections. 
Notice that we are not representing the galaxies connected by PBC.
Also, we are always selecting galaxies more massive than out minimum stellar
mass cut of $M_{\star} = 1.95 \cdot 10^8$ M$_{\odot}/h$.
By removing one spatial component this figure shows main differences compared to the $3$D graphs. 
First, due the fact that we have artificially created galaxies closer to each other
(because we removed one spatial component), we allowed more connections (roughly
speaking, we have something $\sim 50$ connections per galaxy, $5$ times more than in
the usual $3$D case of \cite{deSanti2023}).
This number is still larger because, as already mentioned in the previous paragraph, 
the preferred value for the linking radius is larger than in the $3$D graphs.
This can be indicative of how the GNNs keep information to still provide 
good predictions, even while losing the effects at larger scales.

%%%%%%%%%%%%%%
%%%%%%%%%%%%%%%%%%%%%%%%%%%%%%%%%%%%%%%%%%%%%%%%%%
\section{Complete set of metrics for the different systematics}
\label{sec:complete_set}

In the present appendix we present all the set of scores from Section \ref{sec:scores} for the different observational effects seen in the paper in Tables \ref{tab:results1}, \ref{tab:results2}, and \ref{tab:results3}.
Note that we omit some of these scores when the statistic is ill-defined. This is the case of PCC, which is not well defined when all the predictions have the same true value (e.g., SWIFT-EAGLE boxes).

\begin{table}
 \addtolength{\tabcolsep}{-1.7pt}
 \fontsize{9pt}{9pt}\selectfont
 \caption{\label{tab:results1} \textbf{Score values: RMSE, R$^2$, PCC, $b$, $\epsilon$, $\chi^2$.} We present the results for the best model, masking $10 \%$ and $5 \%$ of the galaxies, and absolute velocity perturbation of $150$km/s and $100$km/s. The scores are taken for all and selected galaxy catalogs (by $\chi^2$ values) for Astrid, SIMBA, IllustrisTNG, SB28, Magneticum, and SWIFT-EAGLE.}
 \begin{center}
  \begin{tabular}{ccccccccccc}
   \hline\hline
    & \multicolumn{2}{c}{\textbf{Best}} & \multicolumn{2}{c}{\textbf{Mask} $\mathbf{10 \%}$} & \multicolumn{2}{c}{\textbf{Mask} $\mathbf{5 \%}$} & \multicolumn{2}{c}{\textbf{Abs} $\mathbf{150 km/s}$} & \multicolumn{2}{c}{\textbf{Abs} $\mathbf{100 km/s}$}\\
   \hline\hline
   {\bf Astrid} & \textbf{All} & \textbf{Sel} &\textbf{All} & \textbf{Sel} & \textbf{All} & \textbf{Sel} & \textbf{All} & \textbf{Sel} & \textbf{All} & \textbf{Sel}\\
   \hline
   RMSE &         $0.043$  & $0.043$    & $0.043$   & $0.041$    & $0.040$    & $0.035$ & $0.043$ & $0.036$ & $0.044$    & $0.040$\\
   $R^2$ &        $0.831$  & $0.835$    & $0.848$   & $0.859$    & $0.859$    & $0.863$ & $0.843$   & $0.892$    & $0.830$    & $0.855$\\
   PCC &          $0.921$  & $0.923$    & $0.929$   & $0.934$    & $0.934$    & $0.951$ & $0.923$   & $0.952$    & $0.919$    & $0.935$\\
   $b$ &        $- 0.0091$ & $- 0.0091$ & $- 0.010$ & $- 0.010$  & $- 0.011$  & $- 0.010$ & $- 0.009$ & $- 0.009$  & $- 0.009$  & $- 0.009$\\   
   $\epsilon$ & $11.80\%$  & $11.82\%$  & $12.00\%$ & $11.73 \%$ & $11.41 \%$ & $10.92 \%$ & $11.34\%$ & $9.88 \%$  & $12.21 \%$ & $11.66 \%$\\
   $\chi^2$ &      $2.02$ & $1.65$ & $1.86$ & $1.52$     & $5.31$     & $1.32$ & $2.30$    & $1.45$     & $1.44$     & $1.13$\\
   \hline
   {\bf SIMBA} & \textbf{All} & \textbf{Sel} & \textbf{All} & \textbf{Sel} & \textbf{All} & \textbf{Sel} & \textbf{All} & \textbf{Sel} & \textbf{All} & \textbf{Sel}\\
   \hline
   RMSE &        $0.030$ & $0.030$   & $0.030$ & $0.029$   & $0.031$   & $0.028$ & $0.032$  & $0.028$   & $0.030$   & $0.030$\\
   $R^2$ &       $0.934$ & $0.935$   & $0.935$ & $0.941$   & $0.928$   & $0.940$ & $0.930$  & $0.947$   & $0.931$   & $0.934$\\
   PCC &         $0.967$ & $0.968$   & $0.967$ & $0.970$   & $0.964$   & $0.970$ & $0.968$  & $0.975$   & $0.969$   & $0.97$\\
   $b$ &        $-0.001$ & $-0.001$  & $0.000$ & $0.000$   & $- 0.003$ & $- 0.003$ & $0.010$  & $0.008$   & $0.009$   & $0.008$\\   
   $\epsilon$ & $8.08\%$ & $7.98\%$  & $8.15\%$ & $7.97 \%$ & $8.38 \%$ & $8.00 \%$ & $9.18\%$ & $7.78 \%$ & $8.17 \%$ & $8.04 \%$\\
   $\chi^2$ &     $1.05$ & $0.96$    & $1.42$ & $1.06$    & $2.43$    & $1.27$ & $4.01$   & $1.86$    & $1.10$    & $1.03$\\
   \hline
   {\bf Illustris} & \textbf{All} & \textbf{Sel} & \textbf{All} & \textbf{Sel} & \textbf{All} & \textbf{Sel} & \textbf{All} & \textbf{Sel} & \textbf{All} & \textbf{Sel}\\
   \hline
   RMSE &         $0.038$ & $0.036$    & $0.041$   & $0.036$    & $0.038$     & $0.035$ & $0.041$   & $0.035$    & $0.039$    & $0.037$\\
   $R^2$ &        $0.878$ & $0.887$    & $0.846$   & $0.883$    & $0.876$     & $0.893$ & $0.881$   & $0.904$    & $0.873$    & $0.888$\\
   PCC &          $0.951$ & $0.955$    & $0.948$   & $0.958$    & $0.951$     & $0.956$ & $0.951$   & $0.959$    & $0.955$    & $0.960$\\
   $b$ &          $0.014$ & $0.013$    & $0.018$   & $0.014$    & $0.013$     & $0.011$ & $0.017$   & $0.013$    & $0.019$    & $0.018$\\   
   $\epsilon$ & $10.49\%$ & $10.03 \%$ & $10.48\%$ & $9.90 \%$  & $10.37 \%$  & $9.86 \%$ & $11.19\%$ & $9.29 \%$  & $10.54 \%$ & $10.16 \%$\\
   $\chi^2$ &      $1.72$ & $1.44$     & $2.07$    & $1.44$     & $11.70$     & $1.65$ & $4.74$    & $2.02$     & $1.72$     & $1.49$\\
   \hline
   {\bf SB28} & \textbf{All} & \textbf{Sel} & \textbf{All} & \textbf{Sel} & \textbf{All} & \textbf{Sel} & \textbf{All} & \textbf{Sel} & \textbf{All} & \textbf{Sel}\\
   \hline
   RMSE &         $0.047$ & $0.043$    & $0.047$ & $0.041$    & $0.046$    & $0.040$ & $0.045$   & $0.036$    & $0.046$    & $0.040$\\
   $R^2$ &        $0.822$ & $0.851$    & $0.824$ & $0.865$    & $0.833$    & $0.871$ & $0.850$   & $0.896$    & $0.833$    & $0.871$\\
   PCC &          $0.916$ & $0.929$    & $0.918$ & $0.936$    & $0.922$    & $0.939$ & $0.927$   & $0.950$    & $0.925$    & $0.941$\\
   $b$ &          $0.007$ & $0.003$    & $0.007$ & $0.005$    & $0.007$    & $0.004$ & $0.007$   & $0.003$    & $0.012$    & $0.008$\\   
   $\epsilon$ & $13.21\%$ & $12.32 \%$ & $12.72\%$ & $11.82 \%$ & $12.59 \%$ & $11.39 \%$ & $12.54\%$ & $10.56 \%$ & $12.40 \%$ & $11.40 \%$\\
   $\chi^2$ &      $2.42$ & $1.79$     & $2.31$ & $1.69$     & $3.42$     & $1.95$ & $6.59$    & $2.06$     & $2.29$     & $1.56$\\
   \hline
   {\bf Magnet} & \textbf{All} & \textbf{Sel} & \textbf{All} & \textbf{Sel} & \textbf{All} & \textbf{Sel} & \textbf{All} & \textbf{Sel} & \textbf{All} & \textbf{Sel}\\
   \hline
   RMSE &         $0.034$ & $0.033$    & $0.045$ & $0.033$    & $0.035$    & $0.032$ & $0.063$   & $0.045$    & $0.036$    & $0.036$\\
   $R^2$ &        $0.914$ & $0.923$    & $0.851$ & $0.913$    & $0.923$    & $0.932$ & $0.679$   & $0.799$    & $0.903$    & $0.905$\\
   PCC &          $0.966$ & $0.968$    & $0.935$ & $0.967$    & $0.964$    & $0.966$ & $0.957$   & $0.971$    & $0.971$    & $0.971$\\
   $b$ &          $0.016$ & $0.014$    & $0.017$ & $0.017$    & $0.007$    & $0.004$ & $0.053$   & $0.033$    & $0.023$    & $0.023$\\   
   $\epsilon$ &  $8.79\%$ & $8.29 \%$  & $12.89\%$ & $11.02 \%$ & $11.00 \%$ & $9.79 \%$ & $21.14\%$ & $13.55 \%$ & $11.13 \%$ & $10.65 \%$\\
   $\chi^2$ &      $2.21$ & $1.74$     & $6.33$ & $2.06$     & $2.78$     & $2.25$ & $27.15$   & $4.44$     & $2.39$     & $2.14$\\
   \hline
   {\bf SWIFT} & \multicolumn{2}{c}{\textbf{All}}  & \multicolumn{2}{c}{\textbf{All}} & \multicolumn{2}{c}{\textbf{All}} & \multicolumn{2}{c}{\textbf{All}} & \multicolumn{2}{c}{\textbf{All}}\\
   \hline
   RMSE &        \multicolumn{2}{c}{$0.015$} & \multicolumn{2}{c}{$0.016$} & \multicolumn{2}{c}{$0.014$} & \multicolumn{2}{c}{$0.018$}        & \multicolumn{2}{c}{$0.014$}\\
   $b$ &         \multicolumn{2}{c}{$-0.003$} & \multicolumn{2}{c}{$0.003$} & \multicolumn{2}{c}{$0.005$} & \multicolumn{2}{c}{$0.007$}         & \multicolumn{2}{c}{$0.010$}\\   
   $\epsilon$ & \multicolumn{2}{c}{$3.99\%$} & \multicolumn{2}{c}{$4.13\%$} & \multicolumn{2}{c}{$3.64 \%$} & \multicolumn{2}{c}{$4.51\%$} & \multicolumn{2}{c}{$4.03 \%$}\\
   $\chi^2$ &     \multicolumn{2}{c}{$0.25$} & \multicolumn{2}{c}{$0.40$} & \multicolumn{2}{c}{$0.38$}& \multicolumn{2}{c}{$0.63$}     & \multicolumn{2}{c}{$0.25$}\\
   \hline
   \end{tabular}
  \end{center}
\end{table}

\begin{table}
 \addtolength{\tabcolsep}{-1.5pt}
 \fontsize{9pt}{9pt}\selectfont
 \caption{\label{tab:results2} \textbf{Score values: RMSE, R$^2$, PCC, $b$, $\epsilon$, $\chi^2$.} We present the results for the relative velocity perturbation of $25$\% and $15$\%, masking $5$\% of the galaxies and absolute velocity perturbation of $100$km/s and relative velocity perturbation of $15$\%, and 2D positions and 1D velocity galaxies. The scores are taken for all and selected galaxy catalogs (by $\chi^2$ values) for Astrid, SIMBA, IllustrisTNG, SB28, Magneticum, and SWIFT-EAGLE.}
 \begin{center}
  \begin{tabular}{ccccccccccc}
   \hline\hline
    & \multicolumn{2}{c}{\textbf{Rel 25 \%}} & \multicolumn{2}{c}{\textbf{Rel 15 \%}} & \multicolumn{2}{c}{\textbf{M5\%V100km/s}} & \multicolumn{2}{c}{\textbf{M5\%V15\%}} & \multicolumn{2}{c}{\bf2Dpos1Dvel}\\
   \hline\hline
   {\bf Astrid} & \textbf{All} & \textbf{Sel} & \textbf{All} & \textbf{Sel} & \textbf{All} & \textbf{Sel} & \textbf{All} & \textbf{Sel} & \textbf{All} & \textbf{Sel}\\
   \hline
   RMSE &       $0.064$   & $0.064$    & $0.058$    & $0.058$ & $0.040$   & $0.040$    & $0.057$    & $0.057$ & $0.040$   & $0.038$\\
   $R^2$ &      $0.452$   & $0.452$    & $0.594$    & $0.594$ & $0.863$   & $0.863$    & $0.609$    & $0.609$ & $0.842$   & $0.859$\\
   PCC &        $0.814$   & $0.814$    & $0.845$    & $0.845$ & $0.933$   & $0.933$    & $0.853$    & $0.853$ & $0.931$   & $0.940$\\
   $b$ &        $- 0.012$ & $- 0.012$  & $0.003$  &   $0.003$ & $- 0.005$ & $- 0.005$  & $- 0.001$  & $- 0.001$ & $- 0.010$ & $- 0.010$\\   
   $\epsilon$ & $17.10\%$ & $17.10 \%$ & $15.87 \%$ & $15.87 \%$ & $10.78\%$ & $10.77 \%$ & $15.40 \%$ & $15.40 \%$ & $10.45\%$ & $10.19 \%$\\
   $\chi^2$ &   $1.03$    & $1.03$     & $1.35$     & $1.35$ & $1.61$    & $1.61$     & $1.29$     & $1.29$ & $1.12$    & $0.93$\\
   \hline
   {\bf SIMBA} & \textbf{All} & \textbf{Sel} & \textbf{All} & \textbf{Sel} & \textbf{All} & \textbf{Sel} & \textbf{All} & \textbf{Sel} & \textbf{All} & \textbf{Sel}\\
   \hline
   RMSE       & $0.081$  & $0.066$   & $0.051$   & $0.049$ & $0.029$  & $0.027$   & $0.060$   & $0.051$ & $0.044$  & $0.041$\\
   $R^2$      & $0.401$  & $0.601$   & $0.714$   & $0.737$ & $0.937$  & $0.944$   & $0.727$   & $0.781$ & $0.848$  & $0.868$\\
   PCC        & $0.847$  & $0.898$   & $0.911$   & $0.918$ & $0.970$  & $0.973$   & $0.911$   & $0.928$ & $0.931$  & $0.941$\\
   $b$        & $0.052$  & $0.042$   & $0.017$ & $0.016$ & $0.006$  & $0.0051$   & $0.035$ & $0.027$ & $-0.012$  & $-0.011$\\   
   $\epsilon$ & $20.62\%$ & $18.11 \%$ & $12.81 \%$ & $12.57 \%$ & $7.72\%$ & $7.52 \%$ & $17.17 \%$ & $13.85 \%$ & $12.83\%$ & $12.17 \%$\\
   $\chi^2$   & $3.14$   & $1.73$    & $1.17$    & $1.04$ & $1.36$   & $1.19$    & $5.61$    & $1.70$ & $1.59$   & $1.32$\\
   \hline
   {\bf Illustris} & \textbf{All} & \textbf{Sel} & \textbf{All} & \textbf{Sel} & \textbf{All} & \textbf{Sel} & \textbf{All} & \textbf{Sel} & \textbf{All} & \textbf{Sel}\\
   \hline
   RMSE       & $0.083$   & $0.070$    & $0.064$    & $0.060$ & $0.037$   & $0.035$    & $0.066$    & $0.059$ & $0.046$   & $0.044$\\
   $R^2$      & $0.312$   & $0.519$    & $0.543$    & $0.595$ & $0.887$   & $0.902$    & $0.608$    & $0.650$ & $0.820$   & $0.838$\\
   PCC        & $0.821$   & $0.873$    & $0.859$    & $0.874$ & $0.960$   & $0.964$    & $0.884$    & $0.899$ & $0.920$   & $0.928$\\
   $b$        & $0.050$   & $0.040$    & $0.023$    & $0.020$ & $0.018$   & $0.016$    & $0.037$    & $0.030$ & $-0.007$  & $-0.007$\\   
   $\epsilon$ & $20.63\%$ & $18.55 \%$ & $16.27 \%$ & $15.74 \%$ & $10.09\%$ & $9.50 \%$ & $18.46 \%$ & $15.84 \%$ & $13.21\%$ & $12.75 \%$\\
   $\chi^2$   & $2.87$    & $1.67$     & $1.73$    & $1.38$ & $2.12$    & $1.68$     & $4.11$    & $1.85$ & $1.78$    & $1.37$\\
   \hline
   {\bf SB28} & \textbf{All} & \textbf{Sel} & \textbf{All} & \textbf{Sel} & \textbf{All} & \textbf{Sel} & \textbf{All} & \textbf{Sel} & \textbf{All} & \textbf{Sel}\\
   \hline
   RMSE       & $0.078$   & $0.071$    & $0.065$    & $0.061$ & $0.046$   & $0.038$    & $0.069$    & $0.061$ & $0.055$   & $0.048$\\
   $R^2$      & $0.397$   & $0.481$    & $0.556$    & $0.600$ & $0.833$   & $0.885$    & $0.551$    & $0.624$ & $0.752$   & $0.808$\\
   PCC        & $0.762$   & $0.800$    & $0.842$    & $0.857$ & $0.925$   & $0.947$    & $0.831$    & $0.865$ & $0.885$   & $0.912$\\
   $b$        & $0.018$   & $0.013$    & $0.017$    & $0.014$ & $0.013$   & $0.008$    & $0.023$    & $0.017$ & $-0.008$   & $-0.008$\\   
   $\epsilon$ & $22.11\%$ & $21.01 \%$ & $18.06 \%$ & $17.57 \%$ & $12.27\%$ & $10.72 \%$ & $19.32 \%$ & $17.20 \%$ & $15.93\%$ & $14.45 \%$\\
   $\chi^2$   & $2.89$    & $1.44$     & $1.81$     & $1.52$ & $3.16$    & $1.77$     & $3.39$     & $1.76$ & $2.33$    & $1.65$\\
   \hline
   {\bf Magnet} & \textbf{All} & \textbf{Sel} & \textbf{All} & \textbf{Sel} & \textbf{All} & \textbf{Sel} & \textbf{All} & \textbf{Sel} & \textbf{All} & \textbf{Sel}\\
   \hline
   RMSE       & $0.219$   & $0.086$    & $0.054$    & $0.045$ & $0.038$   & $0.032$    & $0.136$    & $0.090$ & $0.033$   & $0.033$\\
   $R^2$      & $-0.314$   & $-0.033$    & $0.748$    & $0.818$ & $0.905$   & $0.927$    & $-2.588$    & $-2.283$ & $0.906$   & $0.906$\\
   PCC        & $0.515$   & $0.968$    & $0.883$    & $0.916$ & $0.973$   & $0.976$    & $-0.004$    & $0.634$ & $0.963$   & $0.963$\\
   $b$        & $0.144$   & $0.071$    & $-0.002$    & $0.001$ & $0.025$   & $0.019$    & $0.011$    & $0.045$ & $0.010$   & $0.010$\\   
   $\epsilon$ & $59.87\%$ & $25.10 \%$ & $17.70 \%$ & $13.36 \%$ & $12.30\%$ & $9.39 \%$ & $48.67 \%$ & $26.31 \%$ & $8.21\%$ & $8.21 \%$\\
   $\chi^2$   & $14.32$   & $3.48$     & $2.06$     & $1.35$ & $5.40$    & $2.32$     & $6.84$     & $3.83$ & $1.00$    & $1.00$\\
   \hline
   {\bf SWIFT} & \multicolumn{2}{c}{\bf All} & \multicolumn{2}{c}{\bf All} & \multicolumn{2}{c}{\bf All} & \multicolumn{2}{c}{\bf All} & \multicolumn{2}{c}{\bf All}\\
   \hline
   RMSE & \multicolumn{2}{c}{$0.063$} & \multicolumn{2}{c}{$0.032$} & \multicolumn{2}{c}{$0.015$} & $0.041$ & $0.039$ & \multicolumn{2}{c}{$0.040$}\\
   $b$ & \multicolumn{2}{c}{$0.027$} & \multicolumn{2}{c}{$0.007$} & \multicolumn{2}{c}{$0.009$} & $-0.008$ & $-0.010$ & \multicolumn{2}{c}{$-0.033$}\\   
   $\epsilon$ & \multicolumn{2}{c}{$16.80\%$} & \multicolumn{2}{c}{$8.69 \%$} & \multicolumn{2}{c}{$4.10\%$} & $11.05\%$ & $10.63\%$ & \multicolumn{2}{c}{$11.27\%$}\\
   $\chi^2$ & \multicolumn{2}{c}{$1.74$} & \multicolumn{2}{c}{$0.47$} & \multicolumn{2}{c}{$0.36$} & $0.89$ & $0.70$ & \multicolumn{2}{c}{$1.01$}\\
   \hline
   \end{tabular}
  \end{center}
\end{table}

\begin{table}
 \addtolength{\tabcolsep}{-1.5pt}
 \fontsize{9pt}{9pt}\selectfont
 \caption{\label{tab:results3} \textbf{Score values: RMSE, R$^2$, PCC, $b$, $\epsilon$, $\chi^2$.} We present the results for blue and red, star forming and non star-forming galaxies. The scores are taken for all and selected galaxy catalogs (by $\chi^2$ values) for Astrid, SIMBA, IllustrisTNG, SB28, Magneticum, and SWIFT-EAGLE.}
 \begin{center}
  \begin{tabular}{ccccccccc}
   \hline\hline
   & \multicolumn{2}{c}{\bf Blues} & \multicolumn{2}{c}{\bf Reds} & \multicolumn{2}{c}{\bf StarForm} & \multicolumn{2}{c}{\bf Non StarForm}\\
   \hline\hline
   {\bf Astrid} & \textbf{All} & \textbf{Sel} & \textbf{All} & \textbf{Sel} & \textbf{All} & \textbf{Sel} & \textbf{All} & \textbf{Sel}\\
   \hline
   RMSE &       $0.046$   & $0.046$    & $0.057$    & $0.053$ & $0.038$   & $0.038$    & $0.056$    & $0.052$\\
   $R^2$ &      $0.788$   & $0.788$    & $0.653$    & $0.691$ & $0.864$   & $0.864$    & $0.714$    & $0.723$\\
   PCC &        $0.907$   & $0.907$    & $0.858$    & $0.874$ & $0.936$   & $0.936$    & $0.867$    & $0.879$\\
   $b$ &        $- 0.001$ & $- 0.001$  & $- 0.007$  & $- 0.007$ & $- 0.003$ & $- 0.003$  & $0.004$  & $ 0.004$\\   
   $\epsilon$ & $11.76\%$ & $11.76 \%$ & $14.96 \%$ & $14.27 \%$ & $10.65\%$ & $10.65 \%$ & $14.21 \%$ & $12.88 \%$\\
   $\chi^2$ &   $1.16$    & $1.16$     & $2.11$     & $1.93$ & $0.89$    & $0.89$     & $2.39$     & $1.50$\\
   \hline
   {\bf SIMBA} & \textbf{All} & \textbf{Sel} & \textbf{All} & \textbf{Sel}\\
   \hline
   RMSE       & $0.041$  & $0.040$   & $0.072$   & $0.053$ & $0.043$   & $0.041$    & $0.068$    & $0.061$\\
   $R^2$      & $0.854$  & $0.860$   & $0.516$   & $0.708$ & $0.833$   & $0.842$    & $0.468$    & $0.585$\\
   PCC        & $0.956$  & $0.958$   & $0.830$   & $0.919$ & $0.952$   & $0.956$    & $0.904$    & $0.923$\\
   $b$        & $0.022$  & $0.022$   & $0.032$ & $0.024$ & $0.025$   & $0.024$    & $0.046$    & $0.039$\\   
   $\epsilon$ & $10.35\%$ & $10.20 \%$ & $16.65 \%$ & $14.87 \%$ & $11.16\%$ & $10.35 \%$ & $17.75 \%$ & $16.05 \%$\\
   $\chi^2$   & $1.34$   & $1.26$    & $3.52$    & $2.19$ & $16.70$   & $1.63$     & $4.09$     & $2.43$\\
   \hline
   {\bf IllustrisTNG} & \textbf{All} & \textbf{Sel} & \textbf{All} & \textbf{Sel} & \textbf{All} & \textbf{Sel} & \textbf{All} & \textbf{Sel}\\
   \hline
   RMSE       & $0.044$   & $0.044$    & $0.062$    & $0.053$ & $0.042$   & $0.041$    & $0.066$    & $0.057$\\
   $R^2$      & $0.809$   & $0.816$    & $0.563$    & $0.689$ & $0.846$   & $0.847$    & $0.567$    & $0.683$\\
   PCC        & $0.939$   & $0.941$    & $0.886$    & $0.914$ & $0.938$   & $0.938$    & $0.892$    & $0.916$\\
   $b$        & $0.019$   & $0.018$    & $0.030$    & $0.023$ & $0.013$   & $0.012$    & $0.040$    & $0.032$\\   
   $\epsilon$ & $11.23\%$ & $11.13 \%$ & $16.43 \%$ & $15.11 \%$ & $11.36\%$ & $11.13 \%$ & $16.71 \%$ & $14.97 \%$\\
   $\chi^2$   & $1.29$    & $1.24$     & $3.15$     & $2.09$ & $5.46$    & $1.39$     & $3.92$     & $2.12$\\
   \hline
   {\bf SB28} & \textbf{All} & \textbf{Sel} & \textbf{All} & \textbf{Sel} & \textbf{All} & \textbf{Sel} & \textbf{All} & \textbf{Sel}\\
   \hline
   RMSE       & $0.066$   & $0.052$    & $0.061$    & $0.052$ & $0.060$   & $0.048$    & $0.063$    & $0.054$\\
   $R^2$      & $0.580$   & $0.736$    & $0.645$    & $0.740$ & $0.669$   & $0.788$    & $0.653$    & $0.738$\\
   PCC        & $0.847$   & $0.902$    & $0.850$    & $0.893$ & $0.871$   & $0.916$    & $0.849$    & $0.888$\\
   $b$        & $0.024$   & $0.016$    & $0.006$    & $0.002$ & $0.020$   & $0.013$    & $0.014$    & $0.008$\\   
   $\epsilon$ & $16.59\%$ & $14.64 \%$ & $17.66 \%$ & $15.97 \%$ & $15.19\%$ & $13.26 \%$ & $17.35 \%$ & $15.70 \%$\\
   $\chi^2$   & $3.43$    & $1.79$     & $4.90$     & $1.96$ & $6.63$    & $1.73$     & $4.21$     & $1.93$\\
   \hline
   {\bf Magneticum} & \textbf{All} & \textbf{Sel} & \textbf{All} & \textbf{Sel} & \textbf{All} & \textbf{Sel} & \textbf{All} & \textbf{Sel}\\
   \hline
   RMSE       & $0.042$   & $0.041$    & $0.041$    & $0.037$ & $0.042$   & $0.042$    & $0.059$    & $0.046$\\
   $R^2$      & $0.843$   & $0.851$    & $0.880$    & $0.870$ & $0.857$   & $0.850$    & $0.782$    & $0.870$\\
   PCC        & $0.961$   & $0.964$    & $0.970$    & $0.965$ & $0.961$   & $0.958$    & $0.959$    & $0.976$\\
   $b$        & $0.027$   & $0.026$    & $-0.029$    & $-0.025$ & $0.027$   & $0.027$    & $-0.047$   & $-0.036$\\   
   $\epsilon$ & $11.35\%$ & $10.79 \%$ & $12.96 \%$ & $12.30 \%$ & $11.14\%$ & $10.85 \%$ & $17.30 \%$ & $13.45 \%$\\
   $\chi^2$   & $1.62$    & $1.36$     & $8.46$     & $1.87$ & $4.30$    & $1.88$     & $4.28$     & $2.12$\\
   \hline
   {\bf SWIFT-EAGLE} & \multicolumn{2}{c}{\textbf{All}} & \multicolumn{2}{c}{\textbf{Sel}} & \multicolumn{2}{c}{\textbf{All}} & \multicolumn{2}{c}{\textbf{Sel}}\\
   \hline
   RMSE & \multicolumn{2}{c}{$0.049$}         & \multicolumn{2}{c}{$0.028$} & \multicolumn{2}{c}{$0.040$}         & \multicolumn{2}{c}{$0.042$}\\
   $b$ & \multicolumn{2}{c}{$0.045$}          & \multicolumn{2}{c}{$-0.017$} & \multicolumn{2}{c}{$0.037$}          & \multicolumn{2}{c}{$-0.039$}\\   
   $\epsilon$ & \multicolumn{2}{c}{$15.08\%$} & \multicolumn{2}{c}{$6.58 \%$} & \multicolumn{2}{c}{$12.24\%$} & \multicolumn{2}{c}{$12.90 \%$}\\
   $\chi^2$ & \multicolumn{2}{c}{$2.04$}      & \multicolumn{2}{c}{$0.62$} & \multicolumn{2}{c}{$1.54$}      & \multicolumn{2}{c}{$1.66$}\\
   \hline
   \end{tabular}
  \end{center}
\end{table}

%%%%%%%%%%%%%%%%%%%%%%%%%%%%%%%%%%%%%%%%%%%%%%%%%%
\section{The best model from the previous paper}
\label{sec:best_model}

\begin{figure}[!ht]
    \centering
    \includegraphics[scale=0.31]{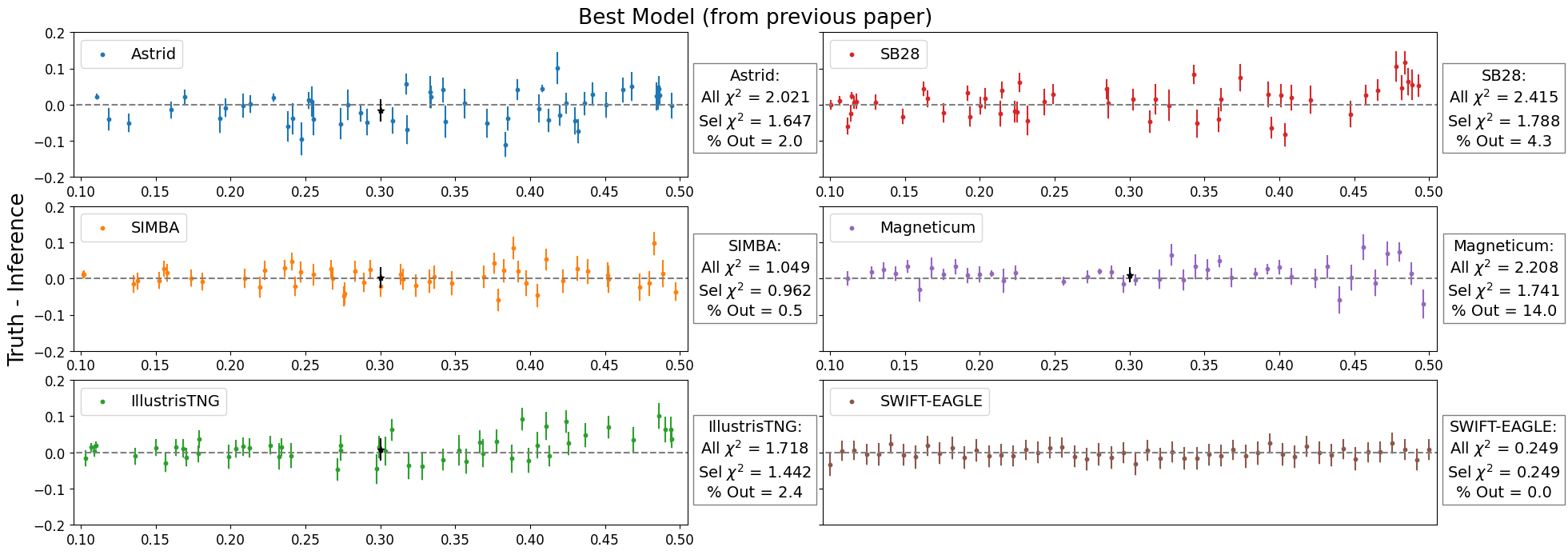}
    \caption{{\bf Truth - Inference of $\Omega_{\rm m}$ -- Best model:} all the catalogs, with all the galaxies were considered. We present the predictions for galaxy catalogs from Astrid, SIMBA, IllustrisTNG, SB28, Magneticum, and SWIFT-EAGLE. For each simulation suite, we indicate the average $\chi^2$ value across all galaxy catalogs in the test set. We also list the $\chi^2$ values after removing outliers, which are selected as catalogs whose predictions exhibit $\chi^2 > 10$ (a different choice from the previous paper \cite{deSanti2023}) and present the percentage of outliers (percentage of catalogs removed after this selection).}
    \label{fig:best_model}
\end{figure}

In this section we re-present the predictions for the best model, found in the
previous work \cite{deSanti2023}, in Figure \ref{fig:best_model}. It is important to notice we have presented the complete set of scores for this model in the Appendix \ref{sec:complete_set} (see Table \ref{tab:results1}) and to stress that we are now considering a selection of $\chi^2 > 10$, the same criteria presented in this paper.

%%%%%%%%%%%%%%

\acknowledgments

We would like to thank Ravi K. Sheth, Michael Strauss, and the CAMELS team for the
enlightening discussions and valuable comments. 
We thank the São Paulo Research Foundation (FAPESP), the Brazilian National Council for
Scientific and Technological Development (CNPq), and the Simons Foundation for financial
support. 
NSMS acknowledges financial support from FAPESP, grants
\href{https://bv.fapesp.br/en/bolsas/187647/cosmological-covariance-matrices-and-machine-learning-methods/}{2019/13108-0} 
and \href{https://bv.fapesp.br/en/bolsas/202438/machine-learning-methods-for-extracting-cosmological-information/}{2022/03589-4}. 
TC is supported by the INFN INDARK PD51 grant and the FARE MIUR grant ``ClustersXEuclid'' R165SBKTMA.
EHM is supported by the grant agreements ANR-21-CE31-0019 / 490702358 from the French Agence Nationale de la Recherche / Deutsche Forschungsgemeinschaft (DFG, German Research Foundation) for the LOCALIZATION project.
KD acknowledges support by the COMPLEX project from the European Research Council (ERC) under the European Union’s Horizon 2020 research and innovation program grant agreement ERC-2019-AdG 882679.
The CAMELS project is supported by the Simons Foundation and NSF grant AST 2108078. 
The training of the GNNs has been carried out using graphic processing units from
Simons Foundation, Flatiron Institute, Center of Computational Astrophysics.

%This is the most common positions for acknowledgments. A macro is
%available to maintain the same layout and spelling of the heading.



\begin{thebibliography}{99}

\bibitem{Feng2010} Feng, J.~L.\ 2010, ARAA, 48, 495. doi:10.1146/annurev-astro-082708-101659

\bibitem{Riess2022} Riess, A.~G., Yuan, W., Macri, L.~M., et al.\ 2022, ApJL, 934, L7. doi:10.3847/2041-8213/ac5c5b

\bibitem{DIVALENTINO2021} Di Valentino, E., Anchordoqui, L.~A., Akarsu, {\"O}., et al.\ 2021, Astroparticle Physics, 131, 102604. doi:10.1016/j.astropartphys.2021.102604

\bibitem{Planck2014} Planck Collaboration, Ade, P.~A.~R., Aghanim, N., et al.\ 2014, AAP, 571, A22. doi:10.1051/0004-6361/

\bibitem{DESI} DESI Collaboration, Aghamousa, A., Aguilar, J., et al.\ 2016, arXiv:1611.00036. doi:10.48550/arXiv.1611.00036

\bibitem{Benitez2014} Benitez, N., Dupke, R., Moles, M., et al.\ 2014, arXiv:1403.5237. doi:10.48550/arXiv.1403.5237

\bibitem{Euclid2022-Tiago_Castro} Euclid Collaboration, Castro, T., Fumagalli, A., et al.\ 2023, AAP, 671, A100. doi:10.1051/0004-6361/202244674

\bibitem{Euclid2016} Racca, G.~D., Laureijs, R., Stagnaro, L., et al.\ 2016, ProscPie, 9904, 99040O. doi:10.1117/12.2230762

\bibitem{Roman2015} Spergel, D., Gehrels, N., Baltay, C., et al.\ 2015, arXiv:1503.03757. doi:10.48550/arXiv.1503.03757

\bibitem{JWST} Pontoppidan, K.~M., Barrientes, J., Blome, C., et al.\ 2022, ApJL, 936, L14. doi:10.3847/2041-8213/ac8a4e

\bibitem{Vogelsberger2020} Vogelsberger, M., Marinacci, F., Torrey, P., et al.\ 2020, Nature Reviews Physics, 2, 42. doi:10.1038/s42254-019-0127-2

\bibitem{Astrid2022} Bird, S., Ni, Y., Di Matteo, T., et al.\ 2022, MNRAS, 512, 3703. doi:10.1093/mnras/stac648

\bibitem{SIMBA2019} Dav{\'e}, R., Angl{\'e}s-Alc{\'a}zar, D., Narayanan, D., et al.\ 2019, MNRAS, 486, 2827. doi:10.1093/mnras/stz937

\bibitem{Pillepich2018} Pillepich, A., Nelson, D., Hernquist, L., et al.\ 2018, MNRAS, 475, 648. doi:10.1093/mnras/stx3112

\bibitem{Vogelsberger2014Nature} Vogelsberger, M., Genel, S., Springel, V., et al.\ 2014, NAT, 509, 177. doi:10.1038/nature13316

\bibitem{Vogelsberger2014MNRAS} Vogelsberger, M., Genel, S., Springel, V., et al.\ 2014, MNRAS, 444, 1518. doi:10.1093/mnras/stu1536

\bibitem{MAGNETICUM2014} Hirschmann, M., Dolag, K., Saro, A., et al.\ 2014, MNRAS, 442, 2304. doi:10.1093/mnras/stu1023

\bibitem{Schaye2015} Schaye, J., Crain, R.~A., Bower, R.~G., et al.\ 2015, MNRAS, 446, 521. doi:10.1093/mnras/

\bibitem{Somerville1999} Somerville, R.~S. \& Primack, J.~R.\ 1999, MNRAS, 310, 1087. doi:10.1046/j.1365-8711.1999.03032.x

\bibitem{Contreras2021} Contreras, S., Angulo, R.~E., \& Zennaro, M.\ 2021, MNRAS, 508, 175. doi:10.1093/mnras/stab2560

\bibitem{Klypin2015} Klypin, A., Prada, F., Yepes, G., et al.\ 2015, MNRAS, 447, 3693. doi:10.1093/mnras/stu2685

\bibitem{Wu2023} Wu, J.~F. \& Kragh Jespersen, C.\ 2023, arXiv:2306.12327. doi:10.48550/arXiv.2306.12327

\bibitem{Rodrigues2023} Rodrigues, N.~V.~N., de Santi, N.~S.~M., Montero-Dorta, A.~D., et al.\ 2023, MNRAS, 522, 3236. doi:10.1093/mnras/stad1186

\bibitem{Cristian2023} Hern{\'a}ndez, C.~A., Gonz{\'a}lez, R.~E., \& Padilla, N.~D.\ 2023, MNRAS, 524, 4653. doi:10.1093/mnras/stad2112

\bibitem{Lovell2023} Lovell, C.~C., Hassan, S., Villaescusa-Navarro, F., et al.\ 2023, Machine Learning for Astrophysics, 21. doi:10.48550/arXiv.2307.06967

\bibitem{deSanti2022MNRAS} de Santi, N.~S.~M., Rodrigues, N.~V.~N., Montero-Dorta, A.~D., et al.\ 2022, MNRAS, 514, 2463. doi:10.1093/mnras/stac1469

\bibitem{Christian2022} Jespersen, C.~K., Cranmer, M., Melchior, P., et al.\ 2022, ApJ, 941, 7. doi:10.3847/1538-4357/ac9b18

\bibitem{Lovell2022} Lovell, C.~C., Wilkins, S.~M., Thomas, P.~A., et al.\ 2022, MNRAS, 509, 5046. doi:10.1093/mnras/stab3221

\bibitem{Shao2021} Shao, H., Villaescusa-Navarro, F., Genel, S., et al.\ 2022, ApJ, 927, 85. doi:10.3847/1538-4357/ac4d30

\bibitem{vonMarttens2021} von Marttens, R., Casarini, L., Napolitano, N.~R., et al.\ 2022, MNRAS, 516, 3924. doi:10.1093/mnras/stac2449

\bibitem{Wadekar2020} Wadekar, D., Villaescusa-Navarro, F., Ho, S., et al.\ 2020, arXiv:2012.00111. doi:10.48550/arXiv.2012.00111

\bibitem{Jo2019} Jo, Y. \& Kim, J.-. hoon .\ 2019, MNRAS, 489, 3565. doi:10.1093/mnras/stz2304

\bibitem{Yip2019} Yip, J.~H.~T., Zhang, X., Wang, Y., et al.\ 2019, arXiv:1910.07813. doi:10.48550/arXiv.1910.07813

\bibitem{Zhang2019} Zhang, X., Wang, Y., Zhang, W., et al.\ 2019, arXiv:1902.05965. doi:10.48550/arXiv.1902.05965

\bibitem{Kamdar2016} Kamdar, H.~M., Turk, M.~J., \& Brunner, R.~J.\ 2016, MNRAS, 457, 1162. doi:10.1093/mnras/stv2981

\bibitem{MCMC2010} Goodman, J. \& Weare, J.\ 2010, Communications in Applied Mathematics and Computational Science, 5, 65. doi:10.2140/camcos.2010.5.65

\bibitem{VI2022} Gunapati, G., Jain, A., Srijith, P.~K., et al.\ 2022, PASA, 39, e001. doi:10.1017/pasa.2021.64

\bibitem{NS2006} Skilling, J.\ 2004, Bayesian Inference and Maximum Entropy Methods in Science and Engineering: 24th International Workshop on Bayesian Inference and Maximum Entropy Methods in Science and Engineering, 735, 395. doi:10.1063/1.1835238

\bibitem{Gualdi2021} Gualdi, D., Gil-Mar{\'\i}n, H., \& Verde, L.\ 2021, JCAP, 2021, 008. doi:10.1088/1475-7516/2021/07/008

\bibitem{Banerjee2021} Banerjee, A. \& Abel, T.\ 2021, MNRAS, 500, 5479. doi:10.1093/mnras/staa3604

\bibitem{Hahn2020} Hahn, C., Villaescusa-Navarro, F., Castorina, E., et al.\ 2020, JCAP, 2020, 040. doi:10.1088/1475-7516/2020/03/040

\bibitem{Uhlemann2020} Uhlemann, C., Friedrich, O., Villaescusa-Navarro, F., et al.\ 2020, MNRAS, 495, 4006. doi:10.1093/mnras/staa1155

\bibitem{CARPool2022} Chartier, N. \& Wandelt, B.~D.\ 2022, MNRAS, 509, 2220. doi:10.1093/mnras/stab3097

\bibitem{Alan2018} Heavens, A.~F., Sellentin, E., de Mijolla, D., et al.\ 2017, MNRAS, 472, 4244. doi:10.1093/mnras/stx2326

\bibitem{deSanti2022JCAP} de Santi, N.~S.~M. \& Abramo, L.~R.\ 2022, JCAP, 2022, 013. doi:10.1088/1475-7516/2022/09/013

\bibitem{Taylor2013} Taylor, A., Joachimi, B., \& Kitching, T.\ 2013, MNRAS, 432, 1928. doi:10.1093/mnras/stt270

\bibitem{lucia2022} Perez, L.~A., Genel, S., Villaescusa-Navarro, F., et al.\ 2023, ApJ, 954, 11. doi:10.3847/1538-4357/accd52

\bibitem{Lemos2023} Lemos, P., Coogan, A., Hezaveh, Y., et al.\ 2023, 40th International Conference on Machine Learning, 202, 19256. doi:10.48550/arXiv.2302.03026

\bibitem{Cranmer2020} Cranmer, K., Brehmer, J., \& Louppe, G.\ 2020, Proceedings of the National Academy of Science, 117, 30055. doi:10.1073/pnas.1912789117

\bibitem{Paco2022} Villaescusa-Navarro, F., Genel, S., Angl{\'e}s-Alc{\'a}zar, D., et al.\ 2022, ApJS, 259, 61. doi:10.3847/1538-4365/ac5ab0

\bibitem{Paco2021} Villaescusa-Navarro, F., Angl{\'e}s-Alc{\'a}zar, D., Genel, S., et al.\ 2021, arXiv:2109.09747. doi:10.48550/arXiv.2109.09747

\bibitem{Kacprzak2022} Kacprzak, T. \& Fluri, J.\ 2022, Physical Review X, 12, 031029. doi:10.1103/PhysRevX.12.031029

\bibitem{Massara2023} Massara, E., Villaescusa-Navarro, F., \& Percival, W.~J.\ 2023, JCAP, 2023, 012. doi:10.1088/1475-7516/2023/12/

\bibitem{pablo-galaxies-2022} Villanueva-Domingo, P. \& Villaescusa-Navarro, F.\ 2022, ApJ, 937, 115. doi:10.3847/1538-4357/ac8930

\bibitem{helen-halos-2022} Shao, H., Villaescusa-Navarro, F., Villanueva-Domingo, P., et al.\ 2022, arXiv:2209.06843. doi:10.48550/arXiv.2209.06843

\bibitem{lucas2022} Makinen, T.~L., Charnock, T., Lemos, P., et al.\ 2022, The Open Journal of Astrophysics, 5, 18. doi:10.21105/astro.2207.05202

\bibitem{Anagnostidis_2022} Anagnostidis, S., Thomsen, A., Kacprzak, T., et al.\ 2022, arXiv:2211.12346. doi:10.48550/arXiv.2211.12346

\bibitem[de Santi et al. (2023)]{deSanti2023} de Santi, N.~S.~M., Shao, H., Villaescusa-Navarro, F., et al.\ 2023, ApJ, 952, 69. doi:10.3847/1538-4357/acd1e2

\bibitem{helen2023} Shao, H., de Santi, N.~S.~M., Villaescusa-Navarro, F., et al.\ 2023, ApJ, 956, 149. doi:10.3847/1538-4357/acee6f

\bibitem{Coupon2018} Coupon, J., Czakon, N., Bosch, J., et al.\ 2018, PASJ, 70, S7. doi:10.1093/pasj/psx047

\bibitem{Coupon2009} Coupon, J., Ilbert, O., Kilbinger, M., et al.\ 2009, AAP, 500, 981. doi:10.1051/0004-6361/200811413

\bibitem{Heymans2012} Heymans, C., Van Waerbeke, L., Miller, L., et al.\ 2012, MNRAS, 427, 146. doi:10.1111/j.1365-2966.2012.21952.x

\bibitem{SLOAN2022} Howlett, C., Said, K., Lucey, J.~R., et al.\ 2022, MNRAS, 515, 953. doi:10.1093/mnras/

\bibitem{SLOAN_catalog-2022} Howlett, C., Said, K., Lucey, J.~R., et al.\ 2022, MNRAS, 515, 953. doi:10.1093/mnras/stac1681

\bibitem{Kourkchi2020} Kourkchi, E., Tully, R.~B., Eftekharzadeh, S., et al.\ 2020, ApJ, 902, 145. doi:10.3847/1538-4357/abb66b

\bibitem{Howlett2017} Howlett, C., Staveley-Smith, L., \& Blake, C.\ 2017, MNRAS, 464, 2517. doi:10.1093/mnras/stw2466

\bibitem{Zehavi2002} Zehavi, I., Blanton, M.~R., Frieman, J.~A., et al.\ 2002, ApJ, 571, 172. doi:10.1086/339893

\bibitem{Madgwick2003} Madgwick, D.~S., Hawkins, E., Lahav, O., et al.\ 2003, MNRAS, 344, 847. doi:10.1046/j.1365-8711.2003.06861.x

\bibitem{EAGLE2015} Schaye, J., Crain, R.~A., Bower, R.~G., et al.\ 2015, MNRAS, 446, 521. doi:10.1093/mnras/stu2058

\bibitem{Ni2023} Ni, Y., Genel, S., Angl{\'e}s-Alc{\'a}zar, D., et al.\ 2023, ApJ, 959, 136. doi:10.3847/1538-4357/

\bibitem{Villaescusa-Navarro2022-relCAMELS} Villaescusa-Navarro, F., Genel, S., Angl{\'e}s-Alc{\'a}zar, D., et al.\ 2023, ApJS, 265, 54. doi:10.3847/1538-4365/acbf47

\bibitem{Paco2021-projCAMELS} Villaescusa-Navarro, F., Angl{\'e}s-Alc{\'a}zar, D., Genel, S., et al.\ 2021, ApJ, 915, 71. doi:10.3847/1538-4357/abf7ba

\bibitem{LH2016} Shields, M. D. and Zhang, J.\ 2016, Reliability Engineering \& System Safety, 148, 96, doi: https://doi.org/10.1016/j.ress.2015.12.002

\bibitem{Yueying2023} Ni, Y., Genel, S., Angl{\'e}s-Alc{\'a}zar, D., et al.\ 2023, ApJ, 959, 136. doi:10.3847/1538-4357/ad022a

\bibitem{MPGadget}Feng, Y., Bird, S., Anderson, L., Font-Ribera, A., \& Pedersen, C. 2018, MP-Gadget/MP-Gadget: A tag for getting a DOI, FirstDOI, Zenodo, doi: 10.5281/zenodo.1451799

\bibitem{Astrid-Y-2022} Ni, Y., Di Matteo, T., Bird, S., et al.\ 2022, MNRAS, 513, 670. doi:10.1093/mnras/stac351

\bibitem{Hopkins2015} Hopkins, P.~F.\ 2015, MNRAS, 450, 53. doi:10.1093/mnras/stv195

\bibitem{Weinberger2020} Weinberger, R., Springel, V., \& Pakmor, R.\ 2020, ApJS, 248, 32. doi:10.3847/1538-4365/ab908c

\bibitem{springel2010} Springel, V.\ 2010, MNRAS, 401, 791. doi:10.1111/j.1365-2966.2009.15715.x

\bibitem{Weinberger2016} Weinberger, R., Springel, V., Hernquist, L., et al.\ 2017, MNRAS, 465, 3291. doi:10.1093/mnras/stw2944

\bibitem{GADGET2} Springel, V.\ 2005, MNRAS, 364, 1105. doi:10.1111/j.1365-2966.2005.09655.x

\bibitem{Tornatore2007} Tornatore, L., Borgani, S., Dolag, K., et al.\ 2007, MNRAS, 382, 1050. doi:10.1111/j.1365-2966.2007.12070.x

\bibitem{Springel2005MNRAS} Springel, V., Di Matteo, T., \& Hernquist, L.\ 2005, MNRAS, 361, 776. doi:10.1111/j.1365-2966.2005.09238.x

\bibitem{DiMatteo2005} Di Matteo, T., Springel, V., \& Hernquist, L.\ 2005, NAT, 433, 604. doi:10.1038/nature03335

\bibitem{Springel2002} Springel, V. \& Hernquist, L.\ 2002, MNRAS, 333, 649. doi:10.1046/j.1365-8711.2002.05445.x

\bibitem{Springel2003MNRAS} Springel, V. \& Hernquist, L.\ 2003, MNRAS, 339, 289. doi:10.1046/j.1365-8711.2003.06206.x

\bibitem{Steinborn2016MNRAS} Steinborn, L.~K., Dolag, K., Comerford, J.~M., et al.\ 2016, MNRAS, 458, 1013. doi:10.1093/mnras/stw316

\bibitem{Hirschmann2014MNRAS} Hirschmann, M., Dolag, K., Saro, A., et al.\ 2014, MNRAS, 442, 2304. doi:10.1093/mnras/stu1023

\bibitem{Fabjan2011MNRAS} Fabjan, D., Borgani, S., Rasia, E., et al.\ 2011, MNRAS, 416, 801. doi:10.1111/j.1365-2966.2011.18497.

\bibitem{Dolag2006MNRAS} Dolag, K., Meneghetti, M., Moscardini, L., et al.\ 2006, MNRAS, 370, 656. doi:10.1111/j.1365-2966.2006.10511.x

\bibitem{Dolag2005MNRAS} Dolag, K., Vazza, F., Brunetti, G., et al.\ 2005, MNRAS, 364, 753. doi:10.1111/j.1365-2966.2005.09630.

\bibitem{Dolag2004ApJ} Dolag, K., Jubelgas, M., Springel, V., et al.\ 2004, ApJL, 606, L97. doi:10.1086/420966

\bibitem{SWIFT2023} Schaller, M., Borrow, J., Draper, P.~W., et al.\ 2024, MNRAS. doi:10.1093/mnras/stae922

\bibitem{Schaller2018} Schaller, M., Gonnet, P., Draper, P.~W., et al.\ 2018, Astrophysics Source Code Library. ascl:1805.020

\bibitem{Schaller2016} Schaller, M., Gonnet, P., Chalk, A.~B.~G., et al.\ 2016, Proceedings of the Platform for Advanced Scientific Computing Conference, 2. doi:10.1145/2929908.2929916

\bibitem{Crain2015} Crain, R.~A., Schaye, J., Bower, R.~G., et al.\ 2015, MNRAS, 450, 1937. doi:10.1093/mnras/stv725

\bibitem{EAGLE-Borrow2022} Borrow, J., Schaller, M., Bah{\'e}, Y.~M., et al.\ 2023, MNRAS, 526, 2441. doi:10.1093/mnras/stad2928

\bibitem{Dolag2009} Dolag, K., Borgani, S., Murante, G., et al.\ 2009, MNRAS, 399, 497. doi:10.1111/j.1365-2966.2009.15034.x

\bibitem{Springel2001} Springel, V., White, S.~D.~M., Tormen, G., et al.\ 2001, MNRAS, 328, 726. doi:10.1046/j.1365-8711.2001.04912.x

\bibitem{velociraptop1} Elahi, P.~J., Ca{\~n}as, R., Poulton, R.~J.~J., et al.\ 2019, PASA, 36, e021. doi:10.1017/pasa.2019.12

\bibitem{velociraptor2} Ca{\~n}as, R., Elahi, P.~J., Welker, C., et al.\ 2019, MNRAS, 482, 2039. doi:10.1093/mnras/sty2725

\bibitem{fight-halo_finders-2022} G{\'o}mez, J.~S., Padilla, N.~D., Helly, J.~C., et al.\ 2022, MNRAS, 510, 5500. doi:10.1093/mnras/stab3661

\bibitem{Tonry1988} Tonry, J. \& Schneider, D.~P.\ 1988, AJ, 96, 807. doi:10.1086/114847

\bibitem{Tully1977} Tully, R.~B. \& Fisher, J.~R.\ 1977, AAP, 54, 661

\bibitem{Hubble1929} Hubble, E.\ 1929, Proceedings of the National Academy of Science, 15, 168. doi:10.1073/pnas.15.3.168

\bibitem{Bahcall1996} Bahcall, N.~A. \& Oh, S.~P.\ 1996, ApJL, 462, L49. doi:10.1086/310041

\bibitem{Lan2023} Lan, T.-W., Tojeiro, R., Armengaud, E., et al.\ 2023, ApJ, 943, 68. doi:10.3847/1538-4357/

\bibitem{Andersen2016} Andersen, P., Davis, T.~M., \& Howlett, C.\ 2016, MNRAS, 463, 4083. doi:10.1093/mnras/stw2252

\bibitem{Zhou2018} Zhou, J., Cui, G., Hu, S., et al.\ 2018, arXiv:1812.08434. doi:10.48550/arXiv.1812.08434

\bibitem{Battaglia2018} Battaglia, P.~W., Hamrick, J.~B., Bapst, V., et al.\ 2018, arXiv:1806.01261. doi:10.48550/arXiv.1806.01261

\bibitem{Gilmer2017} Gilmer, J., Schoenholz, S.~S., Riley, P.~F., et al.\ 2017, arXiv:1704.01212. doi:10.48550/arXiv.1704.01212

\bibitem{Jeffrey2020} Jeffrey, N. \& Wandelt, B.~D.\ 2020, arXiv:2011.05991. doi:10.48550/arXiv.2011.05991

\bibitem{Bronstein2021} Bronstein, M.~M., Bruna, J., Cohen, T., et al.\ 2021, arXiv:2104.13478. doi:10.48550/arXiv.2104.13478

\bibitem{Corso2020} Corso, G., Cavalleri, L., Beaini, D., et al.\ 2020, arXiv:2004.05718. doi:10.48550/arXiv.2004.05718

\bibitem{Li2017} Li, H., Xu, Z., Taylor, G., et al.\ 2017, arXiv:1712.09913. doi:10.48550/arXiv.1712.09913
 
\bibitem{pytorch-geometric} Fey, M. \& Lenssen, J.~E.\ 2019, arXiv:1903.02428. doi:10.48550/arXiv.1903.02428

\bibitem{Adam} Kingma, D.~P. \& Ba, J.\ 2014, arXiv:1412.6980. doi:10.48550/arXiv.1412.6980

\bibitem{optuna_2019} Akiba, T., Sano, S., Yanase, T., et al.\ 2019, arXiv:1907.10902. doi:10.48550/arXiv.1907.10902

\bibitem{Bergstra2011} Bergstra, J., Bardenet, R., Bengio, Y., \& K'egl, B. 2011, in Advances in Neural Information Processing Systems, ed. J. Shawe-Taylor, R. Zemel, P. Bartlett, F. Pereira, \& K. Weinberger, Vol. 24 (Curran Associates, Inc.). https://proceedings.neurips.cc/paper/2011/file/86e8f7ab32cfd12577bc2619bc635690-Paper.pdf

\bibitem{UMAP} Sainburg, T., McInnes, L., \& Gentner, T.~Q.\ 2020, arXiv:2009.12981. doi:10.48550/arXiv.2009.12981

\bibitem{TSNE} van der Maaten, L.J.P.; Hinton, G.E. Visualizing High-Dimensional Data Using t-SNE. Journal of Machine Learning Research 9:2579-2605, 2008.

\bibitem{PCA} Tipping, M. E., and Bishop, C. M. (1999). “Probabilistic principal component analysis”. Journal of the Royal Statistical Society: Series B (Statistical Methodology), 61(3), 611-622.

\bibitem{LOF} Breunig, Markus \& Kröger, Peer \& Ng, Raymond \& Sander, Joerg. (2000). LOF: Identifying Density-Based Local Outliers.. ACM Sigmod Record. 29. 93-104. 10.1145/342009.335388. 

\bibitem{SVM} M. A. Hearst, S. T. Dumais, E. Osuna, J. Platt and B. Scholkopf, "Support vector machines," in IEEE Intelligent Systems and their Applications, vol. 13, no. 4, pp. 18-28, July-Aug. 1998, doi: 10.1109/5254.708428.

\bibitem{NF} Papamakarios, G., Nalisnick, E., Jimenez Rezende, D., et al.\ 2019, arXiv:1912.02762. doi:10.48550/arXiv.1912.02762

\bibitem{Cen1994} Cen, R., Bahcall, N.~A., \& Gramann, M.\ 1994, ApJL, 437, L51. doi:10.1086/187680

\bibitem{kaiser} Kaiser, N.\ 1987, MNRAS, 227, 1. doi:10.1093/mnras/227.1.1

\bibitem{Lai2023} Lai, Y., Howlett, C., \& Davis, T.~M.\ 2023, MNRAS, 518, 1840. doi:10.1093/mnras/stac3252

\bibitem{Howlett2019} Howlett, C.\ 2019, MNRAS, 487, 5209. doi:10.1093/mnras/stz1403

\bibitem{Berlind2003} Berlind, A.~A., Weinberg, D.~H., Benson, A.~J., et al.\ 2003, ApJ, 593, 1. doi:10.1086/376517


%%%%%%%%%%%%
%%%%%%%%%%%%
%%%%%%%%%%%%

\bibitem{Donnari2019} Donnari, M., Pillepich, A., Nelson, D., et al.\ 2019, MNRAS, 485, 4817. doi:10.1093/mnras/stz712

\bibitem{summarizeSAM-2015} Somerville, R.~S. \& Dav{\'e}, R.\ 2015, ARAA, 53, 51. doi:10.1146/annurev-astro-082812-140951

\bibitem{Wang2023} Wang, B., Leja, J., Villar, V.~A., et al.\ 2023, ApJl, 952, L10. doi:10.3847/2041-8213/ace361

\bibitem{Knobel2013} Knobel, C., Lilly, S.~J., Kova{\v{c}}, K., et al.\ 2013, ApJ, 769, 24. doi:10.1088/0004-637X/769/1/24

\bibitem{Sales2015} Sales, L.~V., Vogelsberger, M., Genel, S., et al.\ 2015, MNRAS, 447, L6. doi:10.1093/mnrasl/slu173

\bibitem{color_Illustris-2018} Nelson, D., Pillepich, A., Springel, V., et al.\ 2018, MNRAS, 475, 624. doi:10.1093/mnras/stx3040

\bibitem{Nicola2022} Nicola, A., Villaescusa-Navarro, F., Spergel, D.~N., et al.\ 2022, JCAP, 2022, 046. doi:10.1088/1475-7516/2022/04/046

\bibitem{Pillepich2018MNRAS} Pillepich, A., Springel, V., Nelson, D., et al.\ 2018, MNRAS, 473, 4077. doi:10.1093/mnras/stx2656

\bibitem{Weinberger2017} Weinberger, R., Springel, V., Hernquist, L., et al.\ 2017, MNRAS, 465, 3291. doi:10.1093/mnras/stw2944

\bibitem{Marinacci2018} Marinacci, F., Vogelsberger, M., Pakmor, R., et al.\ 2018, MNRAS, 480, 5113. doi:10.1093/mnras/sty2206

\bibitem{Naiman2018} Naiman, J.~P., Pillepich, A., Springel, V., et al.\ 2018, MNRAS, 477, 1206. doi:10.1093/mnras/sty618

\bibitem{Nelson2018} Nelson, D., Pillepich, A., Springel, V., et al.\ 2018, MNRAS, 475, 624. doi:10.1093/mnras/stx3040

\bibitem{Springel2018} Springel, V., Pakmor, R., Pillepich, A., et al.\ 2018, MNRAS, 475, 676. doi:10.1093/mnras/stx3304

\bibitem{Nelson2019} Nelson, D., Springel, V., Pillepich, A., et al.\ 2019, Computational Astrophysics and Cosmology, 6, 2. doi:10.1186/s40668-019-0028-x

\bibitem{Borrow2022} Borrow, J., Schaller, M., Bower, R.~G., et al.\ 2022, MNRAS, 511, 2367. doi:10.1093/mnras/stab3166

\bibitem{Joop-Daalen2014} van Daalen, M.~P., Schaye, J., McCarthy, I.~G., et al.\ 2014, MNRAS, 440, 2997. doi:10.1093/mnras/stu482

\bibitem{Joop-Hellwing2016} Hellwing, W.~A., Schaller, M., Frenk, C.~S., et al.\ 2016, MNRAS, 461, L11. doi:10.1093/mnrasl/slw081

\bibitem{Joop-Velliscig2014} Velliscig, M., van Daalen, M.~P., Schaye, J., et al.\ 2014, MNRAS, 442, 2641. doi:10.1093/mnras/stu1044

\bibitem{Li2014} Li, Y., Hu, W., \& Takada, M.\ 2014, PrD, 89, 083519. doi:10.1103/PhysRevD.89.083519

\bibitem{Quijote_sims} Villaescusa-Navarro, F., Hahn, C., Massara, E., et al.\ 2020, ApJS, 250, 2. doi:10.3847/1538-4365/ab9d82

\bibitem{Hamilton2006} Hamilton, A.~J.~S., Rimes, C.~D., \& Scoccimarro, R.\ 2006, MNRAS, 371, 1188. doi:10.1111/j.1365-2966.2006.10709.x

\bibitem{Hu2003} Hu, W. \& Kravtsov, A.~V.\ 2003, ApJ, 584, 702. doi:10.1086/345846

\bibitem{biwei2023} Dai, B. \& Seljak, U.\ 2023, Machine Learning for Astrophysics, 10. doi:10.48550/arXiv.2306.04689


% Please avoid comments such as "For a review'', "For some examples",
% "and references therein" or move them in the text. In general,
% please leave only references in the bibliography and move all
% accessory text in footnotes.
% Also, please have only one work for each \bibitem.


\end{thebibliography}
\end{document}